\documentclass[11pt]{article}
\usepackage{jheppub}
\usepackage{graphicx} 
\usepackage{graphbox}
\usepackage[usenames,dvipsnames]{xcolor} 
\usepackage{tikz}	
\usepackage{dsfont}
\usepackage{amsthm}
\usepackage{float}
\usepackage[normalem]{ulem}
\usepackage{slashed}
\usepackage{mathtools}
\usepackage{cancel}
\usepackage{orcidlink}
\usepackage{enumerate}
\usepackage{fontawesome}
\usepackage{adjustbox}
\usepackage{tcolorbox}
\usepackage{shuffle}

\usetikzlibrary{arrows.meta}

\theoremstyle{definition}

\newcommand{\rd}{\mathrm{d}}
\newcommand{\sv}{\text{sv}}

\newcommand{\mm}{\mathfrak{m}}
\newcommand{\dr}{\mathfrak{dr}}

\newcommand{\ep}{\epsilon}

\newcommand{\nn}{\nonumber}

\newcommand{\dd}{\text{d}}
\newcommand{\mot}{\mathfrak{m}}
\newcommand{\BF}{{\rm BF}}

\newcommand{\MMV}[2]{m\! \left[\begin{smallmatrix}#1\\#2\end{smallmatrix}\right]}

\newcommand{\ee}[3]{{\cal E}\! \left[\begin{smallmatrix}#1\\#2\end{smallmatrix};#3\right]}
\newcommand{\eem}[3]{{\cal E^\mm}\! \left[\begin{smallmatrix}#1\\#2\end{smallmatrix};#3\right]}
\newcommand{\eedr}[3]{{\cal E^\dr}\! \left[\begin{smallmatrix}#1\\#2\end{smallmatrix};#3\right]}

\newcommand{\stf}[3]{{\cal X}^{\mm}\! \left[\begin{smallmatrix}#1\\#2\end{smallmatrix};#3\right]}
\newcommand{\stfdr}[3]{{\cal X}^{\dr}\! \left[\begin{smallmatrix}#1\\#2\end{smallmatrix};#3\right]}
\newcommand{\nuker}[3]{\nu\! \left[\begin{smallmatrix}#1\\#2\end{smallmatrix};#3\right]}

\newcommand{\GG}{ {\rm G} }

\newcommand{\Ip}{\mathbb{I}}

\newcommand{\Ieqv}{\mathbb{I}^{\rm eqv}}

\newcommand{\sigmaT}{\sigma}

\newcommand{\csig}{\xi}

\newcommand{\Pexp}{{\rm P\text{-}exp}}

\newcommand{\ant}{{\cal A}}

\title{Towards Motivic Coactions at Genus One from Zeta Generators}
\abstract{}
\author[a,b
]{Axel Kleinschmidt,}
\emailAdd{axel.kleinschmidt@aei.mpg.de}
\affiliation[a]{Max-Planck-Institut f\"ur Gravitationsphysik (Albert-Einstein-Institut) Am M\"uhlenberg 1, 14476 Potsdam, Germany}
\affiliation[b]{International Solvay Institutes, ULB-Campus Plaine CP231, 1050 Brussels, Belgium}

\author[c,d 
]{Franziska Porkert,}
\emailAdd{fporkert@uni-bonn.de}
\affiliation[c]{Bethe Center for Theoretical Physics, Universit\"at Bonn, D-53115, Germany
}
\affiliation[d]{Nordita, Stockholm University and KTH Royal Institute of Technology, Hannes Alfv\'ens v\"ag 12, 10691 Stockholm, Sweden}

\author[e,f 
]{Oliver Schlotterer}
\emailAdd{oliver.schlotterer@physics.uu.se}
\affiliation[e]{Department of Physics and Astronomy, Uppsala University, Box 516, 75120 Uppsala, Sweden}
\affiliation[f]{Department of Mathematics, Centre for Geometry and Physics, Uppsala University, Box 480, 75106 Uppsala, Sweden}

\abstract{The motivic coaction of multiple zeta values and multiple polylogarithms encodes both structural insights on and computational methods for scattering amplitudes in a variety of quantum field theories and in string theory. In this work, we propose coaction formulae for iterated integrals over holomorphic Eisenstein series that arise from configuration-space integrals at genus one. Our proposal is motivated by formal similarities between the motivic coaction and the single-valued map of multiple polylogarithms at genus zero that are exposed in their recent reformulations via zeta generators. The genus-one coaction of this work is then proposed by analogies with the construction of single-valued iterated Eisenstein integrals via zeta generators at genus one. We show that our proposal exhibits the expected properties of a coaction and deduce $f$-alphabet decompositions of the multiple modular values obtained from regularized limits.}

\preprint{UUITP--21/25}

\begin{document}

\setcounter{tocdepth}{2}

\maketitle

\newpage

\section{Introduction}
\label{intro}
Scattering amplitudes are central in many areas of high-energy physics such as particle phenomenology, string theory and more recently also gravitational-wave physics and cosmology. The mathematical structures arising in these objects also provide interesting connections to different areas of pure and applied mathematics such as algebraic geometry and number theory. In particular,  perturbative expansions give rise to iterated integrals, e.g.\ in the $\varepsilon$-expansion of dimensionally regulated Feynman integrals or in the low-energy expansion of string amplitudes. To learn more about mathematical structures of the respective theories as well as for improved numerical evaluations in precision physics, it is pressing to obtain a better understanding of these iterated integrals. 

In the simplest but very prominent case\footnote{That means at all orders in the low-energy expansion of string tree-level amplitudes and in a vast number of the relevant terms for precision computations in particle and gravitational-wave physics as well as cosmology.}, these iterated integrals are \textit{multiple polylogarithms} (MPLs) and their special values include \textit{multiple zeta values} (MZVs). They arise from integrating rational functions on the sphere, i.e.\ Riemann surfaces of genus zero. The Hopf-algebra structures of MPLs and MZVs have unravelled interesting mathematical structures of scattering amplitudes.
Specifically, their motivic coaction---extending the closely related symbol calculus---has been widely studied in the physics literature \cite{Goncharov:2010jf, Duhr:2011zq, Duhr:2012fh, Schlotterer:2012ny, Drummond:2013vz, Schnetz:2013hqa, Abreu:2014cla, Panzer:2016snt, Abreu:2017enx, Abreu:2017mtm, Schnetz:2017bko, Caron-Huot:2019bsq, Abreu:2019xep, Tapuskovic:2019cpr, Gurdogan:2020ppd, Britto:2021prf, Dixon:2021tdw, Borinsky:2022lds, Dixon:2022xqh, Dixon:2023kop, Dixon:2025zwj} as a hallmark of symmetries and as a tool to connect and simplify polylogarithmic expressions, see \cite{Abreu:2022mfk} for a review. The connection between the coaction prescriptions used by physicists and the motivic Galois coaction studied by mathematicians \cite{Cartier:1988, Ihara1989TheGR, Goncharov:2001iea, Goncharov:2005sla, Brown:2011ik, BrownTate, Brown:2015fyf} is well understood for MPLs \cite{Brown:2019jng, Tapuskovic:2019cpr}. The coaction is also interconnected with the notion of \textit{single-valued periods} and the \textit{single-valued map}
\cite{svpolylog, Schnetz:2013hqa, Brown:2013gia, Brown:2018omk, Charlton:2021uhu}\footnote{
Throughout this work, the single-valued map is understood to act on motivic periods. In the realm of MZVs we can use a result of Brown~\cite{Brown:2013gia} that motivic MZVs are isomorphic to the so-called $f$-alphabet and, by this isomorphism, we will also denote the corresponding map in the $f$-alphabet by ${\rm sv}$, see (\ref{eq:Svfalphabet}). When composing the single-valued map of this work with the period map, it matches the map ${\rm sv}^{\mathfrak{m}}$ in section 3.2 of \cite{Charlton:2021uhu} and not the alternative map from de Rham periods to motivic periods denoted by ${\rm sv}$ in the reference.}
which found a similar wealth of recent physics applications 
\cite{Dixon:2012yy, Drummond:2012bg, DelDuca:2013lma, Stieberger:2013wea, Stieberger:2014hba, Zerbini:2015rss, DHoker:2015wxz, Broedel:2016kls, DelDuca:2016lad, Broedel:2018izr, Schlotterer:2018abc, Vanhove:2018elu, Brown:2019wna, DelDuca:2019tur, Gerken:2020xfv, Alday:2022xwz, Alday:2023jdk, Fardelli:2023fyq, Duhr:2023bku, Alday:2024ksp, Alday:2024rjs, Baune:2024uwj, Alday:2025bjp, Baune:2025hfu}.

Iterated integrals beyond multiple polylogarithms arise, for instance, in string amplitudes on higher-genus surfaces~\cite{Green:1987mn, DHoker:1988pdl, Polchinski:1998rq, Witten:2012bh} but have also been found in a variety of Feynman integrals beyond one loop,
see \cite{Bourjaily:2022bwx} for an overview as of early 2022. More recent Feynman-integral calculations at the precision frontier of particle physics and gravity require higher-genus Riemann surfaces \cite{Huang:2013kh, Georgoudis:2015hca, Doran:2023yzu, Marzucca:2023gto, Jockers:2024tpc, Duhr:2024uid} as well as higher-dimensional varieties including Calabi-Yau geometries 
\cite{Brown:2010bw, Bloch:2014qca, Bloch:2016izu, adams2018feynmanintegralsiteratedintegrals,Bourjaily:2018ycu, Bourjaily:2018yfy,Broedel_2019, Bourjaily:2019hmc, Duhr_2020,  Klemm:2019dbm, Vergu:2020uur, Bonisch:2020qmm, Bonisch:2021yfw, Pogel:2022yat, Forum:2022lpz,Duhr:2022pch, Pogel:2022ken, Pogel:2022vat, Duhr:2022dxb, Cao:2023tpx, McLeod:2023doa, Duhr:2023eld,Frellesvig:2023bbf,Klemm:2024wtd,Frellesvig:2024zph,Duhr:2024bzt,Driesse:2024feo,Forner:2024ojj, Frellesvig:2024abc,Duhr:2025tdf,Duhr:2025ppd, Duhr:2025lbz,Maggio:2025jel,Brammer:2025rqo, Duhr:2025kkq, Pogel:2025bca}. 

This motivates the extension of the motivic coaction and the single-valued map to iterated integrals beyond the sphere. The next -- and already widely studied -- step in the context of both string theory and Feynman integrals are Riemann surfaces of genus one, i.e.\ elliptic curves. The associated iterated integrals are \textit{elliptic multiple polylogarithms} (eMPLs) \cite{Lev, Levrac, BrownLev, Broedel:2014vla, Broedel:2017kkb,Broedel:2018iwv,Broedel:2018izr,Broedel:2018qkq, Broedel:2019tlz, Enriquez:2023emp} whose special values are known as \textit{elliptic multiple zeta values} (eMZVs) \cite{Enriquez:Emzv, Broedel:2015hia, Matthes:Thesis, Lochak:2020, Matthes:2017, Zerbini:2018sox}.
The literature on the elliptic case featured different paths towards a single-valued map based on ideas from string theory \cite{Broedel:2018izr, Gerken:2018jrq, Zagier:2019eus, Gerken:2020xfv} and algebraic geometry \cite{Brown:2017qwo2, Brown:2018omk, Dorigoni:2024oft}. It is an open problem to connect these two different approaches and to relate them to the recent double-copy formulae at genus one \cite{Stieberger:2022lss, Stieberger:2023nol, Bhardwaj:2023vvm, Mazloumi:2024wys}.

Moreover, there has been significant effort to construct an elliptic coaction \cite{Broedel:2018iwv,Wilhelm:2022wow, Forum:2022lpz}, see e.g.\ \cite{Kristensson:2021ani, McLeod:2023qdf, Spiering:2024sea} for applications in particle physics and in particular \cite{Tapuskovic:2019cpr, Tapuskovic:2023xiu} for mathematically rigorous evaluations of motivic coactions of specific Feynman integrals. However, it is an open problem to connect the mathematical work \cite{Tapuskovic:2019cpr, Tapuskovic:2023xiu} with the coaction prescriptions proposed in the physics literature. A key ingredient at genus zero is the de Rham projection that allows us to relate the different types of periods in the well-established motivic coaction of MPLs. One of the open challenges at genus one is to construct an analogous de Rham projection that operates on suitably defined periods and relates them to integrals. 

In this paper, we want to put forward a coaction prescription for \textit{iterated Eisenstein integrals} that could be a first step to bridge the gap between the physics and mathematics literature and is at the same time made accessible to explicit evaluations. 
Our proposal possesses several properties one would expect from an elliptic coaction, however, we do not have a first-principles definition of the de Rham periods and the de Rham projection. In particular, we leave it as an open question whether our proposal matches the abstract notion of a motivic coaction in algebraic geometry.

Our logic follows a recent unified reformulation of the genus-zero coaction and single-valued map of MPLs~\cite{Frost:2023stm,Frost:2025lre}. More specifically, our proposal for the elliptic coaction relies on
generating series and certain Lie-algebra structures known as \textit{zeta generators} \cite{DG:2005, Brown:depth3} that take center stage in the genus-zero construction: The single-valued map of MPLs is determined by their motivic coaction \cite{Brown:2013gia, DelDuca:2016lad} as indicated in figure \ref{fig.quadrat} and conveniently derived via zeta generators as will be reviewed below.
Moreover, zeta generators can be adapted to Riemann surfaces beyond genus zero, see \cite{EnriquezEllAss, Brown:depth3, hain_matsumoto_2020, Dorigoni:2024iyt} for the relations between their incarnation on the sphere and the torus, respectively. By their flexibility to connect the motivic coaction with the single-valued map and to accommodate different genera, zeta generators are a promising tool to analyze general structures of iterated integrals on Riemann surfaces.

\subsection{Overview of results}

The main target in this work are iterated integrals of holomorphic Eisenstein series ${\rm G}_k(\tau)$ with integration kernels of the form 
\begin{align}
   \tau^j  {\rm G}_k(\tau)\rd \tau \, ,  \text{ with $k=4,6,8,\ldots$  and $j =0,1,\dots, k{-}2$} 
   \label{gkker}
\end{align}
that one could call \textit{Brown's version of iterated Eisenstein integrals} \cite{brown2017multiple}. Specifically, we do not include the extension $ {\rm G}_0=-1$ seen in other approaches to iterated Eisenstein integrals \cite{Enriquez:Emzv, Broedel:2015hia} into our primary definition which is instead accounted for by admitting
nonzero powers $j\leq k{-}2$ of $\tau^j$ in (\ref{gkker}). The integration path for the modular parameter $\tau$ is taken to connect the cusp $\tau \rightarrow i \infty$ with a generic point in the upper half-plane. Endpoint divergences are regularized using the tangential-base-point method of \cite{brown2017multiple}.

Note that the kernels in (\ref{gkker}) only involve holomorphic Eisenstein of $\text{SL}(2,\mathbb{Z})$, and the coaction properties of iterated integrals of SL$(2,\mathbb{Z})$ cusp forms as well as modular forms of congruence subgroups are left for the future. Moreover, we will still need to extend our prescription to elliptic multiple polylogarithms and compare to the literature on their coactions \cite{Broedel:2018iwv, Tapuskovic:2019cpr, Wilhelm:2022wow, Forum:2022lpz, Tapuskovic:2023xiu}.
 
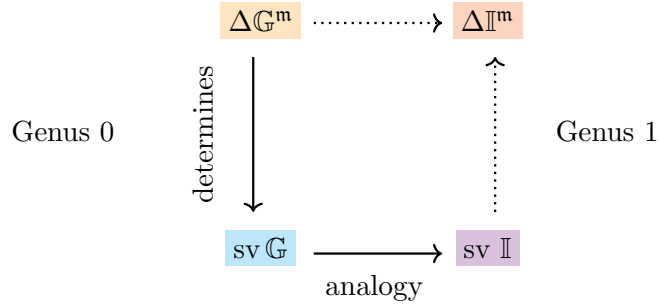
\begin{figure}[H]
\centering
    \begin{tikzpicture}
    \node at (-1.5cm,1.5cm) [above] {\colorbox{YellowOrange!25}{$\Delta \mathbb{G}^{\mathfrak{m}}$}};   
    \node at (0cm,-1.35cm) [below] {analogy};   
 \node at (1.5cm,1.5cm) [above] {\colorbox{RedOrange!25}{$\Delta \mathbb{I}^{\mathfrak{m}}$}};
    \node at (-1.5cm,-1.5cm) [above] {\colorbox{ProcessBlue!25}{$\text{sv} \, \mathbb{G}$}};
    \node at (1.5cm,-1.5cm) [above] {\colorbox{Plum!25}{$\text{sv}\, \,  \mathbb{I}$}};
    \node at (-3.3cm,0.42cm) [left] {Genus 0};
    \node at (3.9cm,0.42cm) [left] {Genus 1};
    \draw[line width=0.3mm,->,dotted] (-0.8cm,1.85cm) -- (0.9cm,1.85cm); 
    \draw[line width=0.3mm,->] (-0.8cm,-1.2cm) -- (0.9cm,-1.2cm); 
    \draw[line width=0.3mm,<-] (-1.6cm,-0.65cm) -- (-1.6cm,1.4cm); 
    \draw[line width=0.3mm,->,dotted] (1.6cm,-0.65cm) -- (1.6cm,1.4cm); 
\node[rotate=90, anchor=south] at (-2cm, 0.375cm) {determines};
\end{tikzpicture}
\caption{\label{fig.quadrat}A pictorial representation of the relations between the known and the proposed coaction and the single-valued map prescriptions for generating series $\mathbb{G}$ and $\mathbb{I}$ of MPLs and iterated Eisenstein integrals. The main result of this work is a proposal for the coaction \colorbox{RedOrange!25}{$\Delta \mathbb{I}^{\mathfrak{m}}$} in the upper-right corner.}
\end{figure}

Our proposed coaction of iterated Eisenstein integrals is inspired by a direct analogy between genus zero and genus one that is illustrated in figure~\ref{fig.quadrat}.  The driving force in these analogies are generating series of zeta generators in their genus-zero and genus-one incarnations. More precisely, we will use explicitly known commutation relations between zeta generators and certain non-commutative variables in the generating series $\mathbb G$ and $\mathbb I$ of MPLs and iterated Eisenstein integrals. The origin of these commutation relations from the action of zeta generators on the fundamental groups of the punctured sphere and the punctured torus is discussed in \cite{EnriquezEllAss, Brown:depth3, hain_matsumoto_2020, Dorigoni:2024iyt}.
Our proposal only concerns those iterated Eisenstein integrals that occur in elliptic multiple zeta values \cite{Enriquez:Emzv, Broedel:2015hia} and are characterised by the fact that the non-commutative variables of the generating series $\mathbb I$ satisfy relations associated with holomorphic cusp forms~\cite{Pollack}.

As already mentioned, the coaction determines the single-valued map \cite{Brown:2013gia, DelDuca:2016lad, Charlton:2021uhu, Frost:2025lre}. An explicit prescription for the generating series \colorbox{Plum!25}{$\text{sv}\, \,  \mathbb{I}$} of single-valued iterated Eisenstein integrals \cite{Brown:2017qwo2, Brown:2018omk} has been put forward in \cite{Dorigoni:2024oft} and it is structurally analogous to the respective prescription \colorbox{ProcessBlue!25}{$\text{sv} \, \mathbb{G}$} for generating series of single-valued MPLs. The close analogy between these two prescriptions for the single-valued map at genus zero and genus one relies on the reformulation \cite{Frost:2023stm, Frost:2025lre} of earlier expressions for \colorbox{ProcessBlue!25}{$\text{sv} \, \mathbb{G}$} \cite{svpolylog, Broedel:2016kls, DelDuca:2016lad} in terms of zeta generators. By adapting the zeta generators in the construction of single-valued MPLs to their genus-one incarnation \cite{Dorigoni:2024iyt}, the composition of generating series in the reformulation of \colorbox{ProcessBlue!25}{$\text{sv} \, \mathbb{G}$} directly carries over to those in \colorbox{Plum!25}{$\text{sv}\, \,  \mathbb{I}$}. 

The essence of our proposed coaction \colorbox{RedOrange!25}{$\Delta \mathbb{I}^{\mathfrak{m}}$} of iterated Eisenstein integrals in (\ref{eq:conj}) relies on a similar analogy between genus zero and genus one, again at the level of generating series. The genus-zero prototype for the prescription of a genus-one coaction is a reformulation \cite{Frost:2023stm, Frost:2025lre} of the Ihara formula \cite{Ihara1989TheGR} for the motivic coaction \colorbox{YellowOrange!25}{$\Delta \mathbb{G}^\mm$} of MPLs in terms of zeta generators. Adapting zeta generators to their genus-one incarnations according to the analogy between \colorbox{ProcessBlue!25}{$\text{sv} \, \mathbb{G}$} and \colorbox{Plum!25}{$\text{sv}\, \,  \mathbb{I}$} translates the composition of generating series in the reformulation of \colorbox{YellowOrange!25}{$\Delta \mathbb{G}^\mm$} into our proposal for \colorbox{RedOrange!25}{$\Delta \mathbb{I}^{\mathfrak{m}}$}.

In all of these four generating-series identities, the appearance of odd zeta values in the coactions and single-valued maps determines the appearance of higher-depth MZVs as well as products of MZVs.
Moreover, all the coactions and single-valued maps of figure~\ref{fig.quadrat} are expressed in terms of bilinears in the generating series of iterated integrals.

In most of this paper, we will analyze the properties of our proposed coaction and find that it meets expectations on its composition with (shuffle) multiplication, derivatives in the modular parameter $\tau$ as well as its limits $\tau \rightarrow i\infty$ and $\tau \rightarrow 0$.

An important aspect of our work will be the implications for so-called multiple modular values (MMVs) that appear as special values of iterated Eisenstein integrals at $\tau = 0$ \cite{brown2017multiple}.
The MMVs under investigation are subject to the same restriction as the iterated Eisenstein integrals in the generating series $\mathbb{I}$ itself through the relation on the non-commutative variables. This implies that they are expressible completely through MZVs and powers of $2\pi i$ with a well-settled motivic coaction \cite{Goncharov:2001iea, Goncharov:2005sla, Brown:2011ik, BrownTate}.
We will find, in agreement with~\cite{brown2017multiple}, that zeta generators fully determine the occurrence of MZVs beyond $\mathbb{Q}[\pi^2]$ in the generating series of MMVs obtained from $\mathbb I$ at $\tau=0$. This reduces the determination of these MMVs to finding a considerably simpler generating series with coefficients in $\mathbb{Q}[\pi^2]$ that is more easily amenable to numerical studies, for instance via PSLQ. Several leading orders of this series defined in~\eqref{ourSm.01} can be found in~\eqref{expxs} with more terms in an ancillary arXiv file accompanying this paper.

\subsection{Outline}

The paper is structured in the following way: In section \ref{sec:2} we review the single-valued and coaction maps at genus zero, zeta generators and our conventions for iterated Eisenstein integrals as well as their modular transformations and single-valued maps. Additionally, we give a very short and conceptual review of the notions of motivic and de Rham periods as well as the motivic coaction, aimed at physicists. Then we present our main proposal, analysing some of its properties in section \ref{sec:3}.
The advertised results on MMVs are contained in section~\ref{sec:4}.
Finally, we study the behavior of our proposed coaction map under modular transformations in section \ref{sec:5} and identify a variant $\Delta^{\rm eqv}$ with improved modular behaviour. In the appendices~\ref{appendixmultiple},~\ref{app_ExamplesEisenstein} and~\ref{appendixapd} we give concrete examples for MMVs, for the proposed coaction of iterated Eisenstein integrals and for their antipodes, respectively. 

\section{Review}
\label{sec:2}

In this section, we review the three known generating series of figure \ref{fig.quadrat} as well as the underlying iterated integrals and algebraic structures. Particular emphasis is placed on a recap of zeta generators at genus zero and genus one as well as motivic and de Rham periods.

\subsection{Genus zero: Multiple polylogarithms}
\label{secgenus0}

We define multiple polylogarithms (MPLs) on the Riemann sphere \cite{GONCHAROV1995197, Goncharov:1998kja, Remiddi:1999ew, Goncharov:2001iea, Vollinga:2004sn}\footnote{For a comprehensive review on conventions for MPLs used in different contexts, see e.g.\ \cite{Duhr:2014woa}.} in conventions where
\begin{align}
\label{eq:MPL}
    G(a_1,\dots, a_w;z)=\int_0^{z} \frac{\rd t}{t-a_1} G(a_2,\dots, a_w; t)\, , 
\end{align}
with labels $a_1,\dots, a_w\in \mathbb{C}$, argument $z\in\mathbb{C}$, weight $w\in \mathbb{N}_0$ and $G(\emptyset;z)=1$.\footnote{We refer to the set of non-negative and positive integers by $\mathbb{N}_0$ and $\mathbb{N}$, respectively.} 
Throughout this work, we  restrict the labels to $a_i \in \{0,1\}$ and speak of \textit{MPLs in one variable}.
Evaluating these functions at $z=1$ gives rise to multiple zeta values  (MZVs)
\begin{align}
    \zeta_{n_1,n_2,\dots, n_r}&=\sum_{0<k_1<\dots <k_r}k_1^{-n_1}k_2^{-n_2}\dots k_r^{-n_r}
    \label{defmzvs}\\
    &=(-1)^r G( \underbracket[0.5pt]{0,\dots, 0}_{n_r-1},1,\dots,\underbracket[0.5pt]{0,\dots, 0}_{n_2-1},1,\underbracket[0.5pt]{0,\dots, 0}_{n_1-1},1;1)
    \notag
\end{align}
with depth $r$ and weight $n_1+\dots+n_r$, where $n_i\in\mathbb{N}$ and $n_r\geq 2$. 
We shall review the motivic coaction and single-valued map of multiple polylogarithms $G(a_1,\ldots,a_w;z)$ in one variable $z$ at the level of their generating series
\begin{align}
\mathbb{G}(e_i;z) &=
 \Pexp\bigg[ {-}\int^0_z \dd t\, \bigg( \frac{e_0}{t}+\frac{e_1}{t{-}1} \bigg)  \bigg]  \label{genpoly}\\
 &= \sum_{w=0}^\infty 
\sum_{a_1,\ldots,a_w =0,1} e_{a_1} e_{a_2} \ldots e_{a_w} G(a_w,\ldots,a_2,a_1;z)
 \notag
\end{align}
which solves the Knizhnik--Zamolodchikov (KZ) equation involving non-commuting variables $e_0, e_1$ (more details on these can be found in \cite{Frost:2023stm} with the notation $\mathbb{G}_{\{0,1\}}(z)$ in the place of $\mathbb{G}(z)$). Throughout this work, our conventions for path-ordered exponentials of Lie-algebra valued one-forms $J(t)$ are
\begin{align}
\Pexp\bigg[ \int^b_a 
J(t) \bigg] = 1 + \int^b_a 
J(t) + \sum_{k=2}^{\infty} \int^b_a 
J(t_1) \int^{t_1}_{a} 
J(t_2) \ldots  \int^{t_{k-1}}_a J(t_k) 
\label{defPOE}
\end{align}
leading to a left-multiplicative factor of $J(b)$ (right-multiplicative factor of $-J(a)$) upon $b$- and $a$-derivative.
The iterated integrals in the expansion of the path-ordered exponential in (\ref{genpoly}) exhibit endpoint divergences which we shall shuffle-regularize with the choices
\begin{align}
G(0;z)=\log(z) \, , \ \ \ \ 
G(z;z)=-\log(z)
\label{regwt1}
\end{align}
at weight one, see \cite{Deligne1989TheGR, Panzer:2015ida, Abreu:2022mfk} for further details, generalizations to higher weights and some background on tangential basepoints.

\subsubsection{$f$-alphabet and coaction / single-valued map on MZVs}

Both multiple polylogarithms and MZVs admit incarnations as motivic periods $G^\mm,\zeta^\mm$ and de Rham periods
$G^\dr,\zeta^\dr$ \cite{Goncharov:2005sla, BrownTate, Brown2014MotivicPA, brown2017notes}. As a convenient way of automatically incorporating all $\mathbb Q$-relations among motivic MZVs, we shall represent them in the $f$-alphabet \cite{Brown:2011ik,BrownTate}.
The $f$-alphabet consists of non-commuting letters $f_{2k+1}$ for $k\in \mathbb{N}$ together with a single commuting letter $f_2$. The non-commuting letters form a Hopf algebra (under shuffle multiplication and the deconcatenation coproduct) and the whole $f$-alphabet is a Hopf algebra comodule of this Hopf algebra. 
The translation between the $f$-alphabet and the zeta values in (\ref{defmzvs}) is achieved by a map $\rho$ from motivic MZVs to the $f$-alphabet.\footnote{This map is not unique and its ambiguities are well-understood as being associated with a basis of irreducible MZVs~\cite{Brown:2011ik,BrownTate}. A canonical choice is described in~\cite{Dorigoni:2024iyt}.} It obeys the following normalisation condition at depth one,
\begin{align}
    \rho(\zeta^\mm_{2k+1}) =f_{2k+1} \text{ and } \rho(\zeta_2^\mm)=f_2\, , 
    \label{normrho}
\end{align}
and is an isomorphism in the motivic realm in the following sense:
The map $\rho$ is taken to be compatible with the various algebraic relations among motivic MZVs over $\mathbb Q$  and their Hopf-algebra-comodule structure. For instance, $\rho$ translates multiplication of MZVs into shuffle products in the $f$-alphabet, e.g~(with $k,\ell \in \mathbb{N}$)
\begin{align}
\rho( \zeta^\mm_{2k+1} \zeta^\mm_{2\ell+1}) = \rho( \zeta^\mm_{2k+1}) \shuffle \rho( \zeta^\mm_{2\ell+1}) = f_{2k+1} \shuffle f_{2\ell+1} = f_{2k+1} f_{2\ell+1} + f_{2\ell+1} f_{2k+1} \, ,
\end{align}and maps the Goncharov--Brown coaction of MZVs \cite{Goncharov:2001iea, Goncharov:2005sla, Brown:2011ik, BrownTate} into a simple deconcatenation formula in the $f$-alphabet (with $n,r \in \mathbb{N}_0$ and $i_1,\ldots,i_r \in 2\mathbb N{+}1$): 
\begin{align}
\label{eq:fcoac}
   \Delta \left(f_2^n f_{i_1}\dots f_{i_r}\right) = \sum_{j=0}^r \left(f_2^n f_{i_1} \dots f_{i_j}\right)\otimes \left(f_{i_{j+1}}\dots f_{i_r}\right)
   \, . 
\end{align}
We will refer to the left entry of the tensor product result of this coaction as the motivic side and the right entry as the de Rham side in analogy with the Goncharov--Brown coaction on (motivic) MZVs. The de Rham version of the MZVs is obtained from the motivic MZVs by setting $\zeta_2^\mot=0$ which translates into $f_2=0$ under $\rho$ and explains why only the first entry on the right-hand side of (\ref{eq:fcoac}) features factors of $f_2$. In absence of $f_2$, the deconcatenation formula (\ref{eq:fcoac}) is the $\rho$ image of the coproduct of the Hopf algebra (over $\mathbb{Q}$) of de Rham MZVs.


The single-valued map of motivic MZVs takes the following simple form when translated into the $f$-alphabet through the $\rho$ isomorphism \cite{Schnetz:2013hqa, Brown:2013gia} ($n,r \in \mathbb{N}_0$ and $i_1,\ldots,i_r \in 2\mathbb N{+}1$)
\begin{align}
\label{eq:Svfalphabet}
    \sv (f_2^n f_{i_1}\dots f_{i_r}) =\delta_{n,0} \sum_{j=0}^r f_{i_j}\dots f_{i_2}f_{i_1} \shuffle f_{i_{j+1}}\dots f_{i_r}
    \,.
\end{align}
The underlying single-valued map of motivic MZVs is then given by the composition $\rho^{-1} \circ \sv \circ \rho$. 
Even though the single-valued map is only well-defined in a motivic setting, we will refer to the outcome of applying the period map to the image of $\rho^{-1} \circ \sv \circ \rho$ as single-valued MZVs for the reasons in the next two paragraphs.


In this paper, we will use the conjecture that the motivic MZVs are in bijection with actual multiple zeta values arising from multiple polylogarithms. This conjecture is discussed for instance in~\cite{BrownTate} and implies in particular that the $\mathbb{Q}$-relations between MZVs are identical to those of motivic MZVs. The former have been studied in detail in~\cite{Blumlein:2009cf, BurgosFresan} and the fact that for instance $\zeta_{3,5}$ can be taken as the only primitive irreducible MZV at transcendental weight 8 follows in the $f$-alphabet since the words $f_3f_5$ and $f_5f_3$ are related by the shuffle $f_3\shuffle f_5$, creating a relation between them.

Using this standard conjecture, we can also define a single-valued map for MZVs which we denote by slight abuse of notation by the same symbol $\sv$ as in (\ref{eq:Svfalphabet}) with for example
$\sv(\zeta_{2n})=0$ or $\sv(\zeta_{2k+1})=2 \zeta_{2k+1}$.

\subsubsection{Generating series in zeta generators}
\label{mwgzero}

The main results of this work are phrased in terms non-commuting zeta generators labelled by odd numbers $w\in 2\mathbb{N}{+}1$ that form a free Lie algebra. Their genus-zero incarnations  are denoted by $M_w$ and characterized by their commutation relations in (\ref{mwcoms}) below with the non-commuting variables $e_0,e_1$ of the polylogarithmic generating series (\ref{genpoly}). The zeta generators $M_w$ with $w\in 2\mathbb{N}{+}1$ are packaged into a generating series
\begin{align}
\label{mseries}
    \mathbb M_0^\mot&= \sum_{r=0}^{\infty} \sum_{i_1,\dots, i_r\in 2\mathbb{N}+1} \rho^{-1}(f_{i_1}\dots f_{i_r})M_{i_1}\dots M_{i_r} \\
    & = 1+ \sum_{i_1\in 2\mathbb{N}+1}\zeta_{i_1}^\mot M_{i_1} +\sum_{i_1,i_2\in 2\mathbb{N}+1} \rho^{-1} (f_{i_1}f_{i_2}) M_{i_1}M_{i_2}+\dots \, . \notag
\end{align}
Terms with a single zeta generator have odd Riemann zeta values $\rho^{-1}(f_{i_1})= \zeta_{i_1}^\mot$ as coefficients, and higher orders in $M_w$ involve all MZVs left after discarding the commutative generator $f_2$ in the $f$-alphabet. The inverse series of (\ref{mseries}) with respect to the concatenation of $M_w$ and shuffle multiplication of $f_w$ are obtained by alternating signs and a reversed concatenation~order,
\begin{align}
\label{Minv}
    (\mathbb M_0^\mot)^{-1}&= \sum_{r=0}^{\infty} (-1)^r\sum_{i_1,\dots, i_r\in 2\mathbb{N}+1} \rho^{-1}(f_{i_1} f_{i_2}\dots f_{i_r})M_{i_r}\dots M_{i_2}M_{i_1} \, .
\end{align}
While the series (\ref{mseries}) itself depends on the choice of $f$-alphabet, this will no longer be the case for the coaction formulae and single-valued maps to be reviewed and proposed below.

We can also project the series $\mathbb{M}_0^\mot$ in~(\ref{mseries}) to the Rham version of the MZVs and then obtain a corresponding generating series $\mathbb M_0^\dr$.
The coaction of the generating series $\mathbb M_0^\mm$ can be derived from the coaction (\ref{eq:fcoac}) of the composing MZVs and takes the simple form
\begin{align}
    \Delta \mathbb M_0^\mm=\mathbb M_0^\mm \mathbb M_0^\dr\, .
    \label{g0coactM}
\end{align}

Unless indicated otherwise, we will typically suppress the $\otimes$ symbol in the image of the map $\Delta$ and instead write
\begin{align}
\label{mmdrlocation}
    X^\mm = X^\mm \otimes {\bf 1}^\dr \text{ and } X^\dr = {\bf 1}^\mot \otimes X^\dr\, ,
\end{align}
i.e.~the superscripts $\mm$ and $\dr$ will be sufficient to distinguish the left- and right-hand sides of the tensor product.

Similar to the generating series involving motivic and de Rham MZVs, one can define a generating series of single-valued MZVs by applying the single-valued map which acts on the $f$-alphabet via (\ref{eq:Svfalphabet}),
\begin{align}
   {\rm{sv}} \, \mathbb M_0&= \sum_{r=0}^{\infty} \sum_{i_1,\dots, i_r\in 2\mathbb{N}+1} \rho^{-1}\big(\text{sv}\,(f_{i_1}\dots f_{i_r})\big)M_{i_1}\dots M_{i_r}  \label{svmm0} \\
    &= 1+ 2\sum_{i_1\in 2\mathbb{N}+1} \zeta_{i_1} M_{i_1} + 2\sum_{i_1,i_2\in 2\mathbb{N}+1} \zeta_{i_1} \zeta_{i_2} M_{i_1} M_{i_2} + \dots\, .
    \notag
\end{align}
Here, we have avoided cluttering the notation by adding a motivic superscript by relying on the standard conjecture that motivic MZVs are in bijection with MZVs. This conjecture also implies that irreducible MZVs at depth $\geq 2$ enter for the first time in the
coefficients of $M_{i_1} M_{i_2} M_{i_3}$ in the ellipsis of~\eqref{svmm0}.

\subsubsection{Coaction and single-valued map of one-variable polylogarithms}
\label{sec:casv0}

The main results of \cite{Frost:2023stm} are closed-form expressions for the coaction and single-valued versions of the (motivic) generating series $\mathbb{G}(e_i;z)$ in (\ref{genpoly}), i.e.\ for single-valued versions and coactions of multiple polylogarithms. 
In the one-variable case of $G(a_1,\ldots,a_w;z)$ with $a_i \in \{ 0,1\}$, these are explicitly given by
\begin{align}
\Delta\mathbb{G}^{\mathfrak{m}}(e_i;z)&=\left(\mathbb{M}_0^{\mathfrak{dr}}\right)^{-1} \mathbb{G}^{\mathfrak{m}}(e_i;z)\mathbb{M}_0^{\mathfrak{dr}} \mathbb{G}^{\mathfrak{dr}}(e_i;z)\, , \label{1varco}\\
    \text{sv}\, \mathbb{G}(e_i;z)&=(\text{sv}\, \mathbb{M}_0)^{-1}\overline{\mathbb{G}(e_i;z)^T} \left(\text{sv}\, \mathbb{M}_0\right) \mathbb{G} (e_i;z)\, , 
    \label{1varsv}
\end{align}
where the notation
$\mathbb{G}^\mot$ and $\mathbb{G}^\dr$ promotes the
MPLs in the expansion (\ref{genpoly}) to their motivic and de Rham versions. The first line is a reformulation of the Ihara formula \cite{Ihara1989TheGR} and the second one is equivalent to Brown's construction \cite{svpolylog} of single-valued polylogarithms in one variable. Moreover, the transposition $T$ in (\ref{1varsv}) indicates that the letters in the complex conjugate of~\eqref{genpoly} have to be reversed according to $(e_i e_j )^T = e_j e_i$.

Even though this is not manifest at the level of the individual generating series, the right-hand sides of (\ref{1varco}) and (\ref{1varsv}) are power series solely in the variables $e_0,e_1$. This can be seen by expanding the adjoint actions in terms of nested commutators such as
\begin{align}
\left(\mathbb{M}_0^{\mathfrak{dr}}\right)^{-1} \mathbb{G}^{\mathfrak{m}}(e_i;z)\mathbb{M}_0^{\mathfrak{dr}}
&= \sum_{r=0}^\infty \sum_{i_1,\ldots,i_r \in 2\mathbb N + 1} \rho^{-1}(f_{i_1} f_{i_2}\ldots f_{i_r})^{\dr}  \label{adjact} \\
&\quad \times \big[ \big[ \ldots [[   
\mathbb{G}^{\mathfrak{m}}(e_i;z), M_{i_1}] , M_{i_2} ],\ldots \big], M_{i_r} \big]\, .
\notag
\end{align}
The multiple commutators can then be carried out by iterative use of the fact that the $M_w$ normalise the power series in the $e_i$ \cite{Ihara:stable, Furusho2000TheMZ}, i.e.
\begin{align}
[M_w,e_0] = 0 \, , \ \ \ \ [M_w,e_1] = \big[ e_1,g_w(e_0,e_1)\big] \, ,
\label{mwcoms}
\end{align}
where $g_w(e_0,e_1)$ are Lie polynomials in the $e_i$ of degree $w$ determined by the Drinfeld associator~\cite{Drummond:2013vz, Frost:2023stm, Dorigoni:2024iyt}.

Note that both the motivic coaction in (\ref{1varco}) and the single-valued map in (\ref{1varsv}) are independent of the choice of the $f$-alphabet isomorphism $\rho$ (subject to the normalization (\ref{normrho}) and compatible with the Hopf-algebra structure of de Rham MZVs): a change of $\rho^{-1}$ affects the MZVs in (\ref{adjact}) in a way that is compensated by the action of the zeta generators in (\ref{mwcoms}) since the Lie polynomials $g_w(e_0,e_1)$ also depend on $\rho$ through the coefficient of $f_w$ in the $f$-alphabet image of the Drinfeld associator \cite{Frost:2025lre}. Canonical choices of $g_w(e_0,e_1)$ and thereby zeta generators $M_w$ are discussed in \cite{Keilthy, Dorigoni:2024iyt}.

\subsection{Genus one: basics of iterated Eisenstein integrals} 
\label{sec:2.2}

Iterated Eisenstein integrals appear prominently in Feynman integrals associated to a genus-one curve 
(see for instance references in \cite{Bourjaily:2022bwx, Weinzierl:2022}) as well as in string perturbation theory (see for instance references in \cite{Gerken:review, Berkovits:2022ivl}).
In this section, we list our conventions for these integrals and their generating series.

We define holomorphic Eisenstein series of ${\rm SL}(2,\mathbb Z)$ by ($k\geq 4$ even) 
\begin{align}
\label{eisensteinseries}
{\rm G}_k(\tau) &=\sum_{(m,n)\in \mathbb{Z}^2\setminus \{(0,0)\}}\frac{1}{(m\tau{+}n)^k }
= 2\zeta_k + \frac{2(2\pi i)^k }{(k{-}1)!} \sum_{n=1}^\infty 
\sum_{m=1}^\infty  m^{k-1} q^{mn}\, ,
\end{align}
where the modular parameter $\tau$ is in the upper half-plane $\mathbb{H}=\{\tau\in\mathbb{C}\, |\, {\rm Im}(\tau)>0\}$ and $q= e^{2\pi i \tau}$ is in the unit disk, $|q|<1$. We will use the  integration kernels 
\begin{align}
    \nuker{j}{k}{\tau}=\left(2\pi i\right)^{1+j-k} \tau^j  {\rm G}_k(\tau)\rd \tau 
    \label{nukerdef}
\end{align}
for integers $0\leq j \leq k{-}2$ to define the iterated integrals 
\begin{align}
\ee{j_1 &j_2 &\ldots &j_\ell}{k_1 &k_2 &\ldots &k_\ell}{\tau} &= \int_\tau^{i \infty} \nuker{j_\ell}{k_\ell}{\tau_\ell} \cdots \int_{\tau_3}^{i\infty} \nuker{j_2}{k_2}{\tau_2}\int_{\tau_2}^{i\infty} \nuker{j_1}{k_1}{\tau_1}\nn\\*
&=(2\pi i)^{1+j_\ell-k_\ell}
 \int^{i\infty}_{\tau}\tau_\ell^{j_\ell} {\rm G}_{k_\ell}(\tau_\ell) \,
 \ee{j_1 &\ldots &j_{\ell-1}}{k_1  &\ldots &k_{\ell-1}}{\tau_\ell}\, \dd \tau_\ell\,   \label{defiteis}
\end{align}
for $k_i \geq 4$ even and integers
$j_i$ in the range
$0\leq j_i \leq k_i{-}2$.
We refer to the number of integrations $\ell$ as the modular depth of the iterated Eisenstein integral and to $\sum_{i=1}^\ell k_i$ as its degree. 
The endpoint divergence at $\tau_k \rightarrow i\infty$ is regularized through the tangential-base-point prescription as in \cite{brown2017multiple} with the net effect $\int^{i\infty}_{\tau}\tau_k^{j} \dd \tau_k = -\frac{1}{j{+}1} \tau^{j+1}$. The $q$-series expansion of
${\rm G}_k$ in (\ref{eisensteinseries}) therefore translates into similar
$q$-series representations of iterated Eisenstein integrals (\ref{defiteis}) with non-negative powers of $\tau$ as coefficients of $q^n$ for any $n\in \mathbb N_0$ \cite{Broedel:2015hia, Broedel:2018izr}.

\subsubsection{Multiple modular values}
\label{sec:2.2.1}

In the regularized limit of iterated Eisenstein integrals as $\tau \rightarrow 0$, we obtain the multiple modular values (MMVs)\footnote{We use the same normalization convention as the MMVs denoted by $\mathfrak{m}[\ldots]$ in \cite{Dorigoni:2021ngn, Dorigoni:2024oft} but use different fonts $m[\ldots]$ in the notation of (\ref{eq:mmvdef}) to avoid clashes with the superscript of motivic periods $X^\mot$.}
\begin{equation}
\MMV{j_1 &j_2 &\ldots &j_\ell}{k_1 &k_2 &\ldots &k_\ell} = \int_0^{i \infty}  \tau_\ell^{j_\ell} \GG_{k_\ell}(\tau_\ell) \,\dd \tau_\ell\,\int_{\tau_\ell}^{i \infty}  \cdots  \int_{\tau_3}^{i \infty}  \tau_2^{j_2} \GG_{k_2}(\tau_2)\,\dd \tau_2 \int_{\tau_2}^{i \infty}  \tau_1^{j_1} \GG_{k_1}(\tau_1) \,\dd \tau_1 \, ,\label{eq:mmvdef}
\end{equation}
which are also said to have modular depth $\ell$ and degree $\sum_{i=1}^\ell k_i$. The underlying regularized limit amounts to mapping the $\tau_i \rightarrow 0$ regime of the integrand in (\ref{eq:mmvdef}) to the $\tau_i \rightarrow i\infty$ by a modular $S$ transformation as detailed in \cite{brown2017multiple, Dorigoni:2021ngn}. In this setting, the $q$-series representations of iterated Eisenstein integrals can be used for an efficient numerical evaluation of MMVs to high precision.

In the normalisation~\eqref{eq:mmvdef}, combinations of MMVs that evaluate to $\mathbb Q[2\pi i]$-linear combinations of MZV are found to have transcendental weight $\sum_{i=1}^\ell k_i$. This can be seen from the appearance of such MMVs in relating the $\tau \rightarrow i\infty$ asymptotics of $A$- and $B$-elliptic associators through the modular $S$-transformation of iterated Eisenstein integrals \cite{EnriquezEllAss, Broedel:2015hia, Broedel:2018izr}. 
MMVs at modular depth one are explicitly given in terms of Riemann zeta values by
\begin{align}
\MMV{j}{k} = \int_0^{i\infty} \tau_1^{j} \GG_k(\tau_1)\,\dd\tau_1  = \left\{\begin{array}{cl}  \displaystyle
- \frac{2\pi i \zeta_{k{-}1}}{k{-}1} &: \ \text{$j=0 \, $,}\\[2mm] \displaystyle
\frac{2 (-1)^{j{+}1} j! (2\pi i)^{k{-}1{-}j}}{(k{-}1)!} \zeta_{j{+}1}\zeta_{j{+}2{-}k} &: \ \text{$0<j\leq k{-}2 \, , $}
\end{array}\right.
\label{mmvd1}
\end{align}
where the $j=k{-}2$ cases simplify to $\MMV{k-2}{k}= \frac{2\pi i \zeta_{k{-}1}}{k{-}1}$.
Examples of MMVs of modular depth two and three that evaluate to MZVs can be found in appendix \ref{appendixmultiple}. Generic MMVs beyond modular depth one involve a considerably wider set of periods including MZVs and (non-critical) L-values of holomorphic cusp forms, see \cite{Brown2019} and appendix~\ref{appendixmultiple} for examples and \cite{Dorigoni:2024oft} for a discussion of more general periods at modular depth three.
In this work, however, we will only consider combinations of MMVs that reduce to $\mathbb Q[2\pi i]$-linear combinations of MZVs. This subclass of MMVs are in principle genus zero objects, that  can therefore be lifted to their motivic and de Rham versions, using the motivic and de Rham versions of the appearing (multiple) zeta values, also see \cite{saad2020multiple} for a discussion in the light of motivic iterated Eisenstein integrals. 

\subsubsection{Generating series for iterated Eisenstein integrals}

A generating series whose coefficients are iterated Eisenstein integrals 
is given by~\cite{Dorigoni:2024oft}
\begin{equation}
\label{eq:GenSerinA}
\Ip(\epsilon_k;\tau) = \Pexp \left[\int_{\tau}^{i\infty} \mathbb{A}(\epsilon_k;\tau_1)\right]\, ,
\end{equation}
where the conventions for path-ordered exponentials are specified in (\ref{defPOE}) and the connection one-form is given in terms of the kernels $\nuker{j}{k}{\tau}$ of (\ref{nukerdef}) by\footnote{Note that $\Ip(\epsilon_k;\tau) $ and $ \mathbb{A} (\epsilon_k;\tau)$ match the objects denoted by $\Ip_+(\epsilon_k;\tau) $ and $ \mathbb{A}_+ (\epsilon_k;\tau)$ in \cite{Dorigoni:2024oft}.}
\begin{align}
\label{eq:DefineA}
    \mathbb{A} (\epsilon_k;\tau)&=\sum_{k=4}^{\infty} \sum_{j=0}^{k-2} (-1)^j \frac{(k{-}1)}{j!}   \nuker{j}{k}{\tau}\epsilon_k^{(j)}  \, .
\end{align}
Explicitly, the expansion of the path-ordered exponential amounts to 
\begin{align}
\label{eq:II}
\Ip(\epsilon_k;\tau) &= 1 
+ \sum_{k_1=4}^{\infty}\sum_{j_1=0}^{k_1-2} (-1)^{j_1} \frac{(k_1{-}1)}{j_1!}
\ee{j_1}{k_1}{\tau} \epsilon_{k_1}^{(j_1)} \\
& \quad + \sum_{k_1=4}^{\infty}\sum_{j_1=0}^{k_1-2}\sum_{k_2=4}^{\infty}\sum_{j_2=0}^{k_2-2} (-1)^{j_1+j_2} \frac{(k_1{-}1)(k_2{-}1)}{j_1!j_2!}
\ee{j_1 & j_2}{k_1 & k_2}{\tau} \epsilon_{k_1}^{(j_1)}\epsilon_{k_2}^{(j_2)}+\dots  \notag
\end{align}
with iterated Eisenstein integrals (\ref{defiteis}) of modular depth $\geq 3$ in the ellipsis and
$\tau$  derivative
\begin{equation}
\label{eq:derivativeIp}
    \partial_{\tau} \Ip(\epsilon_k;\tau) \rd \tau = -\Ip(\epsilon_k;\tau) \mathbb{A}(\epsilon_k;\tau)\, . 
\end{equation}
The non-commuting variables (or \textit{letters})
\begin{align}
   \epsilon_k^{(j)}  = {\rm ad}_{\epsilon_0}^j(\epsilon_k)
   \label{adep0eq}
\end{align}
belong to Tsunogai's derivation algebra with generators $\epsilon_{0},\epsilon_{2},\epsilon_{4},\ldots$ which has been studied from a multitude of perspectives in the mathematics literature \cite{tsunogai1,tsunogai2,tsunogai3,tsunogai4,tsunogai5,tsunogai6,tsunogai7,Broedel:2015hia,Pollack,tsunogai11,Hain,Brown:depth3,tsunogai12,hain_matsumoto_2020}.  
Following the accompanying iterated Eisenstein integrals in (\ref{eq:II}), the derivations $\epsilon_k^{(j)}$ with $k\geq 4$ even are assigned degree~$k$ and modular depth one.\footnote{The derivation $\epsilon_2$ subject to $[\epsilon_2, \epsilon_{2n}]= 0 $ for any $n\in \mathbb N_0$ has been excluded from the generating series (\ref{eq:DefineA}) and won't enter the discussions of this work. Similarly, the derivations $\epsilon_k^{(j)}$ at $ j \geq k{-}1$ which are absent from the connection form (\ref{eq:DefineA}) actually vanish by virtue of $\epsilon_k^{(k-1)} = 0$.}

The Tsunogai derivations in the connection form (\ref{eq:DefineA}) fulfil commutation relations of homogeneous degree, that we will call \textit{Pollack relations} \cite{LNT, Pollack, Broedel:2015hia}. 
These appear starting from degree $14$ and the simplest ones are given by 
\begin{align}
  0 &=   [\epsilon_4,\epsilon_{10}] - 3[\epsilon_6,\epsilon_8]\, ,
 \label{pollrels} \\
0 &= 80 [\ep_4^{(1)}, \ep_{12} ] + 16 [\ep_{12}^{(1)},\ep_4]
- 250 [\ep_6^{(1)},\ep_{10}] -125 [\ep_{10}^{(1)},\ep_6] + 280 [\ep_8^{(1)},\ep_8] \notag \\
&\quad
- 462[\ep_4, [\ep_4,\ep_8]] - 1725 [\ep_6,[\ep_6,\ep_4]] \, .\notag 
\end{align}
By analogy with the accompanying iterated Eisenstein integrals in (\ref{eq:II}), concatenation products $\epsilon_{k_1}^{(j_1)} \epsilon_{k_2}^{(j_2)}\ldots \epsilon_{k_\ell}^{(j_\ell)}$ are said to have modular depth $\ell$. As illustrated by the last two lines of (\ref{pollrels}), generic Pollack relations mix different modular depths. More details on the letters $\epsilon_k^{(j)}$ and the Tsunogai derivations can be found for instance in section 2.3 of \cite{Dorigoni:2024oft}. 
The occurrence of relations of the type~\eqref{pollrels} are known to be associated with holomorphic cusp forms for $\mathrm{SL}(2,\mathbb{Z})$~\cite{Pollack}.

\subsubsection{Zeta generators at genus one}
\label{zwsec}

In preparation for constructing modular transformations, the single-valued map and our conjectured coaction for iterated Eisenstein integrals, we also use versions of the MZV series of (\ref{mseries}) albeit with different sets of generators $\sigmaT_w$ (see~\cite{hain_matsumoto_2020}) accompanying the motivic MZVs that replace the $M_w$, i.e.
\begin{align}
\mathbb M_\sigma^\mot = \mathbb M_0^\mot \, \big|_{M_w \rightarrow \sigma_w} =  \sum_{r=0}^{\infty} \sum_{i_1,\dots, i_r\in 2\mathbb{N}+1} \rho^{-1} (f_{i_1}\dots f_{i_r})\sigma_{i_1}\dots \sigma_{i_r} \, .
\label{msigser}
\end{align}
A de Rham version $\mathbb{M}_\sigma^\dr$ is defined in the same way as in section~\ref{mwgzero} by projecting the motivic MZVs to their de Rham version.
The genus-one zeta generators  $\sigmaT_w$ in turn consist of an arithmetic part $z_w$ and a geometric part $\sigmaT_w^{\rm g} = \sigma_w{-} z_w$ that is  a Lie series in Tsunogai's derivation algebra. Structurally, the generators $\sigmaT_w$ take the form 
\begin{align}
\label{extintr.10}
\sigma_w &= z_w - \frac{1}{(w{-}1)!}\ep_{w+1}^{(w-1)} 
-\frac{1}{2}\sum_{d=3}^{w-2} \frac{\BF_{d-1}}{\BF_{w-d+2}} \sum_{k=d+1}^{w-1} \BF_{k-d+1}\BF_{w-k+1}\BF_{w-k+1} s_{k,w-k+d}^{d} \notag  \\
&\quad -\sum_{d=5}^{w} \BF_{d-1}s_{d-1,w+1}^{d} -\frac{1}{2}\BF_{w+1}s_{w+1,w+1}^{w+2} \notag \\
&\quad+\sum_{k=w+3}^\infty \BF_k \sum_{j=0}^{w-2} \frac{(-1)^j\left(\begin{smallmatrix} k-2\\ j\end{smallmatrix}\right)^{-1}}{j!(w{-}2{-}j)!} \left[\epsilon_{w+1}^{(w{-}2{-}j)},\epsilon_k^{(j)} \right]+\dots\, ,
\end{align}
where the ubiquitous ratios of 
Bernoulli numbers $B_k$ and the respective factorials are denoted~by
\begin{align}
    \BF_k= B_k/k!
    \label{defbf}
\end{align}
and
\begin{align}
\label{224}
    s_{p,q}^{d}= \frac{(d{-}2)!}{(p{-}2)!(q{-}2)!} \sum_{i=0}^{d-2} (-1)^{i} \left[\epsilon_p^{(p-2-i)}, \epsilon_{q}^{(q-d+i)}\right]\, 
\end{align}
is related to projecting on highest-weight vectors in certain $\mathfrak{sl}(2)$ tensor products\footnote{\label{s2foot}The derivations
$\{ \epsilon_k^{(j)} , \ j=0,1,\ldots,k{-}2 \}$ at fixed even $k\geq 2$ form $(k{-}1)$-dimensional
representations of the $\mathfrak{sl}(2)$ algebra spanned by the raising operator $\epsilon_0$ subject to $[\epsilon_0,\epsilon_k^{(j)}] = \epsilon_k^{(j+1)}$ and the lowering operator $\epsilon_0^\vee$ subject to
$[\epsilon_0^\vee,\epsilon_k^{(j)}] = j(k{-}1{-}j) \epsilon_k^{(j-1)}$.}.
As detailed for instance in \cite{Dorigoni:2024oft, Dorigoni:2024iyt}, the ellipsis in (\ref{extintr.10}) refers to an infinite tower of nested brackets of $\ep_k^{(j)}$  of modular depth $\geq 3$.

The arithmetic parts $z_w$ of the zeta generators $\sigma_w$ in (\ref{extintr.10}) commute with $\epsilon_0$ and $\epsilon_0^\vee$ because $z_w$ is a singlet under the $\mathfrak{sl}(2)$ algebra defined in footnote \ref{s2foot} as discussed in \cite{hain_matsumoto_2020, Dorigoni:2024iyt}. Moreover, the adjoint action of $z_w$ normalises series in $\epsilon_k^{(j)}$, i.e.\ commutators $[z_w,\epsilon_k^{(j)}]$ are expressible in terms of nested brackets of $\epsilon_{m}^{(n)}$, for instance \cite{hain_matsumoto_2020}
\begin{align}
[z_w,\ep_k] &= \frac{{\rm BF}_{w+k-1} }{{\rm BF}_k \, (w{+}k{-}3)!} \sum_{i=0}^{w-1} (-1)^i \frac{(k{+}i{-}2)!}{i!} [\ep_{w+1}^{(i)} , \ep_{k+w-1}^{(w-i-1)} ] + \ldots
\label{zepcom}    
\end{align}
with terms of modular depth $\geq 3$ in the ellipsis (see section 7.4 of \cite{Dorigoni:2024iyt} for partial results
at modular depth three).
Accordingly, the brackets $[\sigma_w,\epsilon_k^{(j)}]$ involving the full zeta generators are also expressible via Lie polynomials in $\epsilon_{m}^{(n)}$ as will be crucially used below. 

The $\mathfrak{sl}(2)$ singlet property of the arithmetic parts $z_w$ still leaves ambiguities of redefining them by $\mathfrak{sl}(2)$-invariant\footnote{Lie polynomials in Tsunogai derivations are $\mathfrak{sl}(2)$-invariant if they commute with both $\epsilon_0$ and the derivation $\epsilon_0^\vee$ subject to $[\epsilon_0^\vee, \epsilon_k^{(j)}] = j(k{-}j{-}1) \epsilon_k^{(j-1)}$. The ambiguities of redefining $z_w$ by such $\mathfrak{sl}(2)$  invariants only occur for $w\geq 7$ and Lie polynomials in $\epsilon_k^{(j)}$ of modular depth $\geq 3$. The simplest ambiguity of this type concerns redefinitions of $z_7$ by $\mathfrak{sl}(2)$-invariant combinations of $[\epsilon_4^{(j_1)},[\epsilon_4^{(j_2)}, \epsilon_6^{(j_3)}]]$ with $j_1+j_2+j_3=4$.} Lie polynomials in Tsunogai derivations from the geometric part of~$\sigma_w$. A canonical choice of $z_w$ that resolves all of these ambiguities can be found in Theorem 5.4.1 (vi) of \cite{Dorigoni:2024iyt}.

\subsection{Genus one: Equivariant and single-valued iterated Eisenstein integrals} 
\label{sec:2.3}

In this section, we review a construction of equivariant versions of iterated Eisenstein integrals in terms of generating series, which originate in Brown's work \cite{brown2017multiple, Brown:2017qwo, Brown:2017qwo2} and were made explicit in \cite{Dorigoni_2022, Dorigoni:2024oft}. As a defining property of equivariant iterated Eisenstein integrals, their transformation properties under the modular group ${\rm SL}(2,\mathbb Z)$ are identical to those of the integration kernels $\nuker{j}{k}{\tau}$ in (\ref{nukerdef}). Moreover, they are closely related to Brown's single-valued iterated Eisenstein integrals which furnish a key motivation for our main proprosal.

\subsubsection{$S$-transformation of iterated Eisenstein integrals and MMVs}
\label{sec:2.3.1}

The modular $S$-transformation operates on $\tau$ by $\tau\rightarrow -\frac{1}{\tau}$. It is realized on the generating series $\Ip(\epsilon_k;\tau)$ by the operation $\mathcal{S}$ which acts by 
\cite{brown2017multiple, Dorigoni:2024oft}
\begin{align}
\label{Sfull}
\mathcal{S}\big[\, \Ip\left(\epsilon_k;\tau\right)\big]= \Ip\left(\epsilon_k;-\frac{1}{\tau}\right)&= \mathbb{S}(\epsilon_k) U_S^{-1} \mathbb{I}(\epsilon_k;\tau)U_S\, .
\end{align}
The modular transformation of the integration kernels (\ref{nukerdef}) entering the connection (\ref{eq:DefineA}) are taken into account by the following action of an operator $U_S$ on the letters
\begin{align}
\label{eq:S1}
U_S^{-1}\epsilon_k^{(j)}U_S=(-1)^j (2\pi i)^{k-2-2j} \frac{j!}{(k{-}j{-}2)!} \epsilon_k^{(k-j-2)}\, .
\end{align}
The generating series $\mathbb{S}(\epsilon_k)$ appearing in~\eqref{Sfull} is the tangentially regulated $S$-cocycle
\begin{equation}
 \mathbb{S}(\epsilon_k) = \Pexp \bigg( \int_{0}^{i\infty} \mathbb{A}(\epsilon_k;\tau_1) \bigg)\, .
\label{spluscocyc}
\end{equation} 
Explicitly, it takes the form 
\begin{align}
&\mathbb S(\epsilon_k) = \Ip(e_k;\tau)|_{\tau\rightarrow 0}\label{eq:limitS} \\
&=1 + \sum_{k_1=4}^{\infty}\sum_{j_1=0}^{k_1-2} (-1)^{j_1} \frac{(k_1{-}1)}{j_1!}
( 2\pi i)^{j_1+1-k_1} \MMV{j_1}{k_1} \epsilon_{k_1}^{(j_1)}  \nn
\\
&\quad  + \sum_{k_1=4}^{\infty}\sum_{j_1=0}^{k_1-2}\sum_{k_2=4}^{\infty}\sum_{j_2=0}^{k_2-2} (-1)^{j_1+j_2} \frac{(k_1{-}1)(k_2{-}1)}{j_1!j_2!}
 ( 2\pi i)^{j_1+j_2+2-k_1-k_2} \MMV{j_1 & j_2}{k_1 & k_2}  \epsilon_{k_1}^{(j_1)}{\epsilon_{k_2}^{(j_2)}}+\ldots \notag 
 \end{align}
with the MMVs defined in (\ref{eq:mmvdef}) as coefficients and
terms of modular depth $\geq 3$ in the ellipsis. The expansion coefficients of arbitrary words $\epsilon_{k_1}^{(j_1)} \ldots \epsilon_{k_\ell}^{(j_\ell)}$ in Tsunogai derivations are real since the same is true for the connection form $\mathbb{A}(\epsilon_k;\tau_1)$ in (\ref{eq:DefineA}) on the integration path $(0,i\infty)$ for $\tau_1$.

Since the letters $\epsilon_k^{(j)}$ of the generating series $\mathbb S(\epsilon_k)$ satisfy the Pollack relations, all periods other than MZVs are projected out as can be understood from several perspectives \cite{Enriquez:Emzv, Brown:2017qwo2,saad2020multiple}.
By lifting the zeta values in the MMVs to their motivic versions and matching the conventions for the accompanying bookkeeping variables, the series $\mathbb{S}^\mot(\epsilon_k)$ in \eqref{eq:limitS} can be related to the series $\mathcal{C}_S^\mm$ of \cite{brown2017multiple}, see section \ref{sec:4.1}. 
The $S$-cocycle $\mathcal{C}_S^\mm$  satisfies a number of identities~\cite[\S5]{brown2017multiple} that we will make use of later.

As we will see in section \ref{sec:4}, the $S$-cocycle takes an elegant factorized form (\ref{ourSm.01}) where all the MZVs besides rational polynomials in $2\pi i$ are captured by the series (\ref{msigser}) in zeta generators and which considerably simplifies computations.

\subsubsection{$T$-transformation of iterated Eisenstein integrals}
\label{sec:ttrf}

The modular $T$-transformation operates on $\tau$ by $\tau\rightarrow \tau{+}1$. Its operator-realization $\mathcal{T}$ on the generating series $\Ip(\ep_k;\tau)$ is given by
\cite{brown2017multiple, Dorigoni:2024oft}
\begin{align}
\label{eq:Ttrm}
\mathcal{T}\big[\,\Ip(\ep_k;\tau)\big] =  \Ip(\ep_k;\tau{+}1)&=e^{2\pi i N}\mathbb{I} (\epsilon_k;\tau)U_T \, ,
\end{align}
where the operation $U_T= \exp(2\pi i\, \ep_0)$ on the letters is again determined by modular transformations of the integration kernels (\ref{nukerdef})
\begin{align}
\label{eq:T1}
    U_T^{-1} \epsilon_k^{(j)} U_T = \sum_{p=0}^{k-2-j} \frac{(-2\pi i)^p}{p!} \epsilon_k^{(j+p)}
\end{align}
The analogue $\exp(2\pi i N)$ of the $S$-cocycle (\ref{spluscocyc}) for the $T$-transformation is considerably simpler and expressed in terms of
\cite{brown2017multiple, Dorigoni:2024oft} (see (\ref{defbf}) for the notation ${\rm BF}_k$) 
\begin{align}
\label{defNN}
{N}={N}_+-\epsilon_0\quad\text{ with } \quad {N}_+=\sum_{k=4}^\infty (k{-}1){\rm BF}_k \ep_k \,.
\end{align}

\subsubsection{Equivariant iterated Eisenstein integrals} 
\label{eqvIsec}

The generating series $\mathbb{I}(\epsilon_k;\tau)$ of iterated Eisenstein integrals defined in~\eqref{eq:GenSerinA} can be turned into a generating series of real-analytic functions that transform with definite modular weights under modular transformations as was shown by Brown in~\cite{Brown:2017qwo2}. In the explicit form of~\cite{Dorigoni:2024oft} these \textit{equivariant} iterated Eisenstein integrals arise from\footnote{The generating series $\overline{ \mathbb I(\epsilon_k;\tau)^T}$ of complex conjugate iterated Eisenstein integrals is denoted by $\tilde{\mathbb I}_-$ in \cite{Dorigoni:2024oft}, where the tilde operation reverses words in $\ep_k^{(j)}$.  The kernels in the series $\tilde{\mathbb I}_-$ in the reference depart from the complex conjugate of those in $\mathbb I_+ = \mathbb I$ by alternating signs $\ep_k^{(j)} \rightarrow (-1)^j \ep_k^{(j)}$. Inserting these alternating signs into the word reversal of $\tilde{\mathbb I}_-$ precisely reproduces the action of the $(\ldots)^T$ operation in (\ref{toperation}), that is why (\ref{mainiequv}) and (\ref{mainisv}) match the analogous expressions for $\Ieqv(\epsilon_k;\tau)$ and $\mathbb{I}^{\rm sv}(\epsilon_k;\tau)$ in \cite{Dorigoni:2024oft}.}
\begin{equation}
\Ieqv(\epsilon_k;\tau) = 
(\sv \, \mathbb M_z)^{-1} \,\overline{ \mathbb I(\epsilon_k;\tau)^T}\,
  (\sv \, \mathbb M_\sigma) \, \Ip(\epsilon_k;\tau)
\, .
\label{mainiequv}
\end{equation}
The operation $(\ldots)^T$ acts on the words in derivations $\ep_k^{(j)}$ in the expansion of $\mathbb I(\epsilon_k;\tau)^T$ via
\begin{align}
(\ep_{k_1}^{(j_1)} \ep_{k_2}^{(j_2)}\ldots \ep_{k_r}^{(j_r)} )^T
= (-1)^{j_1+j_2+\ldots+j_r} \ep_{k_r}^{(j_r)} \ldots \ep_{k_2}^{(j_2)}
\ep_{k_1}^{(j_1)} 
    \label{toperation}
\end{align}
which amounts to a reversal of all $\ep_m$ including $m=0$ after exposing all the $\ep_0$ entering the expression (\ref{adep0eq}) for $\ep_k^{(j)}$. The single-valued map of $\sv \, \mathbb M_z$ and $\sv \, \mathbb M_\sigma$ simply acts on the MZVs in the expansion (\ref{msigser}) via (\ref{eq:Svfalphabet}), and the series $\mathbb M_z = \mathbb M_\sigma |_{\sigma_w \rightarrow z_w}$ only retains the arithmetic parts of the genus-one zeta generators. This is crucial to attain the defining property
\begin{align}
\Ieqv \bigg(\epsilon_k;-\frac{1}{\tau}\bigg) = U_S^{-1}\Ieqv(\epsilon_k;\tau) U_S
\, , \ \ \ \ 
\Ieqv(\epsilon_k;\tau{+}1) = U_T^{-1} \Ieqv(\epsilon_k;\tau) U_T
\label{eqv:trf}
\end{align}
of the equivariant series (\ref{mainiequv}) to transform in the same way under ${\rm SL}(2,\mathbb Z)$ as the connection (\ref{eq:DefineA}) does, without any cocycles.
The equivariance property (\ref{eqv:trf}) furthermore relies on the reality $\overline{ \mathbb S(\epsilon_k) } = \mathbb S(\epsilon_k)$ of the $S$-cocycle and its interplay $\mathbb S(\epsilon_k)^T (\sv \, \mathbb M_\sigma)\mathbb S(\epsilon_k) = U_S^{-1} (\sv \, \mathbb M_\sigma) U_S$ with the single-valued series in zeta generators \cite{Brown:2017qwo2, Dorigoni:2024oft}. 
It implies that the coefficients of all independent words $\ep_{k_1}^{(j_1)} \ldots \ep_{k_r}^{(j_r)}$ under Pollack relations can be combined to non-holomorphic modular forms \cite{brown2017multiple, Brown:2017qwo, Brown:2017qwo2} (say through the ${\rm U}_{{\rm SL}_2}(\tau)$-transformation in section 3.1 of \cite{Dorigoni:2024oft}) which reproduce the modular graph forms\footnote{See \cite{Green:2008uj, DHoker:2015gmr, DHoker:2015wxz, DHoker:2016mwo} for the earlier literature and \cite{Gerken:review, Dorigoni:2022iem, DHoker:2022dxx, DHoker:2024book} for overview references on modular graph forms.} in closed-string genus-one amplitudes \cite{Dorigoni_2022}.

Note that we will often write the equivariance condition (\ref{eqv:trf}) in unified form
\begin{align}
\Ieqv (\epsilon_k; \gamma\cdot \tau) = U_\gamma^{-1}\Ieqv(\epsilon_k;\tau) U_\gamma
\label{eqv:uni}
\end{align}
for the full modular group. For $\gamma = \big( \smallmatrix a &b \\ c &d \endsmallmatrix \big) \in {\rm SL}(2,\mathbb Z)$, we use the shorthand $\gamma \cdot \tau = \frac{a\tau + b}{c\tau + d}$, and the SL$(2)$ action of $U_\gamma^{\pm 1}$ on the $\epsilon_k^{(j)}$ of the enclosed series $\Ieqv(\epsilon_k;\tau)$ can be inferred from that of the generators $U_S$ and $U_T$ in (\ref{eq:S1}) and (\ref{eq:T1}), respectively.

\subsubsection{Single-valued iterated Eisenstein integrals} 
\label{svIsec}

Very similarly to (\ref{mainiequv}), one obtains a generating series for single-valued iterated Eisenstein integrals \cite{Brown:2017qwo2, Dorigoni:2024oft} via
\begin{align}
\sv \, \mathbb{I}(\epsilon_k;\tau) = 
  (\sv \, \mathbb M_\sigma)^{-1} \, \overline{ \mathbb I(\epsilon_k;\tau)^T}
\,
  (\sv \, \mathbb M_\sigma) \, \Ip(\epsilon_k;\tau)
\, .
\label{mainisv}
\end{align}
In order to extract the single-valued versions of the 
iterated Eisenstein integrals in (\ref{defiteis}), one needs to expand this composition of generating series in words $\ep_{k_1}^{(j_1)} \ldots \ep_{k_r}^{(j_r)}$ without any separate reference to zeta generators. This can be accomplished by expanding
\begin{align}
(\sv \, \mathbb M_\sigma)^{-1} \, \overline{ \mathbb I(\epsilon_k;\tau)^T}
\,
  (\sv \, \mathbb M_\sigma)
  &= 
  \sum_{r=0}^{\infty} \sum_{i_1,\dots, i_r\in 2\mathbb{N}+1} \rho^{-1}\big( {\rm sv}\,(f_{i_1}\dots f_{i_r}) \big) \label{svconj}\\
  &\quad \times
  \big[ \big[ \ldots [[ \overline{ \mathbb I(\epsilon_k;\tau)^T},  \sigma_{i_1}],\sigma_{i_2}],\ldots ,\sigma_{i_{r-1}} \big],\sigma_{i_r} \big]
\notag
\end{align}
as in (\ref{adjact}) and iteratively converting the brackets $[\sigma_w,\epsilon_k^{(j)}]$ into Lie series in $\epsilon_m^{(n)}$ which is always possible by the arguments in section \ref{zwsec}. Each of the resulting words $\ep_{k_1}^{(j_1)} \ldots \ep_{k_r}^{(j_r)}$ that is independent under Pollack relations such as (\ref{pollrels}) defines a single-valued iterated Eisenstein integral.

Brown's work \cite{Brown:2017qwo2} distinguishes the equivariant iterated Eisenstein integrals in (\ref{mainiequv}) from their single-valued counterparts in (\ref{mainisv}). The latter are unaffected by the ambiguities of $\Ieqv(\epsilon_k;\tau) $ due to $\mathfrak{sl}(2)$-invariant redefinitions of the arithmetic parts $z_w$ of genus-one zeta generators discussed in section \ref{zwsec}. However, since the left-multiplicative series $ (\sv \, \mathbb M_\sigma)^{-1}$ in (\ref{mainisv}) does not commute with the $U_S, U_T$ action in (\ref{eq:S1}) and (\ref{eq:T1}), the series $\sv \, \mathbb{I}(\epsilon_k;\tau)$ does not share the equivariance property (\ref{eqv:uni}).

By arguments similar to those at the end of section \ref{sec:casv0}, $\sv \, \mathbb{I}(\epsilon_k;\tau)$ does not depend on the choice of $f$-alphabet isomorphism: a change of $\rho^{-1}$ in the MZVs of (\ref{svconj}) leads to a compensating modification of the expansion of $\sigma_w$ in Tsunogai derivations which one can ultimately trace back to the coefficient of $f_w$ in the $\rho$-image of the Drinfeld associator \cite{Dorigoni:2024iyt}.

As highlighted in \cite{Dorigoni:2024oft}, the expression (\ref{mainisv}) for the single-valued versions of iterated Eisenstein integrals mirrors the structure of the generating series $\text{sv}\, \mathbb{G}(e_i;z)$ for single-valued MPLs in (\ref{1varsv}) (see also figure \ref{fig.quadrat}). The dictionary between the genus-zero and genus-one formulae amounts to trading MPLs for iterated Eisenstein integrals, the variables $e_0, e_1$ in the expansion (\ref{genpoly}) of $\mathbb{G}(e_i;z)$ for the Tsunogai derivations $\epsilon_k^{(j)}$ in the expansion (\ref{eq:II}) of $\Ip(\epsilon_k;\tau)$ and the genus-zero incarnation $M_w$ of zeta generators for their genus-one counterparts $\sigma_w$.

For the meromorphic MPLs entering $\mathbb{G}(e_i;z)$, we also had a coaction prescription (\ref{1varco}) involving a similar conjugation by zeta generators which matches the Ihara formula for the motivic coaction. Before giving and explaining our proposal for the coaction for iterated Eisenstein, we give a short review on motivic and de Rham periods and the master formula for the motivic coaction.

\subsection{Motivic and de Rham periods for physicists}
\label{sec_motivicderahm}

Here, we give an overview of the notions of \textit{motivic} and \textit{de Rham periods} as well as the \textit{motivic coaction}, whose manifestation in the genus zero case has already appeared in section \ref{secgenus0}. The aim of this section is not to give a complete review, but rather to make the concepts clearer for physicists and specifically to highlight open questions at genus one. For a more in-depth review see \cite{Brown:2013gia, Brown:2015fyf}.

\subsubsection{Generalities and genus zero}

The notion of a \textit{period} commonly used in the physics literature and also above is as a pairing -- \textit{the period pairing} -- between a contour $\Gamma$ (an element of Betti homology) and a differential $\omega$ (an element of algebraic de Rham cohomology) by integration: 
\begin{align}
    P= [ \Gamma , \omega ]=\int_{\Gamma} \omega \, . 
    \label{ppairing}
\end{align}
The period depends only on the equivalence classes and not on the representatives of the (co-)homology groups.

The definition of a \textit{motivic period}, i.e.\ the motivic  version of a period, requires more structure. Starting from an algebraic variety $X$, motivic periods can be realized by writing them as tuples 
\begin{align}
\label{Motivicreal}
    [(H_{\bullet,\rm B}(X), H^{\bullet}_{\text{dR}}(X), c), \Gamma, \omega ]^\mm \, , 
\end{align}
where\footnote{One can also generalise this setup to integrands featuring a multi-valued ``twist'' factor corresponding to an integrable connection $\nabla$ for de Rham cohomology as well as a local system for Betti homology and obtain twisted periods in this way~\cite{Abreu:2022mfk}.}
\begin{itemize}
    \item $H_{\bullet,\rm B}(X)$ is a \textit{Betti homology group} on the space $X$ and $\Gamma$ a representative of some equivalence class of its elements.
    \item $H^{\bullet}_{\text{dR}}(X)$ is an algebraic \textit{de Rham cohomology group} with $\omega$ a representative of some equivalence class. 
    \item $c$ is the \textit{comparison isomorphism} between the groups $H_{\rm B}(X)$ and $H_{\text{dR}}(X)$. In the non-motivic realm, one can think of this isomorphism as the period pairing. 
\end{itemize}
Generally, we do not give all of this information here, i.e.\ do not explicitly specify the \mbox{(co-)}homology groups, but instead write motivic periods as 
\begin{align}
\label{motivicshorthand}
    P^\mm =[\Gamma, \omega]^\mm= \int_\Gamma^\mm \omega\, .  
\end{align}
This is justified by the existence of the period homomorphism, which recovers (families of) $\mathbb C$ numbers from motivic periods according to
\begin{align}
\text{per}\left[P^\mm\right]= P=\int_\Gamma \omega\, . 
\label{permap}
\end{align}
A simple genus zero example, that implicitly appeared already above, is the motivic version of the logarithm: 
\begin{align}
    \text{per}[\log^\mm(x)]=\int_{\Gamma=[1,x]} \frac{\rd t}{t} =\log(x)\, . 
\end{align}
The notion of a \textit{de Rham period} is less straight-forward and more intricate. One can write this, similarly to (\ref{Motivicreal}) as realizations 
\begin{align}
\label{deRhamrealisatuion}
      [(H_{\bullet,\rm dR}(X), H^{\bullet}_{\text{dR}}(X), c), \check{\omega}, \omega ]^\dr \, \text{ or here shortly } [\check{\omega}, \omega]^\dr\,, 
\end{align}
where, in some sense, $\check{\omega}$ is a representative of elements in a dual version of the de Rham cohomology (also called the \textit{de Rham homology}). So instead of pairing Betti and de Rham representatives, we now pair de Rham representatives and dual de Rham representatives. In full generality, however, no natural analogue of a period map is known to relate these representatives to integrals. 

The general version of the \textit{motivic coaction} is given by
\begin{align}
\label{masterformula}
    \Delta_{\text{mot}}[\Gamma, \omega]^\mm= \sum_{e_i} [\Gamma, e_i]^\mm \otimes [\check{e}_i, \omega]^\dr\, , 
\end{align}
where the $e_i$ form a basis of the de Rham cohomology group, $\check{e}_i$ form the dual basis, and the shorthand notations $[\bullet, \bullet]$ are introduced in~\eqref{ppairing} and \eqref{deRhamrealisatuion}. To use \eqref{masterformula} in physics, we want to interpret both sides of the $\otimes$ as iterated integrals (when possible). The motivic coaction \eqref{masterformula} is then loosely translated to the so-called \textit{master formula} \cite{Abreu:2018sat,Abreu:2017mtm}.
The translation of the left-hand side of $\otimes$ into an integral is straightforward due to the period map acting on motivic periods as in \eqref{permap}. The interpretation of the right-hand side as objects useful for physics is more involved:  One can for instance define the so-called \textit{single-valued period map} \cite{Brown:2013gia,Brown:2018omk,Brown2019} which maps a representative $[\check{e}_i, \omega]^\dr$ to a \textit{single-valued period}. But in order to cleanly interpret this single-valued period as the single-valued version of a specific period, i.e.\ a specific integral arising from the period mapping of a motivic period, one needs a prescription that translates between dual de Rham cycles and Betti cycles.

This translation exists for genus zero objects in the form of the so-called {\it de Rham projection} \cite{brown2017notes, Brown:2018omk, Brown:2019wna}.  This identification was also used to obtain canonical forms in the context of positive geometry rigorously~\cite{Brown:2025jjg}. A simple example for the de Rham projection at genus zero is for intervals $[a,b]$ which are mapped to $\rd \log$ forms, interpreted as elements of the de Rham cohomology which are dual to $[a,b]$ as homology elements in the sense of \cite{Schlotterer:2018abc, Brown:2018omk, Brown:2019wna}
\begin{align}
    \Gamma =[a,b]\,\,\longleftrightarrow \,\,\check{\omega}_\Gamma =\rd \log \left(\frac{z-a}{z-b}\right)\, ,
\end{align}
and a simple example for a single-valued period is the single-valued logarithm
\begin{align}
    \log^\sv (z)=\log|z|^2\, . 
\end{align}
 In that sense, all genus-zero objects considered here are on firm ground as motivic and de Rham periods and this includes  the  motivic and de Rham zeta values.
Moreover, the multiple modular values in the $\epsilon_k^{(j)}$-valued generating series~\eqref{eq:limitS} are $\mathbb Q[i\pi]$-linear combinations of MZVs and thus inherit the well-definedness as motivic and de Rham periods.

\subsubsection{Open questions at genus one}

The primary focus of this work is on genus-one objects,  namely the iterated Eisenstein integrals of (\ref{defiteis}), where the situation is less clear. The main obstacle at the time of writing is the lack of a de Rham projection and consequently the lack of a definition for objects that one would naturally call \textit{de Rham versions of iterated Eisenstein integrals}. 
It turns out that the construction of de Rham periods at genus one is not unique and 
it is not clear yet, in the general case,
how to fix the ambiguities through canonical choices.

The simplest examples of the ambiguities in the search for a de Rham projection beyond genus zero can be understood from the geometry of the torus and the cycles and differentials one canonically assigns to it: First, one has considerable freedom in choosing the cycles due to the periodicity, reflected in  SL$(2,\mathbb Z)$ transformations. Second, the  basis of differentials contains an Abelian differential of the second kind (i.e.\ meromorphic with vanishing residue) that is  not unique either: adding any multiple of the holomorphic differential to it still yields a representative of the same class. 
Hence, in the absence of specific choices (such as the proposal of \cite{Gerken:2020xfv} based on the degeneration limit $\tau \rightarrow i\infty$),
 there is no a priori unique way of assigning dual differentials to integration cycles on the torus.

In principle, for the problem of $\tau$ integration at hand, we would  need
a prescription for assigning dual differentials. For the case of iterated Eisenstein integrals, a first step towards their de Rham version could be to consider \textit{motivic iterated Eisenstein integrals} in the light of our shorthand notation (\ref{motivicshorthand}),\footnote{We here think of $\nu^{j_1\ldots j_\ell}_{k_1\ldots k_\ell}$ as an algebraic de Rham cocycle since the holomorphic Eisenstein series replace the rational functions appearing at genus zero (they are in fact polynomials in the branch points in the algebraic description of elliptic curves). Similarly, the powers of $\tau$ in the integration kernels (\ref{nukerdef}) which correspond to the superscripts $j_1\ldots j_\ell$ are algebraically related to the polynomial equations defining elliptic curves.}
\begin{align}
\label{abhiereisensteinstuff}
     \eem{j_1&\dots& j_{\ell}}{k_1&\dots &k_{\ell}}{\tau}= \left[\Gamma_{\tau}\, , \,     {\nu}_{k_1\dots k_{\ell}}^{j_1\dots j_{\ell}} =\nuker{j_{\ell}}{k_{\ell}} {\tau_{\ell}} \dots\nuker{j_1}{k_1}{\tau_1}\right]^\mm \, , 
\end{align}
where $\Gamma_\tau$ is a cycle that can be represented by the contour specified in the iterated integral (\ref{defiteis}).\footnote{The closely related task of setting up motivic elliptic multiple zeta values is for instance discussed in the talk \cite{Matthes:talk} of Nils Matthes.}
In a second step, it is a major question whether the duals $\check {\nu}_{k_1\dots k_{\ell}}^{j_1\dots j_{\ell}}$ of the de Rham representatives can be canonically translated into contours
\begin{align}
\label{specificprojection}
    \check{\nu}_{k_1\dots k_{\ell}}^{j_1\dots j_{\ell}} \stackrel{?}{\mapsto} \Gamma_{k_1\dots k_{\ell}}^{j_1\dots j_{\ell}}
\end{align}
with $k_i \geq 4$ even and $0\leq j_i \leq k_i{-}2$ according to the integration kernels $\nuker{j}{k} {\tau} \sim \tau^j {\rm G}_k(\tau) \dd \tau$ in (\ref{nukerdef}).
If this is the case, one will be tasked with defining objects
\begin{align}
\label{derahmperiodseisenstein}
{\cal E^\dr}\! \left[\begin{smallmatrix}m_1&\dots &m_{\ell}\\ n_1&\dots &n_{\ell}\end{smallmatrix}\Big|\begin{smallmatrix}j_1&\dots& j_{\ell}\\k_1& \dots &k_{\ell}\end{smallmatrix};\tau\right]= \left[\check{\nu}_{n_1\dots n_{\ell}}^{m_1 \dots  m_{\ell}} \, , \, \nu_{k_1\dots k_{\ell}}^{j_1\dots j_{\ell}}\right]^\dr  \,.
\end{align}
Taking this as a choice for the de Rham versions of iterated  Eisenstein integrals, 
one can compute the motivic coaction of motivic iterated Eisenstein integrals with the general formula (\ref{masterformula})\footnote{To be more precise, this formula might not be applicable in exactly that way and one would rather use a version derived using the de Rham fundamental group as done for other cases in \cite{ Brown:2009qja,Tapuskovic:2023xiu}.}. 

Our proposal for a genus-one coaction in the following section uses (for now not rigorously defined) objects 
\begin{align}
\label{EDR}
\eedr{j_1&\dots &j_{\ell}}{k_1&\dots &k_{\ell}}{\tau}
\end{align}
on the right-hand side of the tensor product. They are treated as abstract symbols in this work which follow the labelling as the motivic iterated Eisenstein integrals (\ref{abhiereisensteinstuff}) through a single block of $j_i,k_i$. Moreover, they are taken to inherit the shuffle multiplication,  differential equations and regularized limits $\tau \rightarrow 0,i\infty$ of the iterated Eisenstein integrals (\ref{defiteis}) in the same way as de Rham MPLs at genus zero inherit shuffle multiplication, differential equations and regularized limits $z\rightarrow 0,1$ of the iterated integrals (\ref{eq:MPL}). 
It is not clear at the moment how the tentative objects (\ref{EDR})
are expressed in terms of those in~\eqref{derahmperiodseisenstein}. 
Given our current lack of understanding of all these relations, we leave it to future work to compare the outcome of the general formula~\eqref{masterformula} to our proposed coaction in the next section.

\subsubsection{The labels $\mm$ and $\dr$}

Taking the above into account we take the labels $\mm$ and $\dr$ to be simply indicators for which side of the $\otimes$ in \eqref{masterformula} the respective objects appear (see (\ref{mmdrlocation})), with $\mm$ indicating the left-hand side and $\dr$ indicating the right-hand side. But of course these labels are also alluding to motivic and de Rham objects. We can immediately think of genus-zero objects $X^\mm$ and $X^\dr$ as motivic and de Rham periods.
For genus-one objects including iterated Eisenstein integrals, however,
we consider the $\dr$ labelled versions as formal objects on the right-hand side of the tensor product whose rigorous definition remains to be understood.

\section{The main proposal}
\label{sec:3}

\begin{figure}[H]
\centering
    \begin{tikzpicture}
\node at (-3.9cm,1.5cm) [above] {$(\mathbb{M}_0^{\mathfrak{dr}})^{-1}\mathbb{G}^{\mathfrak{m}}\mathbb{M}_0^{\mathfrak{dr}} \mathbb{G}^{\mathfrak{dr}}=$};
\node at (3.8cm,1.5cm) [above] {$\rightsquigarrow (\mathbb{M}_{\sigma}^{\mathfrak{dr}})^{-1} \mathbb{I}^{\mathfrak{m}}\mathbb{M}_{\sigma}^{\mathfrak{dr}} \mathbb{I}^{\mathfrak{dr}}$};
    \node at (-1.5cm,1.5cm) [above] {\colorbox{YellowOrange!25}{$\Delta \mathbb{G}^{\mathfrak{m}}$}};   
    \node at (0cm,-1.35cm) [below] {analogy};   
 \node at (1.5cm,1.5cm) [above] {\colorbox{RedOrange!25}{$\Delta \mathbb{I}^{\mathfrak{m}}$}};
    \node at (-1.5cm,-1.5cm) [above] {\colorbox{ProcessBlue!25}{$\text{sv} \, \mathbb{G}$}};
        \node at (-4.25cm,-1.5cm) [above] {$\left(\text{sv}\,\mathbb{M}_0\right)^{-1}\overline{\mathbb{G}^T}\left(\text{sv}\,\mathbb{M}_0\right) \mathbb{G}=$};
    \node at (4.1cm,-1.5cm) [above] {$=\left(\text{sv}\, \mathbb{M}_{\sigma}\right)^{-1} \overline{\mathbb{I}^T}\left(\text{sv}\,\mathbb{M}_\sigma\right)\mathbb{I}$};
        \node at (1.5cm,-1.5cm) [above] {\colorbox{Plum!25}{$\text{sv}\, \,  \mathbb{I}$}};
    \node at (-3.3cm,0.42cm) [left] {Genus 0};
    \node at (3.9cm,0.42cm) [left] {Genus 1};
    \draw[line width=0.3mm,->,dotted] (-0.8cm,1.85cm) -- (0.9cm,1.85cm); 
    \draw[line width=0.3mm,->] (-0.8cm,-1.2cm) -- (0.9cm,-1.2cm); 
    \draw[line width=0.3mm,<-] (-1.6cm,-0.65cm) -- (-1.6cm,1.4cm); 
    \draw[line width=0.3mm,->,dotted] (1.6cm,-0.65cm) -- (1.6cm,1.4cm); 
\node[rotate=90, anchor=south] at (-2cm, 0.375cm) {determines};
\end{tikzpicture}
\caption{A pictorial representation of the relations between the known and the proposed coaction and the single-valued map prescriptions at genus zero (MPLs) and genus one (iterated Eisenstein integrals). Note that the coaction \textit{determines} the single valued map as reviewed in section \ref{sec:3.3} below.}
\label{fig.quadrat2}
\end{figure}
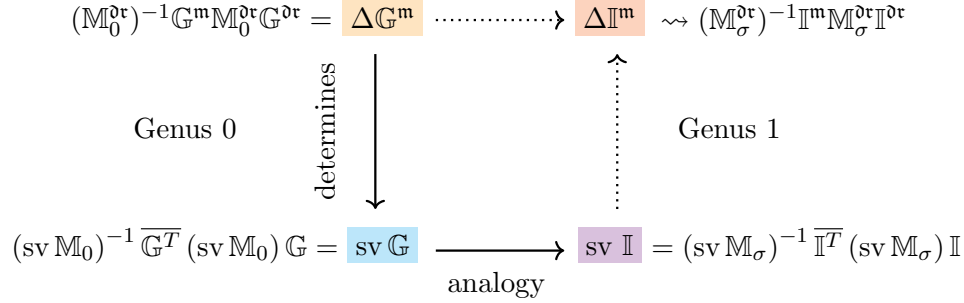

As reviewed in section \ref{svIsec}, the generating series (\ref{mainisv}) of single-valued iterated Eisenstein integrals mirrors the structure of that of single-valued MPLs in (\ref{1varsv}). The genus-one construction (\ref{mainisv}) can be formally obtained from (\ref{1varsv}) by exchanging the role of genus-zero and genus-one quantities as follows: replace meromorphic MPLs by iterated Eisenstein integrals (\ref{defiteis}), the non-commutative $e_0, e_1$ of the KZ connection by Tsunogai derivations $\epsilon_k^{(j)}$ and zeta generators $M_w$ at genus zero by their genus-one counterparts $\sigma_w$. This analogy is represented by the solid horizontal arrow at the bottom of figure \ref{fig.quadrat2}.

The main proposal of this work arises from applying the same analogy to the motivic coaction (\ref{1varco}) of MPLs as in the dashed horizontal arrow in the top line of figure \ref{fig.quadrat2}, resulting in the following proposal for the coaction of generating series of iterated Eisenstein integrals:
\begin{tcolorbox}[colframe=gray, colback=white, sharp corners]
{\bf Proposal.}\vspace{-5mm}
    \begin{align}
    \label{eq:conj}
    \Delta \Ip^{{\mm}}(\epsilon_k;\tau)= (\mathbb{M}^{\dr}_\sigma)^{-1}\Ip^\mm (\epsilon_k;\tau)\mathbb{M}^{\dr}_\sigma \Ip^{\dr}(\epsilon_k;\tau) 
    \end{align}
\end{tcolorbox}
In order to understand whether this proposed coaction is an explicit version of the motivic coaction for iterated Eisenstein integrals, one first needs to develop a better understanding of the de Rham versions of iterated Eisenstein integrals as discussed in section \ref{sec_motivicderahm}. For now, we interpret (\ref{eq:conj}) as a map onto tensor products according to $X^\mot Y^\dr \cong X \otimes Y$, where the iterated Eisenstein integrals in the expansion (\ref{eq:II}) of $\Ip^\mm $ and $ \Ip^{\dr}$ are attributed to the first and second entry, respectively.
Factors of $i\pi$ in the second entry $Y$ are discarded according to de Rham periods at genus zero, also when multiplying rational multiples of iterated Eisenstein integrals (\ref{defiteis}). The linear independence result of \cite{Nilsnewarticle} on iterated Eisenstein integrals ${\cal E}$ with different entries $j_i,k_i$ are carried over to both ${\cal E}^\mot$ and~${\cal E}^\dr$.

The coaction of individual iterated Eisenstein integrals (\ref{defiteis}) can be extracted 
from (\ref{eq:conj}) only when its right-hand side is expanded solely in words $\ep_{k_1}^{(j_1)} \ldots \ep_{k_r}^{(j_r)}$ without any separate reference to zeta generators. This can be accomplished by organizing the zeta generators into nested brackets as in \eqref{adjact} and (\ref{svconj})
and iteratively converting all instances of $[\sigma_w,\epsilon_k^{(j)}]$ to Lie series in $\epsilon_{m}^{(n)}$ as detailed in sections \ref{zwsec} and \ref{svIsec}. 

However, the Pollack relations among Lie polynomials in $\ep_k^{(j)}$ imply that not all the $\Delta \eem{j_1 &j_2 &\ldots &j_\ell}{k_1 &k_2 &\ldots &k_\ell}{\tau} $ occur individually in the coaction (\ref{eq:conj}) of the generating series. Instead, our proposal determines the coaction of those iterated Eisenstein integrals which are realized in elliptic multiple zeta values \cite{Broedel:2015hia, Matthes:Thesis, Matthes:2017,Zerbini:2018sox,Lochak:2020}. The first instance of this phenomenon occurs at modular depth two and degree $k_1{+}k_2 = 14$, where the Pollack relation $  [\epsilon_4,\epsilon_{10}] = 3[\epsilon_6,\epsilon_8] $ implies that (\ref{eq:conj}) only contains the coaction for three linear combinations of $\eem{0 &0}{4 &10}{\tau}$, $\eem{0 &0}{10 &4}{\tau}$, $\eem{0 &0}{6 &8}{\tau}$, $\eem{0 &0}{8 &6}{\tau}$. The investigation of coaction formulae for the missing linear combination at degree 14 and more general iterated Eisenstein integrals that drop out from $\Ip^{{\mm}}(\epsilon_k;\tau)$ is left for the future.

Note that our proposal (\ref{eq:conj}) for the coaction of $\Ip^{{\mm}}(\epsilon_k;\tau)$ is independent of the choice of $f$-alphabet isomorphism $\rho$
for the same reason that the series $\sv \, \mathbb{I}(\epsilon_k;\tau)$ in (\ref{mainisv}) is independent of $\rho$, see the discussion in section \ref{svIsec}.

\subsection{Examples at modular depth one and two}
\label{sec:3.1}

The proposed coaction $\Delta$ of iterated Eisenstein integrals of \textit{modular depth one} simply evaluates to 
    \begin{align}
    \label{deltadepthone}
    \Delta    \eem{j}{k}{\tau}&= \eem{j}{k}{\tau}+ \eedr{j}{k}{\tau} \,.
    \end{align}
This follows from the fact that the 
contributions involving MZVs to the series expansion  of 
$(\mathbb{M}^{\dr}_\sigma )^{-1} \Ip^\mm \mathbb{M}^{\dr}_\sigma = \Ip^\mm+ \sum_{k=1}^\infty \zeta_{2k+1}^\dr [\Ip^\mm, \sigma_{2k+1}]+\ldots$
starts to contribute only from modular depth two since the expansion of $[\ep_k^{(j)}, \sigma_w]$ does.

The simplest
non-trivial contributions from zeta generators to the proposed coaction are captured by the following closed formula at \textit{modular depth two}
\begin{align}
  \Delta    \eem{j_1&j_2}{k_1&k_2}{\tau}&= \eem{j_1&j_2}{k_1&k_2}{\tau}+\eedr{j_2}{k_2}{\tau}\eem{j_1}{k_1}{\tau}+\eedr{j_1&j_2}{k_1&k_2}{\tau} \label{clform} \\
  &\quad + \delta_{j_1,k_1-2} \frac{\zeta^\dr_{k_1-1}}{k_1{-}1} \eem{j_2}{k_2}{\tau} - \delta_{j_2,k_2-2} \frac{\zeta^\dr_{k_2-1}}{k_2{-}1} \eem{j_1}{k_1}{\tau}  \notag \\
  &\quad - \theta_{k_1<k_2} \zeta^\dr_{k_1-1} \stf{j_1&j_2}{k_1&k_2}{\tau}+  \theta_{k_2<k_1} \zeta^\dr_{k_2-1} \stf{j_2&j_1}{k_2&k_1}{\tau} \, ,
  \notag
\end{align}
where the terms with the Kronecker deltas in the second line are due to the geometric contributions $\sigma_w = -\frac{1}{(w{-}1)!}\ep_{w+1}^{(w-1)}+\ldots$ at lowest degree and modular depth. The arithmetic part $z_w$ of $\sigma_w$ in turn gives rise to the following terms in the third line\footnote{This expression can be derived from the modular-depth-two contributions to the brackets $[z_w,\ep_k]$ in (\ref{zepcom}), and its dependence on $j_1,j_2,k_1,k_2$ is analogous to (4.20) of \cite{Dorigoni_2022} up to a factor of $-2$ (the factor of 2 from the single-valued MZVs entering equivariant iterated Eisenstein and the minus sign from the fact that the ordering conventions of the $\epsilon_{k_1}^{(j_1)} \epsilon_{k_2}^{(j_2)}\ldots$ in the generating series $J^{\rm eqv}$ of the reference are opposite to those of $\Ip(\epsilon_k;\tau)$).
A similar separation of terms arises in the construction of iterated integrals called $\beta^{\rm sv}$ in~\cite[(3.21)]{Dorigoni_2022}, where two types of contributions to purely antiholomorphic integration constants, called $\overline{\alpha_{\text{easy}}}$ and $\overline{\alpha_{\text{hard}}}$, were identified that have the same origin.}:
\begin{align}
\stf{j_1&j_2}{k_1&k_2}{\tau} &= 
\frac{(-1)^{j_1}  (k_2 {-} k_1 {+} 1)!  j_2! (k_2 {-} j_2 {-} 
      2)!\, {\rm BF}_{k_2}}{(k_1 {-} 
        1)   (k_2 {-} 1)! (k_2 {-} j_1 {-} j_2 {-} 2)! (j_1 {+} j_2 {-} k_1 {+} 
         2)! \, {\rm BF}_{k_2 - k_1 + 2}} \eem{j_1 {+} j_2 {-} k_1 {+} 2}{k_2 {-} 
      k_1 {+} 2}{\tau} \, .
\label{aritd2}
 \end{align}
This expression is only defined for $k_1<k_2$, consistently with the factors of $\theta_{k_i<k_j}$ in (\ref{clform}) which are defined to be 1 for $k_i<k_j$ and 0 otherwise.
By the absence of the arithmetic contributions (\ref{aritd2}) at $k_1=k_2 = k$, the general coaction formula (\ref{clform}) in such ``diagonal'' cases simplifies to
\begin{align}
  \Delta    \eem{j_1&j_2}{k&k}{\tau}&= \eem{j_1&j_2}{k&k}{\tau}+\eedr{j_2}{k}{\tau}\eem{j_1}{k}{\tau}+\eedr{j_1&j_2}{k&k}{\tau} \notag \\
  &\quad + \frac{\zeta^\dr_{k-1}}{k{-}1} \bigg( \delta_{j_1,k-2} \eem{j_2}{k}{\tau} - \delta_{j_2,k-2}  \eem{j_1}{k}{\tau} \bigg) \, .
  \label{cldiag}
\end{align} 
Note that (\ref{clform}) and (\ref{cldiag}) are understood to only apply to those linear combinations of iterated Eisenstein integrals that occur in the generating series (\ref{eq:conj}) after taking all Pollack relations into account.

In general, de Rham MZVs represented by a total of $r$ letters $f_{i_1}\ldots f_{i_r}$ in the $f$-alphabet with odd $i_k$ firstly appear in the proposed coaction at modular depth $r{+}1$. Concrete examples of the proposed coaction
at modular depth two and three can be found in appendix \ref{app_ExamplesEisenstein}. 

As exemplified by the first line in (\ref{clform}), the contributions to the 
proposed coaction without
any factors of $\zeta^\dr_{n_1,\ldots,n_r}$ 
can be given in closed form for arbitrary modular depth
\begin{align}
\Delta    \eem{j_1 &\ldots &j_\ell}{k_1 &\ldots &k_\ell}{\tau}&= 
\sum_{i=0}^{\ell} \eem{j_1 &\ldots &j_i}{k_1 &\ldots &k_i}{\tau}
\eedr{j_{i+1} &\ldots &j_\ell}{k_{i+1} &\ldots &k_\ell}{\tau} + {\cal O}( \zeta^\dr_{n_1,\ldots,n_r} ) \, ,
\label{nozeta}
\end{align}
setting $\eem{\emptyset}{\emptyset}{\tau}=1= \eedr{\emptyset}{\emptyset}{\tau}$. This simple
deconcatenation formula follows from expanding
the generating series $\Ip^\mm (\epsilon_k;\tau) \Ip^{\dr}(\epsilon_k;\tau)$ which is left after setting $\mathbb{M}^{\dr}_\sigma \rightarrow \textbf{1}^\dr$ in (\ref{eq:conj}).

\subsection{Properties of the coaction proposal}
\label{sec:3.2}

We shall here describe how our proposal (\ref{eq:conj}) for the coaction of iterated Eisenstein integrals composes with their shuffle multiplication, $\tau$-derivatives and (regularized) limits where $\tau$ tends to $i\infty$ or $0$. As we will see, all composition properties follow the expectations that one may have from the properties of the genus-zero coaction (\ref{1varco}) of MPLs.

\subsubsection{Compatibility with shuffle multiplication}
\label{sec:3.2.1}

First of all, the proposed coaction (\ref{eq:conj}) of
the generating series is compatible with the shuffle multiplication of the iterated integrals in its coefficients,
\begin{align}
    \Delta \left( \eem{j_1&\ldots&j_i}{k_1&\ldots & k_i}{\tau} \eem{j_{i+1}&\ldots&j_\ell}{k_{i+1}&\ldots & k_\ell}{\tau}\right)
    = \left(\Delta \eem{j_1&\ldots&j_i}{k_1&\ldots & k_i}{\tau} \right)  \left( \Delta \eem{j_{i+1}&\ldots&j_\ell}{k_{i+1}&\ldots & k_\ell}{\tau}\right)
    \label{shufflmul}
\end{align}
for all of $\ell\geq 2$ and $i=1,\ldots,\ell{-}1$. This follows from the fact that the right-hand side of (\ref{eq:conj}) takes values in the Lie group generated by the Tsunogai derivations, i.e.\ that its logarithm is a Lie series in $\epsilon_k^{(j)}$. The group-valuedness in turn relies on the expansion
of $(\mathbb M^\dr_\sigma)^{-1} \mathbb I^\mm(\epsilon_k;\tau) \mathbb M^\dr_\sigma$
in terms of Lie polynomials in zeta generators (cf.\ (\ref{svconj})) and the fact that $[\sigma_w,\epsilon_k^{(j)}]$ are Lie series in $\epsilon_m^{(n)}$. Further discussions and proofs for the link between group-valuedness and shuffle multiplication can be found in \cite{Rtnau:1993rq, Brown:unpubl}.

We have not written out the shuffle product on either side of (\ref{shufflmul}) since it acts in the standard way on the iterated integrals (\ref{defiteis}), and their tangential base-point regularization preserves their shuffle multiplication \cite{brown2017multiple}. On the right-hand side of (\ref{shufflmul}), it is also understood that the shuffle product operates separately on the ${\cal E}^\mot$ and ${\cal E}^\dr$ factors of the tensor product produced by the proposed coaction.

One can readily check (\ref{shufflmul}) to hold for the closed formula (\ref{clform}) at modular depth two since both its second and third line are antisymmetric under the exchange $(j_1,k_1) \leftrightarrow (j_2,k_2)$. As a consequence, all $\zeta^\dr_{k_i-1}$ cancel from $ \Delta    \eem{j_1&j_2}{k_1&k_2}{\tau}+ \Delta    \eem{j_2&j_1}{k_2&k_1}{\tau}$, and one is left with a shuffle product of the expressions (\ref{deltadepthone}) for coactions at modular depth one.

\subsubsection{$\tau$-derivatives}
\label{sec:3.2.2}

At genus zero, the motivic coaction of MPLs (\ref{eq:MPL}) composes with derivatives in the variable $z$ (and in fact also in the labels $a_i$) according to $\Delta \partial_z = (\textbf{1}\otimes \partial_z)\Delta$ \cite{Duhr:2012fh}\footnote{\label{otimesft}Here and in later sections, we reinstate the $\otimes$ symbol to indicate that an operator ${\cal O}$ only acts on one of the entries of the tensor product (\ref{mmdrlocation}), i.e.\ $\textbf{1}\otimes {\cal O}$ for action on the de Rham entry and ${\cal O} \otimes \textbf{1}$ for the motivic entry. Hence, $\textbf{1}\otimes \partial_z$ and $\textbf{1}\otimes \partial_\tau$ are understood to only act on $G^\dr(\cdots;z)$ and $\eedr{\ldots}{\ldots}{\tau}$ but not on $G^\mot(\cdots;z)$ and $\eem{\ldots}{\ldots}{\tau}$, respectively.}. Our proposed coaction (\ref{eq:conj}) at genus one will now be shown to have the same property that derivatives (here in the modular parameter $\tau$ instead of points $z,a_i$ on the sphere) solely act on the de Rham entry: using (\ref{eq:derivativeIp}), we find that:
\begin{align}
    \Delta \partial_{\tau} \Ip^\mm(\epsilon_k;\tau)& =- \Delta \big(\Ip^\mm(\epsilon_k;\tau)  \mathbb{A}(\epsilon_k;\tau)\big)=- \big(\Delta \mathbb{I}^\mm(\epsilon_k;\tau)  \big)\mathbb{A}(\epsilon_k;\tau)\notag\\
    &= -\left(\mathbb{M}_{\sigmaT}^\dr\right)^{-1} \mathbb{I}^\mm(\epsilon_k;\tau)  \mathbb{M}_{\sigmaT}^\dr\mathbb{I}^\dr (\epsilon_k;\tau)  \mathbb{A}(\epsilon_k;\tau) \notag  \\
    &= \left(\mathbb{M}_{\sigmaT}^\dr\right)^{-1} \mathbb{I}^\mm(\epsilon_k;\tau)  \mathbb{M}_{\sigmaT}^\dr \partial_\tau \mathbb{I}^\dr (\epsilon_k;\tau) 
        \label{dtaucomp} \, . 
\end{align}
We have assigned a trivial coaction to the holomorphic Eisenstein series in the connection $\mathbb{A}(\epsilon_k;\tau)$ and identified the period images of the motivic and de Rham versions of holomorphic Eisenstein series ${\rm G}_k(\tau) \cong {\rm G}^\mot_k(\tau) \cong {\rm G}^\dr_k(\tau)$ in analogy with rational functions on the sphere.\footnote{This is consistent with the fact that, in the
Weierstra\ss\ form of the elliptic curve, holomorphic Eisenstein series are symmetric polynomials in the branch points and therefore rational functions.}
Since the series $\mathbb{M}_{\sigmaT}^\dr$ does not depend on $\tau$, the last line of (\ref{dtaucomp}) features the $\tau$-derivative of the entire de Rham entry of $\Delta \mathbb{I}^\mm(\epsilon_k;\tau)$, and we conclude that
\begin{align}\Delta\partial_{\tau}\Ip^\mm (\epsilon_k;\tau)=\left(\textbf{1}\otimes \partial_{\tau}\right) \Delta \Ip^\mm(\epsilon_k;\tau)\, .
\end{align}
In other words, the proposal (\ref{eq:conj}) extends the genus-zero property $\Delta \partial_z = (\textbf{1}\otimes \partial_z)\Delta$ of the coaction of MPLs to $\Delta\partial_{\tau}=\left(\textbf{1}\otimes \partial_{\tau}\right) \Delta$ when acting on  arbitrary combinations of iterated Eisenstein integrals that occur in the expansion of $\Ip^\mm(\epsilon_k;\tau)$.

\subsubsection{Limits $\tau\rightarrow i\infty$ and $\tau\rightarrow 0$}
\label{sec:tauzero}

We shall here investigate the behaviour of the proposed coaction in the two regularized limits $\tau\rightarrow i\infty$ and $\tau\rightarrow 0$ of the series $\Ip^\mm (\epsilon_k;\tau)$. The first limit $\tau\rightarrow i\infty$ is readily found to commute with the proposed coaction, and the commutativity of $\tau\rightarrow 0$ with $\Delta$ is tied to a condition on the $S$-cocycle which is expected to follow from a result in the mathematics literature \cite{brown2017multiple}.

\paragraph{\underline{$\tau \rightarrow i\infty$:}}
All the non-trivial iterated Eisenstein integrals (\ref{defiteis}) entering the generating series $\Ip(\epsilon_k;\tau)$ vanish in the regularized limit $\tau\rightarrow i\infty$, leading to 
\begin{align}
   \lim_{\tau \rightarrow i\infty} \Ip(\epsilon_k;\tau)= \textbf{1}
   \label{tauinfy}
\end{align}
with the obvious coaction
\begin{align}
    \Delta \left(\lim_{\tau \rightarrow i\infty}\Ip^\mm(\epsilon_k;\tau)\right)= \Delta \textbf{1}^\mm =\textbf{1}^\mm \cdot \textbf{1}^\dr \, . 
\end{align}
On the other hand, inserting the limit
(\ref{tauinfy}) into the right-hand side of (\ref{eq:conj}) gives rise to
\begin{align}
  \lim_{\tau \rightarrow i\infty}  \big(\Delta  \Ip^\mot(\epsilon_k;\tau)\big)&=\left(\mathbb{M}_\sigma^\dr\right)^{-1} \cdot \textbf{1}^\mm\cdot \mathbb{M}^\dr_\sigma  \cdot \textbf{1}^\dr =\textbf{1}^\mm \cdot \textbf{1}^\dr \,  .
\end{align}
Thus, $\Delta$ commutes with evaluation in the regularized limit $\tau\rightarrow i\infty$.

\paragraph{\underline{$\tau \rightarrow 0$:}}
The opposite regularized limit $\tau \rightarrow 0$ reduces the generating series of iterated Eisenstein integrals to that of MMVs in (\ref{spluscocyc}) and (\ref{eq:limitS}),
\begin{align}
\lim_{\tau \rightarrow 0}\Ip(\epsilon_k;\tau) = \mathbb{S}(\epsilon_k)\, .
\end{align}
Hence, commutativity of the regularized $\tau \rightarrow 0$ limit with our proposal (\ref{eq:conj}),
\begin{align}
    \Delta\left(\lim_{\tau \rightarrow 0}\Ip^\mm(\epsilon_k;\tau)\right) =\lim_{\tau \rightarrow 0}\Delta\big(\Ip^\mm(\epsilon_k;\tau)\big)  
    \label{limconj0}
\end{align}
translates into the following coaction property of the $S$-cocycle:
\begin{align}
 \Delta \mathbb{S}^\mm(\epsilon_k)= \left(\mathbb{M}_{\sigma}^\dr\right)^{-1} \mathbb{S}^\mm(\epsilon_k) \mathbb{M}_{\sigmaT}^\dr\mathbb{S}^\dr(\epsilon_k)\, . 
    \label{limconj1}
\end{align}
By the expansion (\ref{eq:limitS}) of the $S$-cocycle, (\ref{limconj1}) generates the motivic coaction of all combinations of MMVs that enter (\ref{eq:limitS}) as the coefficients of words $\epsilon_{k_1}^{(j_1)}\ldots \epsilon_{k_\ell}^{(j_\ell)}$ that are independent under Pollack relations. These combinations of MMVs are $\mathbb Q[2\pi i]$-linear combinations of MZVs, and $\mathbb{S}^\dr(\epsilon_k)$ is obtained from $\mathbb{S}^\mot(\epsilon_k)$ by passing from motivic to de Rham MZVs, see
sections \ref{sec:2.2.1}, \ref{sec:2.3.1}
and in particular \cite{saad2020multiple} for a discussion in the light of motivic iterated Eisenstein integrals. 
As the relation~\eqref{limconj1} reduces to a statement about MZVs, it is amenable to direct study.

In section \ref{sec:4.1} below, we will translate an abstract statement on the motivic coaction from \cite{Brown:2017qwo} to our notation, finding that it agrees with (\ref{limconj1}) for a specific, natural definition of ~$\mathbb{S}^\mm$ and $\mathbb{S}^\dr$. Then we explain the computational advantages of an equivalent of the explicit expression (\ref{limconj1}).

\medskip
 
In summary, we have identified the conditions for the proposal (\ref{eq:conj}) for $\Delta$ to commute with evaluation at the points $\tau \rightarrow i\infty$ and $\tau \rightarrow 0$. Commutativity with $\tau \rightarrow i\infty$ follows solely from the tangential-base-point regularization of the integrals of $\Ip^\mm (\epsilon_k;\tau)$. Commutativity with $\tau \rightarrow 0$ is additionally tied to the
coaction property (\ref{limconj1}) of the $S$-cocycle which we shall establish under a certain assumption in section \ref{sec:4.1} below. It would be interesting to check if the same commutativity with $\Delta$ applies to evaluations at other specific or even at arbitrary points in the upper half-plane.

\subsection{The single-valued map from the coaction}
\label{sec:3.3}

This section is dedicated to the construction of single-valued maps from  motivic coactions, using the antipode of the de Rham Hopf algebra as an intermediate step. 
We shall review in the one-variable case how this construction efficiently relates the formulation of the coaction and the single-valued map of MPLs via zeta generators \cite{Frost:2023stm, Frost:2025lre} and discuss its extension to genus one: deriving the established single-valued iterated Eisenstein integrals (\ref{mainisv}) from our proposal (\ref{eq:conj}) for their motivic coaction, and obtaining a tentative antipode of iterated Eisenstein integrals as a byproduct.

\subsubsection{The antipode of de Rham MZVs and MPLs from their coaction}
\label{sec:3.3.1}

Given that only de Rham versions of MPLs and MZVs form a Hopf algebra (as opposed to the Hopf-algebra comodule formed by their motivic versions), the antipode only makes sense on de Rham periods\footnote{We are grateful to Hadleigh Frost and Deepak Kamlesh for valuable discussions on this point.} which we shall obtain from motivic periods through the projection\footnote{Not to be confused with the de Rham projection of section \ref{sec_motivicderahm} which translates between Betti cycles and dual de Rham cycles.}
\begin{align}
\Pi^\dr( \mathbb M_0^\mot) = \mathbb M_0^\dr \, , \ \ \ \ \Pi^\dr\big( \mathbb G^\mot(e_i;z) \big) = \mathbb G^\dr(e_i;z) \, .
\label{svde.01}
\end{align}
With this definition in place, one can uniquely determine the antipode $\ant$ of MPLs from the condition~\cite{Goncharov:2001iea}
\begin{align}
\textbf{1} = \mu \circ
(\ant \circ \Pi^\dr \otimes \textbf{1}) \circ \Delta
\label{svde.02}
\end{align}
on arbitrary group-like generating series such as $\mathbb M_0^\mot$, $\mathbb G^\mot(e_i;z)$ or their generalizations to multiple variables. The notation $\ant \circ \Pi^\dr \otimes \textbf{1}$ instructs us to only apply the map $\ant \circ \Pi^\dr$ (i.e.\ the projection (\ref{svde.01}) followed by the antipode to be determined) to the motivic entry of the image of $\Delta$, see footnote \ref{otimesft}. Finally, the leftmost operation $\mu$ in (\ref{svde.02}) is taken to multiply the de Rham periods in the two entries of the tensor product produced by $(\ant \circ \Pi^\dr \otimes \textbf{1}) \circ \Delta$ while leaving the concatenation order of the non-commutative variables $e_i,M_w,\epsilon_k^{(j)},\sigma_w$ inert.

When applied to the generating series $\mathbb M_0^\mot$ of MZVs, the motivic coaction (\ref{g0coactM}) together with (\ref{svde.02}) results in
\begin{align}
\textbf{1} = \mu \circ
(\ant \circ \Pi^\dr \otimes \textbf{1}) \mathbb M_0^\mot \mathbb M_0^\dr = (\ant \mathbb M_0^\dr) \mathbb M_0^\dr \, ,
\label{svde.03}
\end{align}
which can be straightforwardly solved for (see (\ref{Minv}) for the expansion of $(\mathbb M_0^\dr )^{-1}$)
\begin{align}
\ant \mathbb M_0^\dr = (\mathbb M_0^\dr )^{-1}\, .
\label{svde.04}
\end{align}
Similarly, (\ref{svde.02}) applied to the
generating series $\mathbb G^\mot(e_i;z)$ of MPLs  in one variable with motivic coaction (\ref{1varco}) implies the condition
\begin{align}
\textbf{1} &= \mu \circ
(\ant \circ \Pi^\dr \otimes \textbf{1}) ( \mathbb M_0^\dr)^{-1} \mathbb G^\mot(e_i;z) \mathbb M_0^\dr \mathbb G^\dr(e_i;z) \notag \\
&= ( \mathbb M_0^\dr)^{-1} \big( \ant \mathbb G^\dr(e_i;z) \big) \mathbb M_0^\dr \mathbb G^\dr(e_i;z)
\label{svde.05}
\end{align}
which fixes its antipode to be \cite{Frost:2025lre}
\begin{align}
\ant \mathbb G^\dr(e_i;z) = \mathbb M_0^\dr \big( \mathbb G^\dr(e_i;z) \big)^{-1} ( \mathbb M_0^\dr)^{-1}\, .
\label{svde.06}
\end{align}

\subsubsection{The single-valued map of MZVs and MPLs from their coaction}
\label{sec:3.3.2}

With the above antipodes (\ref{svde.04}) and (\ref{svde.06}) in place, we can next take advantage of the general prescription for the single-valued map at genus zero \cite{Brown:2013gia, DelDuca:2016lad} (also see \cite{Charlton:2021uhu})
\begin{align}
{\rm sv} = \mu \circ
(\tilde \ant \circ \Pi^\dr \otimes \textbf{1}) \circ \Delta\, .
\label{svde.08}
\end{align}
The right-hand side only departs from that of
the antipode condition (\ref{svde.02}) by the extra tilde of $\tilde \ant $ which instructs us to take the complex conjugate of the above antipode $\ant$ and to multiply by the parity of the transcendental weight $w$ in
\begin{align}
\tilde \ant G^\dr(a_1,\ldots,a_w;z) = (-1)^w \overline{ \ant G^\dr(a_1,\ldots,a_w;z) }\, .
\label{svde.09}
\end{align}
At the level of the generating series $\mathbb G^\dr(e_i;z)$ and $\mathbb M_0^\dr$, the minus sign in the action of $\tilde \ant$ on MPLs and MZVs of odd transcendental weights can be simply implemented via $e_i \rightarrow -e_i$ and $M_w \rightarrow -M_w$. This translates the antipodes (\ref{svde.04}) and (\ref{svde.06}) into
\begin{align}
\tilde \ant  \mathbb M_0^\dr  &= (  \mathbb M_0^\dr  )^T \, ,
\label{svde.10} \\
\tilde \ant  \mathbb G^\dr(e_i;z) &=
\big( (  \mathbb M_0^\dr  )^T \big)^{-1} 
\overline{ \mathbb G^\dr(e_i;z)^T  }
(  \mathbb M_0^\dr  )^T \, ,
\notag
\end{align}
where the superscript $T$ reverses the concatenation order of both the zeta generators in $ \mathbb M_0^\dr $ and the $e_i$ in $\mathbb G^\dr(e_i;z)^T$.\footnote{By the odd degree of the Lie polynomial $g_w(e_0,e_1)$ in (\ref{mwcoms}), the reversal $(\ldots)^T$ of the expression $\mathbb M_0^\dr \big( \mathbb G^\dr(e_i;z) \big)^{-1} ( \mathbb M_0^\dr)^{-1}$ for the antipode (\ref{svde.06}) commutes with the conversion of zeta generators to Lie polynomials in $e_0,e_1$ via (\ref{mwcoms}), i.e.\ one can consistently perform the reversal via $(\ldots M_w e_i \ldots)^T= \ldots  e_i M_w \ldots$ before applying any commutation relations between $M_w$ and $e_i$.}

Upon insertion into the general genus-zero prescription (\ref{svde.08}) for the single-valued map, the $\tilde \ant $ actions (\ref{svde.10}) imply that 
\begin{align}
{\rm sv} \, \mathbb M_0^\mot = \mu \circ
(\tilde \ant \circ \Pi^\dr \otimes \textbf{1})\mathbb M_0^\mot \mathbb M_0^\dr = ( \mathbb M_0^\dr)^T  \mathbb M_0^\dr
\label{svde.11}
\end{align}
as well as
\begin{align}
{\rm sv} \, \mathbb G^\mot(e_i;z) &= \mu \circ
(\tilde \ant \circ \Pi^\dr \otimes \textbf{1}) ( \mathbb M_0^\dr)^{-1} \mathbb G^\mot(e_i;z) \mathbb M_0^\dr \mathbb G^\dr(e_i;z) \label{svde.12}\\
&= ( \mathbb M_0^\dr )^{-1} \big( (  \mathbb M_0^\dr  )^T \big)^{-1} 
\overline{ \mathbb G^\dr(e_i;z)^T  }
(  \mathbb M_0^\dr  )^T  \mathbb M_0^\dr  \mathbb G^\dr(e_i;z)\, .
\notag
\end{align}
After removing the $^\mot$ and $^\dr$ superscripts\footnote{Instead of defining the single-valued map as a map from de Rham to motivic periods (which is most common in the literature), we employ the symbol sv for the equivalent map from motivic periods to $\mathbb C$-valued functions which matches the map sv$^\mot$ in \cite{Charlton:2021uhu} (shortly above Example 3.1 in the reference).},
the expression for ${\rm sv} \, \mathbb M_0$ in (\ref{svde.11}) gives an alternative representation of (\ref{svmm0}) which simplifies the second line of (\ref{svde.12}) to $( {\rm sv}\, \mathbb M_0 )^{-1} 
\overline{ \mathbb G(e_i;z)^T  }$ $
({\rm sv} \, \mathbb M_0)  \mathbb G(e_i;z)$ and thus reproduces ${\rm sv} \, \mathbb G(e_i;z) $ in (\ref{1varco}).

\subsubsection{Genus-one antipode and single-valued map from the coaction proposal}
\label{sec:3.3.3}

We shall now extend the prescriptions (\ref{svde.02}) and (\ref{svde.08}) for the antipode and the single-valued map to genus one and apply them to our proposed coaction (\ref{eq:conj}) of iterated Eisenstein integrals. Most of the above steps in the derivation of $\ant \mathbb G(e_i;z)$ and ${\rm sv} \, \mathbb G(e_i;z)$ carry over to the generating series $\mathbb I(\epsilon_k;\tau)$ of iterated Eisenstein integrals for two reasons: First, the above manipulations of $\mathbb M_0$ apply in identical form to $\mathbb M_\sigma$ in (\ref{msigser}) (both incarnations of zeta generators $M_w$ and $\sigma_w$ form a free algebra). Second, the proposed coaction for $\mathbb I(\epsilon_k;\tau)$ is constructed to mirror the structure of $\Delta \mathbb G^\mot(e_i;z)$ which fixes the antipode and the single-valued map via (\ref{svde.02}) and~(\ref{svde.08}).

More specifically, inserting the proposal (\ref{eq:conj}) for $\Delta \mathbb I^\mot(\epsilon_k;\tau)$ into (\ref{svde.02}) results in the following antipode of iterated Eisenstein integrals:
\begin{align}
\ant \mathbb I^\dr(\epsilon_k;\tau) = \mathbb M_\sigma^\dr \big( \mathbb I^\dr(\epsilon_k;\tau) \big)^{-1} ( \mathbb M_\sigma^\dr)^{-1}  \, .
\label{altsv.06}
\end{align}
This is obtained by following the steps in (\ref{svde.05}) with $\mathbb M_\sigma$, $\mathbb{I}(\epsilon_k;\tau)$ in the place of $\mathbb M_0$, $ \mathbb G(e_i;z)$ and lines up with (\ref{svde.06}) under these replacements. The resulting antipode formulae for $\ant \eedr{j_1}{k_1}{\tau}$ and $\ant \eedr{j_1 &j_2}{k_1 &k_2}{\tau}$ at modular depth one and two can be found in appendix \ref{appendixapd}.

Before inserting the proposal (\ref{eq:conj}) for $\Delta \mathbb I^\mot(\epsilon_k;\tau)$ into the prescription (\ref{svde.08}) for the single-valued map, it remains to generalize the action of $\tilde \ant$ in (\ref{svde.09}) and (\ref{svde.10}) from MPLs to iterated Eisenstein integrals. Our guiding principle is to impose
\begin{align}
\tilde \ant  \mathbb I^\dr(\epsilon_k;\tau) &=
\big( (  \mathbb M_\sigma^\dr  )^T \big)^{-1} 
\overline{ \mathbb I^\dr(\epsilon_k;\tau)^T  }
(  \mathbb M_\sigma^\dr  )^T
\label{svde.13}
\end{align}
by analogy with (\ref{svde.10}). In view of the antipode (\ref{altsv.06}) and the expansion (\ref{eq:II}) of the series $\mathbb I^\dr(\epsilon_k;\tau)$ as well as the prescription (\ref{toperation}) for the reversal $(\epsilon_{k_1}^{(j_1)}\ldots \epsilon_{k_\ell}^{(j_\ell)})^T$, the desired $\tilde \ant $ action (\ref{svde.13}) on the generating series fixes\footnote{Note that (\ref{altde.09}) is compatible with applying Pollack relations to the expansion (\ref{eq:II}) of $\mathbb I$ in terms of iterated Eisenstein integrals: the combinations of $\ee{j_1  &\ldots &j_\ell}{k_1  &\ldots &k_\ell}{\tau}$ occurring with the independent words in
$\epsilon_k^{(j)}$ under Pollack relations have a uniform
value of $\ell+ \sum_{i=1}^\ell j_i$.} 
\begin{align}
\tilde \ant \eedr{j_1 &\ldots &j_\ell}{k_1 &\ldots &k_\ell}{\tau} = (-1)^{\ell+ \sum_{i=1}^\ell j_i} \overline{ \ant \eedr{j_1 &\ldots &j_\ell}{k_1 &\ldots &k_\ell}{\tau} }\, .
\label{altde.09}
\end{align}
With the assignment of transcendental weight $\ell+ \sum_{i=1}^\ell j_i$ to $\ee{j_1 &\ldots &j_\ell}{k_1 &\ldots &k_\ell}{\tau}$ \cite{Gerken:2020yii}, this also
lines up with the parity of the transcendental weight in (\ref{svde.09}).

Finally, the steps in (\ref{svde.12}) can be adapted to genus one to obtain
\begin{align}
{\rm sv} \, \mathbb I^\mot(\epsilon_k;\tau) &= \mu \circ
(\tilde \ant \circ \Pi^\dr \otimes \textbf{1}) ( \mathbb M_\sigma^\dr)^{-1} \mathbb I^\mot(\epsilon_k;\tau) \mathbb M_\sigma^\dr \mathbb I^\dr(\epsilon_k;\tau) \label{svde.14}\\
&= ( \mathbb M_\sigma^\dr )^{-1} \big( (  \mathbb M_\sigma^\dr  )^T \big)^{-1} 
\overline{ \mathbb I^\dr(\epsilon_k;\tau)^T  }
(  \mathbb M_\sigma^\dr  )^T  \mathbb M_\sigma^\dr  \mathbb I^\dr(\epsilon_k;\tau)
\notag
\end{align}
which upon removing the $^\mot$, $^\dr$ superscripts and recalling $(  \mathbb M_\sigma )^T  \mathbb M_\sigma = {\rm sv}\,\mathbb M_\sigma$ reproduces the generating series of single-valued iterated Eisenstein integrals in (\ref{mainisv}).

In summary, we have shown that the construction of the single-valued map from the motivic coaction in (\ref{svde.02}) and (\ref{svde.08}), when uplifted to genus one, maps our proposal (\ref{eq:conj}) for $\Delta \mathbb I^\mot(\epsilon_k;\tau)$ to the established
expression for ${\rm sv} \, \mathbb I(\epsilon_k;\tau)$. This can be viewed as another validation of the proposed coaction to the extent that the uplift of (\ref{svde.02}) and (\ref{svde.08}) beyond genus zero makes sense. If this is the case, then it would be interesting to explore in more generality how the information on single-valued periods or closely related modular forms (see section \ref{eqvIsec}) may feed into constructions of unknown coactions, ideally reversing the direction of the solid horizontal arrow in figure \ref{fig.quadrat2}.

\section{Implications for multiple modular values} 
\label{sec:4}

Starting from (\ref{limconj1}), we study in this section the motivic coaction 
\begin{align}
\label{mmvcoacgen}
    \Delta \mathbb{S}^\mm= \left(\mathbb{M}_{\sigma}^\dr\right)^{-1} \mathbb{S}^\mm \mathbb{M}_{\sigmaT}^\dr\mathbb{S}^\dr\, ,
\end{align}
of the $S$-cocycle $\mathbb{S} = \mathbb{S}(\epsilon_k)$ defined in (\ref{spluscocyc}) and its implications for the MMVs in its expansion (\ref{eq:limitS}). Given that the MMVs entering the $S$-cocycle $\mathbb{S}$ are $\mathbb Q[2\pi i]$-linear combinations of MZVs, the motivic coaction at each order of its expansion is fixed by that of MZVs and settles the interpretation of $\mathbb{S}^\dr$ as a series of de Rham MZVs. We encountered (\ref{mmvcoacgen}) in the discussion of section \ref{sec:tauzero} as a condition for the proposed coaction (\ref{eq:conj}) to commute with the regularized $\tau \rightarrow 0$ limit of iterated Eisenstein integrals.

The key result of this section is the following explicit solution to (\ref{mmvcoacgen}):
\begin{tcolorbox}[colframe=gray, colback=white, sharp corners]
{\bf Conjecture.}\vspace{-5mm}
\begin{align}
\label{ourSm.01}
    \mathbb{S}^\mm= \left(\mathbb{M}^\mm_\sigma\right)^{-1} \mathbb{X}_S^\mm U_S^{-1}\mathbb{M}_\sigma^\mm
U_S  \, . 
\end{align}
\end{tcolorbox}
\noindent
This casts the $S$-cocycle into a factorized form.
The action of $U_S$ can be found in (\ref{eq:S1}) and essentially acts by reflection $\epsilon_k^{(j)} \rightarrow \epsilon_k^{(k-j-2)}$ on the geometric parts of the zeta generators in the expansion (\ref{msigser}) of $\mathbb M_\sigma^\mm$, while leaving the arithmetic parts invariant, $U_S^{-1} z_w
U_S = z_w$. The second factor $\mathbb{X}_S^\mm$ on the right side of (\ref{ourSm.01}) is a group-like series in Tsunogai derivations with $\mathbb Q[\pi^2]$ coefficients. Hence, the series $\mathbb M_\sigma^\mot$ in
zeta generators exposes all non-trivial MZVs (i.e.\ all odd $f$-alphabet generators $f_{2k+1}$) in the MMVs that enter the $S$-cocycle~$\mathbb{S}$ at arbitrary degree and modular depth. The right side of
(\ref{ourSm.01}) is a series in $\epsilon_k^{(j)}$ since the $\mathfrak{sl}(2)$-invariant arithmetic parts $z_w$ of the zeta generators commute with $U_S^{-1}$ and normalize the $\epsilon_k^{(j)}$ in the expansion of~$\mathbb{X}_S^\mm$.

In section \ref{sec:4.inf}, we
shall present an informal derivation of the key result 
(\ref{ourSm.01}) from the coaction (\ref{mmvcoacgen}) based on certain assumptions.
We will then relate (\ref{mmvcoacgen}) and (\ref{ourSm.01}) in section~\ref{sec:4.1} to a construction of the motivic coaction for MMVs of \cite{brown2017multiple,Brown2019,saad2020multiple}, proposing the explicit form of certain ingredients in the reference.
The comments in section \ref{sec4.com} among other things add credence to our main proposal (\ref{eq:conj}) through its regularized limit $\tau \rightarrow 0$. Finally, section~\ref{sec:4.2} describes the advantages of (\ref{ourSm.01}) for analytical and numerical computations of MMVs. 

\subsection{Informal derivation of (\ref{ourSm.01})}
\label{sec:4.inf}

In preparation for an informal derivation of (\ref{ourSm.01}), we notice that the coaction formula (\ref{mmvcoacgen}) can be simplified by the redefinition of $\mathbb S$
\begin{align}
\label{ourSm.02}
 \mathbb H = \mathbb M_\sigma \mathbb S
\ \ \ \Rightarrow \ \ \ 
\Delta \mathbb H^\mm = \mathbb H^\mm \mathbb H^\dr 
\end{align}
using the coaction (\ref{g0coactM}) of $\mathbb M_\sigma^\mm$.
A large class of solutions to the equivalent (\ref{ourSm.02}) of (\ref{mmvcoacgen}) is parametrized by\footnote{We do not attempt to present the most general solution to (\ref{ourSm.02}) in our ansatz (\ref{ourSm.03}) and for instance anticipate the explicit form (\ref{clXone}) of the expansion to low modular depth by prescribing a conjugation $U^{-1}\mathbb M^\mm_\sigma U$ instead of a more general composition $U^{-1}\mathbb M^\mm_\sigma V$.} 
\begin{align}
\label{ourSm.03}
\mathbb H^\mm = \mathbb X^\mm  U^{-1}\mathbb M^\mm_\sigma U 
\end{align}
where $\mathbb X^\mm$ and $U$ are taken to be a series of Tsunogai derivations and an SL$(2)$ transformation with a trivial coaction
\begin{align}
\label{ourSm.04}
\Delta \mathbb X^\mm = \mathbb X^\mm \, , \ \ \ \ \ \  \Delta U = U \, .
\end{align}
These conditions on $\mathbb X^\mm $ and $U$ are sufficient for the series $\mathbb H^\mm$ in (\ref{ourSm.03}) to exhibit the simplified coaction (\ref{ourSm.02}). In order to further constrain the series $\mathbb X^\mm$ and $U$, we exploit the cocycle relation \cite{brown2017multiple} 
\begin{align}
\label{ourSm.05}
\mathbb S^{-1} = U_S^{-1} \mathbb S U_S
\end{align}
which follows from the property $\Pexp( \int_{0}^{i\infty} \mathbb{A}(\epsilon_k;\tau_1) )^{-1} = \Pexp( \int_{i\infty}^0 \mathbb{A}(\epsilon_k;\tau_1) )$ of the path-ordered exponential in (\ref{spluscocyc}) together with $\mathbb{A}(\epsilon_k;-1/\tau) = U_S^{-1} \mathbb{A}(\epsilon_k;\tau) U_S$ and is equivalent to the following reflection property of MMVs (that also follow from (\ref{eq:mmvdef}))
\begin{align}
\label{ourSm.06}
\MMV{j_1 &j_2 &\ldots  & j_\ell}{k_1 & k_2 &\ldots &k_\ell}  = (-1)^{\ell + j_1+\ldots+j_\ell} 
\MMV{k_\ell-j_\ell-2 &\ldots &k_2-j_2-2 &k_1- j_1-2}{k_\ell  &\ldots & k_2 &k_1}\, .
\end{align}
More specifically, (\ref{ourSm.05}) with the ansatz for $\mathbb S$ encoded in (\ref{ourSm.02}) and (\ref{ourSm.03}) requires
\begin{align}
\textbf{1}^\mm &=  \mathbb S^\mm U_S^{-1} \mathbb S^\mm U_S
\label{ourSm.07} \\
&= (\mathbb M_\sigma^\mm)^{-1}
\mathbb X^\mm
U^{-1} \mathbb M_\sigma^\mm
U U_S^{-1} (\mathbb M_\sigma^\mm)^{-1} \mathbb X^\mot U^{-1} \mathbb M_\sigma^\mm U U_S\, .
\notag
\end{align}
Given that $U_S$ in (\ref{eq:S1}) associated with the idempotent modular $S$ transformation obeys $(U_S)^2 = \textbf{1}$, a simple solution to (\ref{ourSm.07}) is found by setting
\begin{align}
\label{ourSm.08}
U = U_S \, , \ \ \ \ \ \ 
(\mathbb X^\mot)^{-1} = U_S^{-1} \mathbb X^\mot U_S
\end{align}
In view of (\ref{ourSm.02}) and (\ref{ourSm.03}), this is equivalent to the conjecture (\ref{ourSm.01}) where the series
$\mathbb X^\mot_S$ in Tsunogai derivations with $\mathbb Q[\pi^2]$ coefficients is found to obey the reflection property of (\ref{ourSm.08}),
\begin{align}
\label{ourSm.09}
(\mathbb X^\mot_S)^{-1} = U_S^{-1} \mathbb X_S^\mot U_S
\end{align}
Note that our conjecture (\ref{ourSm.01}) is consistent with the necessary condition  $\mathbb S(\epsilon_k)^T (\sv \, \mathbb M_\sigma)\mathbb S(\epsilon_k) = U_S^{-1} (\sv \, \mathbb M_\sigma) U_S$ for $S$-equivariance (\ref{eqv:trf}) of $\Ieqv(\epsilon_k;\tau)$ provided that the series $\mathbb X_S^\mot$ also obeys 
\begin{align}
    (\mathbb X_S^\mot)^T \mathbb X_S^\mot = \textbf{1}^\mm\,,
\end{align}
see (\ref{toperation}) for $(\ldots)^T$ acting on words in~$\epsilon_k^{(j)}$.

Instead of arguing that the ansatz (\ref{ourSm.03}) for $\mathbb S^\mm$ was sufficiently general to qualify the present discussion as a rigorous proof of the conjecture (\ref{ourSm.01}), we shall see in the next section that it follows from Brown's work \cite{brown2017multiple} under milder assumptions.
We hope that the informal reasoning in this section will be useful to anticipate unknown coaction formulae along similar lines, for instance for elliptic MPLs or generalizations to higher genus.

\subsection{Coaction for MMVs from \cite{brown2017multiple}}
\label{sec:4.1}

In Brown's work \cite{brown2017multiple} a variant of the $S$-cocycle and generating series of MMVs $\mathbb S^\mm$ is defined and denoted by $\mathcal{C}^\mm_S$.
Upon projection to the MMVs that are expressible in terms of MZVs, the series $\mathcal{C}^\mm_S$ is equivalent to our generating series $\mathbb{S}^\mm$ albeit expressed in a different form using an additional SL$(2)$-doublet of commutative variables $X,Y$. A second and more superficial difference in our conventions is the order of integrations, which leads to a reversed order in the concatenation of generating series when compared to Brown's work. The conjectural translation to our objects and conventions is summarized in table \ref{table.tab1}.
\begin{table}[H]
\centering
\renewcommand{\arraystretch}{1.5}
\begin{tabular}{|c|c|}
\hline
Object in 
\cite{brown2017multiple} & Object in this paper's framework\\
\hline
$\mathcal{C}_S^\mm$&$\mathbb{S}^\mm$\\ \hline
$    \phi^\mm( \mathbb Y ) $&$\left(\mathbb{M}_z^\mm\right)^{-1} \mathbb Y  \mathbb{M}^\mm_z$\\ \hline
 $  b^\mm $&$ \left(\mathbb{M}_{\sigma}^\mm\right)^{-1}\mathbb{M}_z^\mm $\\ \hline
  $  \mathbb Y |_\gamma$&$ U_\gamma^{-1} \mathbb Y U_\gamma$
  \\ \hline
  $s_S$& $\mathbb{X}_S$\\
\hline
\end{tabular}
\caption{A conjectural translation between the objects in \cite{brown2017multiple} and the objects discussed in this paper, where $\mathbb Y$ is a placeholder for a generic series in $\epsilon_k^{(j)}$ or its equivalent upon translation into the conventions of \cite{brown2017multiple}. The action of $U_\gamma$ for modular transformations $\gamma \in {\rm SL}(2,\mathbb Z)$ is defined by the transformation $\mathbb A(\epsilon_k;\gamma\cdot \tau) = U_\gamma^{-1} \mathbb A(\epsilon_k; \tau) U_\gamma$ of the connection form in (\ref{eq:DefineA}). For arbitrary $\gamma \in {\rm SL}(2,\mathbb Z)$, the action of $U_\gamma$ on $\epsilon_k^{(j)}$ can be constructed from that of $U_S$ and $U_T$ for the generators $S$ and $T$ given by (\ref{eq:S1}) and (\ref{eq:T1}), respectively.}
\label{table.tab1}
\end{table}

The Galois action on $\mathcal{C}_S^\mm$ is given in section 15.3.3 of \cite{brown2017multiple} in terms of two group-like series
$b^\mm$ and $\phi^\mm$, the latter being an automorphism acting on the bookkeeping variables of $\mathcal{C}_S^\mm$.
Upon projection to MMVs that are expressible in terms of MZVs, both of $b^\mm$ and $\phi^\mm$ reduce to series in motivic MZVs with $\mathbb Q[(2\pi i)^{- 1}]$ coefficients as given in table \ref{table.tab1}.
Similar translations for the single-valued versions $b^{\rm sv}$ and $\phi^{\rm sv}$ of these objects were also discussed in detail in section 4.1 of \cite{Dorigoni_2022} and section 3.2 of \cite{Dorigoni:2024oft}.

We shall now apply the dictionary in table \ref{table.tab1} to the representation of the $S$-cocycle in Remark 15.12 of \cite{brown2017multiple}
\begin{align}
\label{FBsSm}
\mathcal{C}_S^\mm =( b^\mm)^{-1}|_S \phi^\mm(s_S) b^\mm \, .
\end{align}
With the translation between 
$b^\mm, \phi^\mm$ and
the series $\mathbb M_\sigma^\mm$, $\mathbb M_z^\mm$ in zeta generators of table \ref{table.tab1}, and keeping in mind the reversed order of the letters $\epsilon_k^{(j)}$, the right-hand side corresponds to
\begin{align}
\label{FBsSm.2}
\underbrace{ (\mathbb M_\sigma^\mm)^{-1} \mathbb M_z^\mm }_{ \sim \, b^\mm }  \, \underbrace{ (\mathbb M_z^\mm)^{-1}
\mathbb{X}_S 
\mathbb M_z^\mm  }_{\sim \,  \phi^\mm(s_S) }\,
\underbrace{ U_S
(\mathbb M_z^\mm)^{-1}
\mathbb M_\sigma^\mm
U_S^{-1} }_{ \sim \, ( b^\mm)^{-1}|_S }
= 
(\mathbb M_\sigma^\mm)^{-1}
\mathbb{X}_S  U_S  \mathbb M_\sigma^\mm U_S^{-1} \, ,
\end{align}
where we used SL$(2)$-invariance of the arithmetic parts $z_w$ of zeta generators to commute $U_S^{-1}$ with $\mathbb M_z^\mm$. Finally, taking the reversed concatenation orders between the series of \cite{brown2017multiple} and those of this work into account, (\ref{FBsSm}) with the translation (\ref{FBsSm.2}) of the right-hand side reproduces our conjecture (\ref{ourSm.01}) for the expansion of $\mathbb S^\mm$.

From the information on its counterpart $s_S$ in~\cite[\S 15.6]{brown2017multiple}, the object $\mathbb{X}_S^\mm$ is trivial under the coaction, i.e.\ $\Delta \mathbb{X}_S^\mm=\mathbb{X}_S^\mm$.
This can be seen from the fact that the expansion of $\mathbb{X}_S^\mm$ as a series in $\epsilon_k^{(j)}$ subject to Pollack relations has coefficients in $\mathbb Q[\pi^2]$ as noted below (\ref{ourSm.01}). We reiterate that the whole information about non-trivial MZVs in (\ref{FBsSm.2}) is carried by the series $\mathbb{M}^\mm_\sigma$ in the genus-one zeta generators.

In summary, the conjectural translations of table \ref{table.tab1} lead us to deduce the factorized form (\ref{ourSm.01}) of the $S$-cocycle from the results of Brown's work \cite{brown2017multiple}. This adds an alternative derivation to the informal reasoning of section \ref{sec:4.inf}.

\subsection{Comments on the factorized form}
\label{sec4.com}

This section gathers three comments on the factorized form (\ref{ourSm.01}) of the $S$-cocycle.

\subsubsection{Simplified coaction and de Rham version of the $S$-cocycle}
\label{sec4.com.1}

There are two immediate corollaries of (\ref{ourSm.01}). First, by the $\mathbb Q[\pi^2]$ coefficients in  $\mathbb{X}_S^\mm$ together with
$\zeta_2^\dr=0$, we find a trival de Rham series $\mathbb{X}_S^\dr=\textbf{1}^\dr$, and the de Rham version of (\ref{ourSm.01}) reduces to
\begin{align}
\label{eq:Sdr}
    \mathbb{S}^\dr =\left(\mathbb{M}_\sigma^\dr\right)^{-1} U_S^{-1}\mathbb{M}_\sigma^\dr U_S\, . 
\end{align} 
Second, taking the coaction of (\ref{ourSm.01}) using $\Delta \mathbb{M}^\mm_\sigma = \mathbb{M}_\sigma^\mm\mathbb{M}_\sigma^\dr$ results in
\begin{align}
\label{mmvcoactioncorollary}
    \Delta \mathbb{S}^\mm 
     = (\mathbb{M}_{\sigma}^\dr)^{-1}\mathbb{S}^\mm  U_S^{-1} \mathbb{M}_{\sigma}^\dr U_S\, . 
\end{align}
Consistency with the earlier coaction formula (\ref{mmvcoacgen}) for the $S$-cocycle readily follows from~(\ref{eq:Sdr}).

\subsubsection{Relation to the proposed coaction of iterated Eisenstein integrals}
\label{sec4.com.2}

We shall here make contact with the proposed coaction (\ref{eq:conj}) of iterated Eisenstein integrals.

First, the regularized limit $\tau\rightarrow 0$ relates our main proposal (\ref{eq:conj}) to the motivic coaction of MMVs encoded in (\ref{mmvcoacgen}). One can reverse the logic and derive (\ref{mmvcoacgen}) from the factorized form (\ref{ourSm.01}) of the $S$-cocycle through the corollaries in section \ref{sec4.com.1}. With this viewpoint, Remark 15.12 in Brown's work \cite{brown2017multiple} together with the conjectural translations of table \ref{table.tab1} imply that the proposed coaction (\ref{eq:conj}) of iterated Eisenstein integrals commutes with the regularized $\tau \rightarrow 0$ limit.

Secondly, and in some way conversely, comparing our proposal in this limit to the statement of 
\cite{brown2017multiple} is equivalent to understanding what the de Rham versions of MMVs are. As elaborated in section \ref{sec_motivicderahm}, a similar comparison and understanding of the de Rham versions of iterated Eisenstein integrals should be obtained in the future.

\subsubsection{$T$-cocycle}
\label{sec4.com.3}

Since Remark 15.12 of 
\cite{brown2017multiple} is not only valid for $\mathcal{C}^\mm_S$, but in fact for the cocycle $\mathcal{C}^\mm_\gamma$ associated with arbitrary modular transformations $\gamma \in\text{SL}(2,\mathbb{Z})$, we can also write down a generating series for any $\gamma$ and specifically for $T$: Defining
the $T$-cocycle $\mathbb T$ via
\begin{align}
\Ip(\epsilon_k; \tau{+}1)= \mathbb{T}(\epsilon_k) U_T^{-1} \mathbb{I}(\epsilon_k;\tau)U_T
\label{Tsec.01}
\end{align}
by analogy with (\ref{Sfull}) for the $S$-cocycle and matching with the modular $T$ transformation (\ref{eq:Ttrm}) identifies\footnote{This does not fully line up with the conventions for the quantity $\mathbb T$ in (4.13) of \cite{Dorigoni:2024oft} since
its defining equation (4.8) in the reference differs from our defining equation (\ref{Tsec.01}).}
\begin{align}
\mathbb{T}(\epsilon_k) = e^{2\pi i N} e^{2\pi i \ep_0}
\label{Tsec.02}
\end{align}
with $U_T = \exp(2\pi i \ep_0)$.
Since the expression (\ref{defNN}) for the series $N$ in Tsunogai derivations has rational coefficients, the expansion of the $T$-cocycle (\ref{Tsec.02}) in $\epsilon_k^{(j)}$ only involves rational powers of $2\pi i$ as coefficients. Accordingly, the right side of (\ref{Tsec.02}) is a valid expression for the series
\begin{align}
\mathbb{X}_T(\epsilon_k)  =  e^{2\pi i N} e^{2\pi i \ep_0}
\label{Tsec.03}
\end{align}
in the second factor on the right side of our translation
of Remark 15.12 of \cite{brown2017multiple}: 
\begin{align}
    \mathbb{T}^\mm = (\mathbb{M}^\mm_\sigma)^{-1} \mathbb{X}_T^\mm U_T^{-1} \mathbb{M}_\sigma^\mm U_T\, . 
    \label{ttmm}
\end{align}
Indeed, $\mathbb{X}_T U_T^{-1} = \exp(2\pi i N)$ together with the fact that $N$ commutes with zeta generators \cite{hain_matsumoto_2020, Brown:2017qwo2}, i.e.\ $(\mathbb{M}^\mm_\sigma)^{-1} \exp(2\pi i N)\mathbb{M}_\sigma^\mm = \exp(2\pi i N)$, reproduces the right side of (\ref{Tsec.02}) from (\ref{ttmm}).
Similarly, we can also find the motivic coaction or equivalently the de Rham version of this series in the same way we did for $\mathbb{S}^\mm$. 

\subsection{Computational advantages}
\label{sec:4.2}

The factorized form~\eqref{ourSm.01} can also be tested numerically by using the standard conjecture that motivic MZVs are in bijection with real MZVs.

For MMVs $\MMV{j_1 &\ldots & j_\ell}{k_1 &\ldots & k_\ell}$ at modular depth $\ell\leq 3$ and
degree $\sum_{i=1}^\ell k_i \leq 16$, we can insert their $f$-alphabet representations into the expansion (\ref{eq:limitS}) of $\mathbb{S}^\mm$ and test (\ref{ourSm.01}) after inserting the Pollack relations (\ref{pollrels}). Instead of matching with the expression (\ref{ourSm.01}) for $\mathbb{S}^\mm$ with the unspecified series $\mathbb X_S^\mm$ in $\zeta_2^\mm$, we have in a first pass checked the coaction formula (\ref{mmvcoactioncorollary}) order by order in $\epsilon_k^{(j)}$. The  deconcatenation coaction (\ref{eq:fcoac}) of MZVs on the left-hand side was found to match the coefficients of $\epsilon_{k_1}^{(j_1)}\ldots \epsilon_{k_\ell}^{(j_\ell)}$ on the right-hand side of (\ref{mmvcoactioncorollary}) for all degrees $(k_1,\cdots,k_\ell)$ that we tested, namely:

\begin{center}
\begin{tabular}{c|l}
mod.\ depth $\ell$ & degrees $(k_1,\cdots,k_\ell)$\\ \hline 
2 & (4,4),$\, $ (4,6),$\, $ (6,6),$\, $ (4,10),$\, $ (6,8),$\, $ (4,12),$\, $ (6,10),$\, $ (8,8)\\
3&(4,4,4),$\, $ (4,4,6),$\, $ (4,4,8),$\, $ (4,6,6)
\end{tabular}
\end{center}

In a second pass, we have used the known MMVs at these degrees to determine the series $\mathbb{X}^\mot_S$ by formally setting $f_{2k+1}\rightarrow 0$. The contributions up to degree $10$ are given by 
\begin{align}
\rho\left(\mathbb{X}^\mot_S\right) &= 
 1 + \frac1{12} f_2 \ep_4^{(1)} - \frac1{1440} f_2 \ep_6^{(1)}+ \frac1{144} f_4 \ep_6^{(3)}  +
 \frac{1}{90720}f_2 \ep_8^{(1)} - \frac{1}{43200} f_4 \ep_{8}^{(3)} + \frac1{4320} f_6 \ep_8^{(5)} 
  \nn\\
&\quad + \frac{209}{14400} f_4 \left[\ep_4, \ep_4^{(2)}\right]+ \frac{5}{576} f_4 \ep_4^{(1)} \ep_4^{(1)}  
 -\frac1{4838400}f_2 \ep_{10}^{(1)}  + \frac{1}{5080320}f_4 \ep_{10}^{(3)} \nn\\
&\quad - \frac1{2419200} f_6 \ep_{10}^{(5)} + \frac1{241920} f_8 \ep_{10}^{(7)}  -\frac{2337}{1792627200} f_2 \left[\ep_4 ,\ep_6\right] - \frac{1187}{15240960}f_4 \left[\ep_4 ,\ep_6^{(2)} \right]\nn\\
&\quad   -\frac{107}{2540160} f_4 \ep_4^{(1)}\ep_6^{(1)}  + \frac{437}{5080320}f_4 \left[\ep_4^{(2)} ,\ep_6\right]  -\frac{521}{5080320} f_4 \ep_6^{(1)} \ep_4^{(1)} \nn\\
&\quad + \frac{437}{1451520} f_6 \left[\ep_4,\ep_6^{(4)} \right] + \frac{521}{725760}f_6 \ep_4^{(1)} \ep_6^{(3)}  + \frac{107}{362880} f_6 \ep_6^{(3)} \ep_4^{(1)} - \frac{1187}{725760}f_6 \left[\ep_4^{(2)},\ep_6^{(2)} \right]\nn\\
&\quad - \frac{2377}{853632} f_8 \left[\ep_4^{(2)} ,\ep_6^{(4)} \right]+ \ldots\,,
\label{expxs}
\end{align}
and the analogous terms at degree $\le 16$ and modular depth $\leq 3$ can be found in an ancillary file in the arXiv submission of this work. On the right-hand side of (\ref{expxs}), we have used the notation $f_{2n} = \frac{\zeta_{2n}^\mot}{(\zeta_2^\mot)^n} f_2^n$. 
One can easily check from the expansion in (\ref{expxs}) and the ancillary file that $\mathbb{X}_S^\mot$ is group-like (i.e.\ $\log \mathbb{X}_S^\mot$ is Lie-algebra valued) and obeys $(\mathbb{X}_S^\mot)^T\mathbb{X}_S^\mot = \textbf{1}^\mm$ to the respective degrees and modular depths. 

The terms of modular depth one are in fact available at arbitrary degree,
\begin{align}
\mathbb{X}^\mot_S &=  1 + 
2 \sum_{k=4}^\infty \sum_{j=1}^{k-3}  \frac{\zeta^\mot_{j+1}}{(k{-}2)!}  \, \frac{B_{k-j-1}}{(k{-}j{-}1)} \, \epsilon_k^{(j)}
+ \ldots
\label{clXone}
\end{align}
This expression is in agreement with~\cite[Eq.~(15.12)]{brown2017multiple}. We do not have a method of predicting closed expressions for $\mathbb{X}_S^\mot$ at arbitrary modular depth and degree at present.

Note that $\mathbb{X}_S^\mot$ depends on the choice of the $f$-alphabet isomorphism $\rho$ since the images of $\mathbb Q$-independent indecomposable motivic MZV $\zeta^\mm_{n_1,\ldots,n_r}$ of depth $r\geq 2$ can be shifted by rational multiples of $f_{n_1+\ldots+n_r}$. This firstly occurs at weight $n_1+\ldots+n_r=8$ where the representative $\zeta^{\mm}_{3,5}$ of indecomposable MZVs enters the series $U_S^{-1} \mathbb M_\sigma^\mm U_S$ and $\mathbb M_\sigma^\mm$ with rational multiples of $[\ep_4 ,\ep_6] $ and $ [\ep_4^{(2)} ,\ep_6^{(4)} ]$, respectively, as well as higher-degree brackets. The expansion (\ref{expxs}) is tailored to the choice $\rho(\zeta_{3,5}^\mm) = -5 f_3 f_5+ \frac{100471}{35568} f_8$, but alternative choices of $\rho$ may have a different rational coefficient of $f_8$ which would then modify the above terms $ -\frac{2337}{1792627200} f_2 [\ep_4 ,\ep_6] - \frac{2377}{853632} f_8  [\ep_4^{(2)} ,\ep_6^{(4)} ]$ in $\mathbb{X}_S^\mot$. The full composition $(\mathbb{M}^\mm_\sigma)^{-1} \mathbb{X}_S^\mm U_S^{-1}\mathbb{M}_\sigma^\mm
U_S $ in the factorized form (\ref{ourSm.01}) of $\mathbb S^\mm$ is of course independent of $\rho$ since a change of $\rho^{-1}(f_{i_1}\cdots f_{i_r})$ induces a compensating change in the zeta generators if $r$ is odd and in $\mathbb{X}_S^\mot$ for even $r$.

\subsubsection{Reduction to fixing rational numbers}

The factorized form \eqref{ourSm.01} represents an enormous computational advantage in determining the generating series (\ref{spluscocyc}) of MMVs. For every word in the $\ep_k^{(j)}$ that is independent under Pollack relations, there is no more than a single rational unknown to be fixed if the series $\mathbb M_\sigma$ and
all lower-degree terms of the series $\mathbb X_S$ in \eqref{ourSm.01} are taken into account.

The transcendental weight of a coefficient of a given term $\ep_{k_1}^{(j_1)}\cdots \ep_{k_\ell}^{(j_\ell)}$ in both series $\mathbb X_S$ and $\mathbb S$ is $w=\sum_{i=1}^\ell (j_i{+}1)$. Since $\mathbb{X}_S$ is a series in $\mathbb{Q}[\pi^2]$, this restricts $w$ to be even and therefore the allowed values of $j_i$. We can determine the series $\mathbb{X}_S$ iteratively in degree.
%
In order to fix the rational coefficient of a new term in $\mathbb{X}_S$, one needs to find a corresponding MMV at lowest degree where this term contributes for the first time, and we rely on the accessibility of zeta generators to arbitrary degrees \cite{Brown:depth3, hain_matsumoto_2020, Dorigoni:2024iyt}.

It is convenient to determine the single unknown rational number in an MMV by numerical methods. The $q$-series of Eisenstein series in (\ref{nukerdef}) together with a reparametrization of the integration contour in the definition (\ref{eq:mmvdef}) of MMVs into two copies of $(i , i\infty)$ give access to very high numerical precision. The unkown rational number can then be determined to the same precision by applying PSLQ. This in turn fixes higher orders of zeta generators or the series $\mathbb X_S$ and can be used to determine higher-degree MMVs up to a single rational number. 

\subsubsection{Practicalities of using the factorized form of $\mathbb S$}
\label{sec.expS}

We shall here elaborate on the concrete steps towards determining MMVs modulo Riemann zeta values through the factorized form of $\mathbb S$. For this purpose, it is convenient to expand the right side of (\ref{ourSm.01}) in the odd generators $f_{2k+1}$ of the $f$-alphabet and to isolate the arithmetic parts $z_w$ of zeta generators in nested commutators. The reflected versions of the zeta generators in the expansion of $ U_S^{-1} \mathbb M^\mm_\sigma U_S $ will be denoted by
\begin{align}
\csig_w = U_S^{-1} \sigma_w U_S =\sigma_w \, \big|_{\epsilon_k^{(j)} \, \rightarrow \, (-1)^j \frac{(2\pi i)^{k-2-2j} \, j!}{ (k{-}j{-}2)! } \epsilon_k^{(k-j-2)} }\, .
\label{exp.01}
\end{align}
By SL$(2)$-invariance of the arithmetic terms, we have $U_S^{-1}z_w U_S = z_w$ and therefore
\begin{align}
\csig_w = z_w - \frac{\epsilon_{w+1}}{(2\pi i)^{w-1}} + \ldots
\label{exp.00}
\end{align}
with a Lie series in $\epsilon_k^{(j)}$ of modular depth $\geq 2$ in the ellipsis. Since all terms of modular depth two of $\sigma_w$ are known from (\ref{extintr.10}), we can straightforwardly obtain those of $\csig_w$ from the reflection in (\ref{exp.01}).

In order to assemble contributions to $\mathbb S$ with a fixed number of $f_{2k+1}$ and combine the $z_w$ into nested brackets, we
expand
\begin{align}
\rho(\mathbb S^\mm) &= 
\bigg\{ 1 - \sum_{i_1 \in 2\mathbb N+1} f_{i_1} \sigma_{i_1}
+ \sum_{i_1,i_2 \in 2\mathbb N+1} f_{i_1} f_{i_2} \sigma_{i_2} \sigma_{i_1} + \ldots \bigg\}
\label{exp.03} \\
&\quad \mathbb X^\mm \shuffle \bigg\{ 1 + \sum_{r_1 \in 2\mathbb N+1} f_{r_1} \csig_{r_1}
+ \sum_{r_1,r_2 \in 2\mathbb N+1} f_{r_1} f_{r_2} \csig_{r_1} \csig_{r_2} + \ldots \bigg\} \notag
\end{align}
and split $\sigma_w,\csig_w$ into their common arithmetic term $z_w$ and
different geometric terms $\sigma_w^{\rm g}, \csig_w^{\rm g}$
\begin{align}
\sigma_w = z_w + \sigma_w^{\rm g} \, , \ \ \ \ \ \
\csig_w = z_w + \csig_w^{\rm g}\, .
\label{exp.02}
\end{align}
After some regrouping, we obtain
\begin{align}
\rho(\mathbb S^\mm) &= \mathbb X^\mm + \sum_{i_1 \in 2\mathbb N+1} f_{i_1} \Big( \mathbb X^\mm   \csig^{\rm g}_{i_1}
- \sigma^{\rm g}_{i_1} \mathbb X^\mm 
+ [\mathbb X^\mm,z_{i_1}] \Big)  \label{exp.04}
\\
&\quad + \sum_{i_1,i_2 \in 2\mathbb N+1} f_{i_1} f_{i_2} \Big(
\sigma^{\rm g}_{i_2} \sigma^{\rm g}_{i_1} \mathbb X^\mm
- \sigma^{\rm g}_{i_1} \mathbb X^\mm \csig^{\rm g}_{i_2}
- \sigma^{\rm g}_{i_2} \mathbb X^\mm \csig^{\rm g}_{i_1}
+ \mathbb X^\mm \csig^{\rm g}_{i_1} \csig^{\rm g}_{i_2}
 \notag \\
&\quad \quad \quad \quad \quad \quad
+ \sigma^{\rm g}_{i_1} [z_{i_2} , \mathbb X^\mm] 
+ \sigma^{\rm g}_{i_2} [z_{i_1} , \mathbb X^\mm]
+ [\mathbb X^\mm, z_{i_1}] \csig^{\rm g}_{i_2}
+ [\mathbb X^\mm, z_{i_2}] \csig^{\rm g}_{i_1} \notag \\
 &\quad \quad \quad \quad \quad \quad
+ [z_{i_2}, \sigma^{\rm g}_{i_1} ] \mathbb X^\mm
+ \mathbb X^\mm [ \csig^{\rm g}_{i_1} , z_{i_2} ]
+ \big[ [\mathbb X^\mm , z_{i_1}], z_{i_2} \big]
\Big) +\ldots \, , \notag
\end{align}
where all the arithmetic parts can be eliminated in favor of Lie polynomials
in $\epsilon_k^{(j)}$ by repeated use of the bracket relations (\ref{zepcom}). 

One can separately expand the coefficients of $f_{i_1}$ or $f_{i_1} f_{i_2}$ in (\ref{exp.04}) according to modular depth of the series in $\epsilon_k^{(j)}$: 
\begin{itemize}
\item For the coefficient of $f_{i_1}$, terms of modular depth one are obtained by setting $\mathbb X^\mm \rightarrow \textbf{1}^\mm$ in the first line of (\ref{exp.04}) and replacing $\csig^{\rm g}_{i_1},\sigma^{\rm g}_{i_1}$ by the simplest geometric terms $\epsilon_{i_1+1} , \epsilon_{i_1+1}^{(i_1-1)}$. This reproduces the odd zeta values in the expressions (\ref{mmvd1}) for $\MMV{0}{k} ,\MMV{k-2}{k} $.
\item Terms with $f_{i_1}$ at modular depth two are still accessible to arbitrary degree, starting with the contributions of (\ref{extintr.10}) to zeta generators at modular depth two while setting $\mathbb X^\mm \rightarrow \textbf{1}^\mm$. A second class of terms stems from the closed form (\ref{clXone}) for $\mathbb X^\mm$ at modular depth one which is either accompanied by the simplest geometric terms of $\csig^{\rm g}_{i_1},\sigma^{\rm g}_{i_1}$ or enters the bracket with $z_{i_1}$ (see (\ref{zepcom}) for their contributions at modular depth two).
\item The coefficient of $f_{i_1} f_{i_2}$ starts at modular depth two with the $\mathbb X^\mm \rightarrow \textbf{1}^\mm$ contributions $\sigma^{\rm g}_{i_2} \sigma^{\rm g}_{i_1} 
- \sigma^{\rm g}_{i_1} \csig^{\rm g}_{i_2}
- \sigma^{\rm g}_{i_2}\csig^{\rm g}_{i_1}
+  \csig^{\rm g}_{i_1} \csig^{\rm g}_{i_2} + [z_{i_2}, \sigma^{\rm g}_{i_1} ]  +[ \csig^{\rm g}_{i_1} , z_{i_2} ]$, restricted to the simplest geometric terms of zeta generators and the modular-depth-two term of $[z_{i_1},\epsilon_{i_2+1}]$ in (\ref{zepcom}).
\item Terms with $f_{i_1} f_{i_2}$ at modular depth three first of all require the expansion of the zeta generators and brackets from the earlier $\mathbb X^\mm \rightarrow \textbf{1}^\mm$ terms to subleading modular depth (see section 7.4 of \cite{Dorigoni:2024iyt} for partial results on $[z_{i_1},\epsilon_{i_2+1}]$ at modular depth three). A second class of contributions arises from inserting the terms (\ref{clXone}) of $\mathbb X^\mm$ at modular depth one into the last three lines of (\ref{exp.04}), again using the leading modular depth of zeta generators and brackets with $z_w$.
\end{itemize}
More generally, the coefficients  of $f_{i_1} \ldots f_{i_\ell}$ in $\rho(\mathbb S^\mm)$ start with terms at modular depth $\ell$ which can still be determined from the terms of leading modular depth in $\sigma_w,\csig_w$ and $[z_w,\epsilon_k]$. For MMVs $\MMV{j_1 &\ldots &j_\ell}{k_1 &\ldots & k_\ell} $ with extremal values $j_i \in \{0, k_i{-}2 \}$ for all of $i=1,2,\ldots,\ell$, the arithmetic terms do not contribute, and one can derive closed formulae such as \cite{saad2020multiple}
\begin{align}
\MMV{0 &\ldots &0}{k_1 &\ldots & k_\ell} = \frac{(-2\pi i)^\ell \, \rho^{-1}(f_{k_1 - 1} \ldots f_{k_\ell - 1})}{(k_1{-}1) \ldots (k_\ell{-}1)} + \ldots\, ,
\end{align}
where the terms in the ellipsis are MZVs with $<\ell$ letters $f_{2k+1}$ and periods beyond MZVs that drop out from $\mathbb S$ upon dressing with Tsunogai derivations and inserting Pollack relations.

\section{Composition with modular transformations}
\label{sec:5}

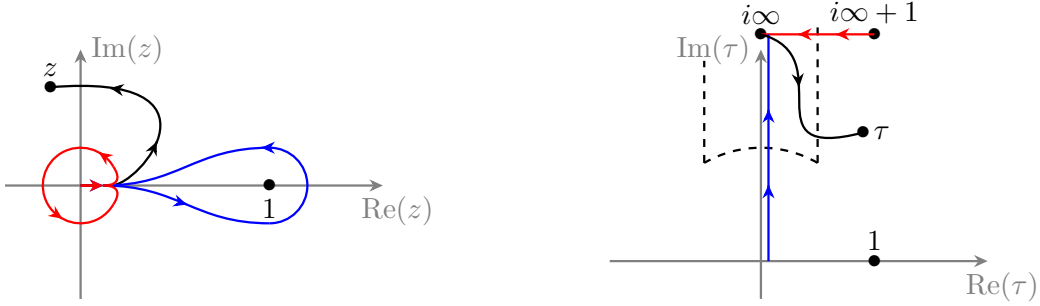
\begin{figure}[H]
\centering
    \begin{tikzpicture}[line width=0.30mm]
\def\thyck{0.5mm}
\begin{scope}[xshift=-4cm]
  \draw [arrows={-Stealth[width=1.6mm, length=1.8mm]}, color=gray] (-1,0) -- (4.0,0) node[below]{$\quad{\rm Re}(z)$};
    \draw [arrows={-Stealth[width=1.6mm, length=1.8mm]}, color=gray] (0,-1.5) -- (0,1.8) node[right]{${\rm Im}(z)$};
\draw (0.4,0) .. controls (0.8,0) and (2,1.5) .. (-0.4,1.3);
\draw(-0.4,1.3)node{$\bullet$}node[above]{$z$};
 \draw[arrows={-Stealth[width=1.6mm, length=1.8mm]}](0.4,1.288)--(0.39,1.2895);
 \draw[arrows={-Stealth[width=1.6mm, length=1.8mm]}](1.008,0.58)--(1.00845,0.581);
      \draw[blue](2.5,-.5) arc (-90:90:.5);
\draw[blue,arrows={-Stealth[width=1.6mm, length=1.8mm]}](0,0)--(0.3,0);
\draw(0,0)--(0.2,0);
  \draw (2.5,0)node{$\bullet$} ;
  \draw(2.5,-0.3)node{$1$};
\draw[blue]  (0.3,0) .. controls (1.3,-0.0) and (1.5,-.5) .. (2.5,-.5);
\draw[blue]  (0.3,0) .. controls (1.3,0.0) and (1.5,.5) .. (2.5,.5);
%
%
 \draw[blue,arrows={-Stealth[width=1.6mm, length=1.8mm]}](1.4,-0.248)--(1.41,-0.251);
 %
  %
 \draw[blue,arrows={-Stealth[width=1.6mm, length=1.8mm]}](2.4,.5)--(2.39,.5);
 \draw[red] (0.433,0.25) arc (30:330:0.5); 
\draw[red] (0.3,0) .. controls (0.55,0.0) and (0.5,0.15) .. (0.433,0.25);
\draw[red] (0.3,0) .. controls (0.55,0.0) and (0.5,-0.15) .. (0.433,-0.25);
\draw[red,arrows={-Stealth[width=1.6mm, length=1.8mm]}]
  (-0.25,-0.433)--(-0.24,-0.44);
  \draw[red,arrows={-Stealth[width=1.6mm, length=1.8mm]}]
  (0.25,0.433)--(0.24,0.44);
\draw[red,arrows={-Stealth[width=1.6mm, length=1.8mm]}](0,0)--(0.3,0);
\draw[red](0,0)--(0.2,0);
\end{scope}
\begin{scope}[xshift=4cm]
  \draw [arrows={-Stealth[width=1.6mm, length=1.8mm]}, color=gray] (-1,-1) -- (4.0,-1) node[below]{$\quad{\rm Re}(\tau)$};
    \draw [arrows={-Stealth[width=1.6mm, length=1.8mm]}, color=gray] (1,-1.5) -- (1,1.8) node[left]{${\rm Im}(\tau)$};
\draw[dashed] (1+1.5*0.5,-1+1.5*0.866) arc (60:120:1.5);
\draw[dashed](1+1.5*0.5,-1+1.5*0.866) -- (1+1.5*0.5,-1+1.5*0.866+1.8);
\draw[dashed](1-1.5*0.5,-1+1.5*0.866) -- (1-1.5*0.5,-1+1.5*0.866+1.35);
\draw(1,2)node{$\bullet$}node[above]{$i\infty$};
\draw(1+1.5,2)node{$\bullet$}node[above]{$i\infty+1$};
\draw(1+1.5,-1)node{$\bullet$}node[above]{$1$};
\draw[red](1,2) -- (1+1.5,2);
 \draw[red,arrows={-Stealth[width=1.6mm, length=1.8mm]}](2,2)--(1.99,2);
  \draw[red,arrows={-Stealth[width=1.6mm, length=1.8mm]}](1.5,2)--(1.49,2);
\draw[blue](1.1,2) -- (1.1,-1);
 \draw[blue,arrows={-Stealth[width=1.6mm, length=1.8mm]}](1.1,1)--(1.1,1.01);
  \draw[blue,arrows={-Stealth[width=1.6mm, length=1.8mm]}](1.1,0)--(1.1,0.01);
\draw (1,2) .. controls (2.1,1.7) and (0.8,0.3) .. (2.35,0.7)node{$\bullet$}node[right]{$\tau$};
 \draw[arrows={-Stealth[width=1.6mm, length=1.8mm]}](1.504,1.31)--(1.505,1.3);
    \end{scope}
\end{tikzpicture}
\caption{Left panel: Discontinuities of MPLs arise from incorporating the loops around points \textcolor{red}{$z=0$} and \textcolor{blue}{$z=1$} on the Riemann sphere into the integration path in their definition (\ref{eq:MPL}). Right panel: The integration paths in the upper half-plane corresponding to generators \textcolor{blue}{$S$} and \textcolor{red}{$T$} of the modular group and arising in the cocycle factors of the generating series (\ref{eq:GenSerinA}).}
\label{fig:disc}
\end{figure}

At genus zero, the motivic coaction has a distinct behaviour under discontinuities of MPLs $G(a_1,\ldots,a_w;z)$ corresponding to deformations of the integration paths in (\ref{eq:MPL}) by an extra loop around the singular points $a_i$ of the integrand (here restricted to $a_i\in\{0,1\}$)  as visualized in the left panel of figure \ref{fig:disc}: The discontinuities with respect to any of these loops around $a_i$ solely act in the first entry \cite{Duhr:2012fh} (see footnote \ref{otimesft} for the notation ${\rm Disc} \otimes \mathbf{1}$) 
\begin{align}
\Delta \,{\rm Disc} \,  G^\mm(a_1,\ldots,a_w;z) = ({\rm Disc} \otimes \mathbf{1}) \, \Delta \, G^\mm(a_1,\ldots,a_w;z)
\, ,
\label{MPLdisk}
\end{align}
reflecting the fact that the de Rham periods in the second entry are only defined up to discontinuities.
In this section, we investigate possible echos of the composition property (\ref{MPLdisk}) at genus one.

Since iterated Eisenstein integrals depend solely on the modular parameter $\tau$ of the torus and not on any marked points, their genus-one analogue of discontinuities of MPLs are modular transformations. Instead of the loops in the integration path for the argument $z$ of MPLs, iterated Eisenstein integrals receive contributions from the paths in the right panel of figure \ref{fig:disc} upon modular $T$ and $S$ transformations of $\tau$.
Hence, we will start in section \ref{sec:mod.1} by determining the composition of the proposed coaction (\ref{eq:conj}) of iterated Eisenstein integrals with their modular transformations.

In contrast to the single-valued KZ connection in the construction (\ref{genpoly}) of MPLs, the connection form $ \mathbb{A} (\epsilon_k;\tau)$ in (\ref{eq:DefineA}) transforms equivariantly under $\gamma \in {\rm SL}(2,\mathbb{Z})$ (see section \ref{eqvIsec} for the notation $\gamma\cdot\tau$),
\begin{align}
\mathbb{A} (\epsilon_k;\gamma\cdot\tau)
 = U_\gamma^{-1} \mathbb{A} (\epsilon_k;\tau) U_\gamma
\label{Aequiv}
\end{align}
with $U_T = \exp(2\pi i \epsilon_0)$ and $U_S$ action given by (\ref{eq:S1}). Accordingly, the genus-one analogue of single-valued MPLs are the equivariant iterated Eisenstein integrals of section \ref{eqvIsec} whose generating series $\mathbb I^{\rm eqv} (\epsilon_k;\tau)$ transforms with the same factor of $U_\gamma$ as the connection by (\ref{eqv:uni}).

On these grounds, the best one can ask for in a genus-one coaction is that the de Rham entry shares the equivariant ${\rm SL}(2,\mathbb{Z})$ transformation of (\ref{Aequiv}) as an echo of the absence of discontinuities in the second entry of the genus-zero coaction (\ref{MPLdisk}). 
The more general statement including this special case is that \textit{any} de Rham period is defined from two de Rham representatives (as in (\ref{deRhamrealisatuion})), so it \textit{doesn't know} how an integration cycle changes under discontinuities in marked points or modular transformations of period matrices. The cocycles $\mathbb S(\epsilon_k)$ and $e^{2\pi i N} e^{2\pi i \ep_0}$ of the modular $S$- and $T$-transformation reflect such a change of integration cycle for $\tau$ in the upper half-plane, so they are expected to drop out from the de Rham entry of a genus-one coaction.

To express these expectations on modular transformation properties, we define the following operation  $\gamma_{\text{\sout{path}}}$ that only transforms the connection of iterated Eisenstein integrals as in (\ref{Aequiv}) and not their integration path in (\ref{eq:GenSerinA}):
\begin{align}
\label{presdef}
\gamma_{\text{\sout{path}}}\big[ \mathbb I(\epsilon_k;\tau) \big] = U_\gamma^{-1} \mathbb I(\epsilon_k;\tau)  U_\gamma
\, , \ \ \ \ \ \ 
\gamma \in {\rm SL}(2,\mathbb{Z})
\end{align}
The action of ${\cal S}_{\text{\sout{path}}}$ and
${\cal T}_{\text{\sout{path}}}$ on the expansion in $\epsilon_k^{(j)}$ via $U_S$ and $U_T$ can be found in (\ref{eq:S1}) and (\ref{eq:T1}), respectively. For $\tau$-independent quantities including $\mathbb M_\sigma$ or $\mathbb M_z$, the
operation $\gamma_{\text{\sout{path}}}$ is taken to act trivially since (\ref{presdef}) aims to implement the transformation $\tau \rightarrow \gamma \cdot \tau$ up to the cocycle factors that were argued not to fit in the de Rham entry. This leads e.g.\ to $\gamma_{\text{\sout{path}}}[ \mathbb M_\sigma ] = \mathbb M_\sigma$, and for the connection $ \mathbb{A} (\epsilon_k;\tau)$ in (\ref{eq:DefineA}), the two transformations agree, 
$\gamma_{\text{\sout{path}}}[ \mathbb{A} (\epsilon_k;\tau) ]  = U_\gamma^{-1}\mathbb{A} (\epsilon_k; \tau) U_\gamma = \mathbb{A} (\epsilon_k;\gamma\cdot \tau)$.

After showing in section \ref{sec:mod.1} that the proposed coaction (\ref{eq:conj}) does not compose in an intuitive way with modular transformations, we shall introduce an alternative map $\Delta^{\rm eqv}$ in section \ref{sec:mod.2} which has improved ${\rm SL}(2,\mathbb Z)$ behaviour but lacks other desirable  properties.

\subsection{The proposed coaction versus modular transformations}
\label{sec:mod.1}

By the analogies and differences between discontinuities of MPLs and modular transformations of iterated Eisenstein integrals, it is tempting to compare the two compositions $\Delta \mathbb I^\mm(\epsilon_k;\gamma \cdot\tau)$ and $(\gamma \otimes \gamma_{\text{\sout{path}}} )\Delta \mathbb I^\mm(\epsilon_k;\tau)$ of modular transformations $\gamma \in {\rm SL}(2,\mathbb Z)$
and the proposed coaction (\ref{eq:conj}).

\subsubsection{$S$-transformation} 

With the modular $S$-transformation (\ref{Sfull}) of the series $\Ip^\mm(\epsilon_k;\tau)$ and the simplified coaction (\ref{mmvcoactioncorollary}) of the $S$-cocycle, we can rewrite
\begin{align}
    \Delta \big(\mathcal{S}\big[\Ip^\mm(\epsilon_k;\tau) \big] \big) &= \Delta \big(  \mathbb{S}^\mm(\epsilon_k) U_S^{-1} \Ip^\mm(\epsilon_k;\tau)  U_S \big) \label{eq.314a} \\
 &= 
    (\mathbb{M}_{\sigma}^\dr)^{-1} \mathbb{S}^\mm(\epsilon_k) U_S^{-1} 
    \mathbb M_\sigma^\dr  U_S U_S^{-1} (\mathbb M_\sigma^\dr )^{-1}
    \Ip^\mm(\epsilon_k;\tau)  \mathbb M_\sigma^\dr \Ip^\dr(\epsilon_k;\tau) U_S  
    \notag \\
    &= 
    (\mathbb{M}_{\sigma}^\dr)^{-1} \mathbb{S}^\mm(\epsilon_k) U_S^{-1} \Ip^\mm(\epsilon_k;\tau)  \mathbb M_\sigma^\dr \Ip^\dr(\epsilon_k;\tau) U_S  \, .
    \notag
\end{align}
This has numerous common terms with
\begin{align}
     \left(\mathcal{S} \otimes {\cal S}_{\text{\sout{path}}}\right) \Delta \Ip^\mm(\epsilon_k;\tau) &= (\mathbb{M}_{\sigma}^\dr)^{-1} \mathbb{S}^\mm(\epsilon_k) U_S^{-1} \Ip^\mm(\epsilon_k;\tau)
     U_S\mathbb{M}_\sigmaT^\dr U_S^{-1}  \Ip^\dr(\epsilon_k;\tau) U_S\, ,  \label{eq.314b}
\end{align}
but the second series in zeta generators differs from (\ref{eq.314a}) by an $\mathrm{SL}(2)$ transformation
\begin{align}
   \big[ \Delta  \mathcal{S} \! - \! 
   (\mathcal{S} \! \otimes \! {\cal S}_{\text{\sout{path}}}) \Delta \big] \Ip^\mm(\epsilon_k;\tau) &= (\mathbb{M}_{\sigma}^\dr)^{-1} \mathbb{S}^\mm(\epsilon_k) U_S^{-1} \Ip^\mm(\epsilon_k;\tau)  \left(\mathbb{M}_{\sigma}^\dr-U_S\mathbb{M}_\sigmaT^\dr U_S^{-1} \right) \Ip^\dr(\epsilon_k;\tau) U_S\, . \label{eq.314}
\end{align}
Note that the mismatch between (\ref{eq.314a}) and (\ref{eq.314b}) cannot be compensated by a non-trivial transformation of $\mathbb{M}_{\sigma}^\dr$ under ${\cal S}_{\text{\sout{path}}}$, for instance since the left-multiplicative factors of $(\mathbb{M}_{\sigma}^\dr)^{-1}$ would otherwise no longer align.

\subsubsection{$T$-transformation} 

Similarly, the $T$-transformation (\ref{eq:Ttrm}) of the series $\Ip^\mm(\epsilon_k;\tau)$
and the fact that $\mathbb{M}_{\sigma}^\dr$ and $e^{2\pi i N}$ commute by
$[\sigma_w, N]=0$ lead to
\begin{align}
\Delta\big(\mathcal{T}\big[\Ip^\mm(\epsilon_k;\tau)\big]\big)
&= \left(\mathbb M_{\sigma}^\dr\right)^{-1} e^{2\pi i N} \Ip^\mm(\epsilon_k;\tau) \mathbb{M}_\sigmaT^\dr \Ip^\dr(\epsilon_k;\tau) U_T  
\end{align}
which we shall compare with
\begin{align}
\left(\mathcal{T}\otimes \mathcal{T}_{\text{\sout{path}}} \right) \Delta \Ip^\mm(\epsilon_k;\tau)
&=  \left(\mathbb{M}_{\sigmaT}^\dr \right)^{-1} e^{2\pi i N} \Ip^\mm(\epsilon_k;\tau) U_T \mathbb{M}_{\sigma}^\dr U_T^{-1}\Ip^\dr(\epsilon_k;\tau) U_T  \, . 
\end{align}
The two expressions again differ by an $\mathrm{SL}(2)$ transformation
\begin{align}
   \big[ \Delta  \mathcal{T} -  
   (\mathcal{T}  \otimes  {\cal T}_{\text{\sout{path}}}) \Delta \big] \Ip^\mm(\epsilon_k;\tau) &=\left(\mathbb{M}_{\sigma}^\dr\right)^{-1}  e^{2\pi i N} \Ip^\mm(\epsilon_k;\tau)    
    \left(\mathbb{M}_{\sigma}^\dr-U_T\mathbb{M}_\sigmaT^\dr U_T^{-1} \right) \Ip^\dr(\epsilon_k;\tau) U_T \, . \label{eq.318}
\end{align}

\subsubsection{Conclusion} 

The composition rule (\ref{MPLdisk}) for the coaction and discontinuities of MPLs does not uplift to a simple relation between $\Delta \gamma$ and $(\gamma \otimes \gamma_{\text{\sout{path}}}) \Delta$ for $\gamma \in {\rm SL}(2,\mathbb{Z})$. For both $S$- and $T$-transformations (\ref{eq.314}) and (\ref{eq.318}) of iterated Eisenstein integrals we encounter the obstruction
\begin{align}
   \big[ \Delta  \gamma - 
   (\gamma \otimes \gamma_{\text{\sout{path}}}) \Delta \big] \Ip^\mm(\epsilon_k;\tau)\sim
     \mathbb{M}_{\sigma}^\dr-U_\gamma\mathbb{M}_\sigmaT^\dr U_\gamma^{-1} 
     \label{remainder}
\end{align}
which does not have any counterpart for MPLs since the zeta generators $M_w$ at genus zero in section \ref{mwgzero} and (\ref{mwcoms}) do not have any SL$(2)$ structure.

\subsection{An equivariantized alternative}
\label{sec:mod.2}

The contamination of modular properties through $\mathrm{SL}(2,\mathbb{Z})$ transformations of zeta generators as in (\ref{remainder}) is familiar from the single-valued iterated Eisenstein integrals in section \ref{svIsec} \cite{Brown:2017qwo2, Dorigoni:2024oft}: 
\begin{align}
   \mathbb I^{\rm sv} (\epsilon_k;\gamma \cdot \tau)
   = (\sv \, \mathbb M_\sigma)^{-1} U_\gamma^{-1} (\sv \, \mathbb M_\sigma )\mathbb I^{\rm sv} (\epsilon_k;\tau) U_\gamma\, .
     \label{gaisv}
\end{align}
The equivariant transformation of $\Ieqv (\epsilon_k;\tau)$ under $\gamma \in {\rm SL}(2,\mathbb{Z})$ is attained by replacing the zeta generators in one of the series $\sv \, \mathbb M_\sigma$ in $\mathbb I^{\rm sv}(\epsilon_k;\tau)$ by their arithmetic and SL$(2)$-invariant parts $\sigma_w \rightarrow z_w$, see (\ref{mainiequv}) and (\ref{svconj}). One can anticipate from (\ref{remainder}) that the composition of the proposed coaction (\ref{eq:conj}) with modular transformations similarly improves by passing to an alternative, \textit{equivariantized} map with a substitution $\sigma_w \rightarrow z_w$ in one of the series $\mathbb M_\sigma^\dr$:\footnote{The second option of substituting $\sigma_w \rightarrow z_w$ in one of the series $\mathbb M_\sigma^\dr$ in (\ref{eq:conj}) results in the
alternative map $\Delta^{\rm alt}\Ip^\mm (\epsilon_k;\tau)=(\mathbb{M}^{\dr}_z)^{-1}\Ip^\mm (\epsilon_k;\tau)\mathbb{M}^{\dr}_\sigma \Ip^{\dr}(\epsilon_k;\tau)$. However, $\Delta^{\rm alt}$ does not obey any simple analogue of the composition rule (\ref{eq:DeltaEQvar}) as one may anticipate from the non-commutativity $(\mathbb{M}^{\dr}_z)^{-1} e^{2\pi i N} \neq e^{2\pi i N} (\mathbb{M}^{\dr}_z)^{-1}$ relevant for the modular $T$ transformation.}
\begin{equation}
     \label{eq:DeltaEQvar}
    \Delta^{{\rm eqv}} \Ip^{{\mm}}(\epsilon_k;\tau)= (\mathbb{M}^{\dr}_\sigma)^{-1}\Ip^\mm (\epsilon_k;\tau)\mathbb{M}^{\dr}_z \Ip^{\dr}(\epsilon_k;\tau)\, .
    \end{equation}
The coefficients of words in $\epsilon_k^{(j)}$ that are independent under Pollack relations specify the action of $\Delta^{{\rm eqv}}$ on $\mathbb Q$-linear combinations of iterated Eisenstein integrals (\ref{defiteis}) including
  \begin{align}
        \Delta^{\rm eqv}   \eem{j}{k}{\tau}&= 
    \eem{j}{k}{\tau}+ \eedr{j}{k}{\tau} +\frac{\delta_{j,k-2}}{k-1}\zeta^\dr_{k-1}
   \label{deqv.01} 
    \end{align}
with an extra zeta value as compared to the depth-one formula (\ref{deltadepthone}) for $\Delta \eem{j}{k}{\tau}$.     
We extend the action of $\Delta^{{\rm eqv}}$ to linear combinations of iterated Eisenstein integrals with $\mathbb Q[(2\pi i)^{ \pm 1}]$ multiples of motivic MZVs as coefficients by defining
 \begin{align}
        \Delta^{\rm eqv} \Big( \big[ (2\pi i)^\mm \big]^N \zeta^\mm_{n_1,\ldots,n_r}  \eem{j_1 &\ldots &j_\ell}{k_1 &\ldots &k_\ell}{\tau} \Big)&=
\big[ (2\pi i)^\mm \big]^N \,
\big( \Delta \zeta^\mm_{n_1,\ldots,n_r} \big)   
\Delta^{\rm eqv}   \eem{j_1 &\ldots &j_\ell}{k_1 &\ldots &k_\ell}{\tau} \, ,
   \label{deqv.02} 
    \end{align}
where $N\in \mathbb Z, \ n_i \in \mathbb N, \ n_r\geq 2$, and we will sometimes drop the superscript of $(2\pi i)^\mm $. The factor of $\Delta \zeta^\mm_{n_1,\ldots,n_r} $ on the right side is the motivic coaction of MZVs \cite{Goncharov:2001iea, Goncharov:2005sla, Brown:2011ik, BrownTate}, and $\Delta^{\rm eqv}   \eem{j_1 &\ldots &j_\ell}{k_1 &\ldots &k_\ell}{\tau}$ is determined by (\ref{eq:DeltaEQvar}) for the $\mathbb Q$-linear combinations of iterated Eisenstein integrals selected by Pollack relations.

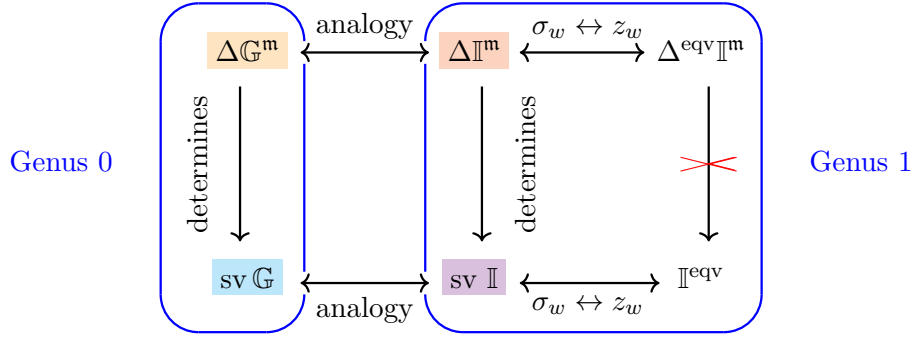
\begin{figure}[H]
\centering
    \begin{tikzpicture}
    \node at (-1.5cm,1.5cm) [above] {\colorbox{YellowOrange!25}{$\Delta \mathbb{G}^{\mathfrak{m}}$}};      
 \node at (1.5cm,1.5cm) [above] {\colorbox{RedOrange!25}{$\Delta \mathbb{I}^{\mathfrak{m}}$}};
    \node at (-1.5cm,-1.5cm) [above] {\colorbox{ProcessBlue!25}{$\text{sv} \, \mathbb{G}$}};
    \node at (1.5cm,-1.5cm) [above] {\colorbox{Plum!25}{$\text{sv}\, \,  \mathbb{I}$}};
 \node at (4.5cm,1.5cm) [above] {\colorbox{white}{$\Delta^{\rm eqv} \mathbb{I}^{\mathfrak{m}}$}};
  \node at (4.5cm,-1.5cm) [above] {\colorbox{white}{$ \mathbb{I}^{\rm eqv}$}};
    \node[blue] at (-3.15cm,0.42cm) [left] {Genus 0};
    \node[blue] at (5.8cm,0.42cm) [right] {Genus 1};
\draw[line width=0.3mm,blue] (-0.75cm,2.0cm) arc (0:90:0.5) ;
\draw[line width=0.3mm,blue] (-0.75cm,-1.35cm) arc (0:-90:0.5) ;
\draw[line width=0.3mm,blue] (-0.75cm,-1.05cm) -- (-0.75cm,1.7cm) ;
\draw[line width=0.3mm,blue] (-2.65cm,-1.35cm) -- (-2.65cm,2cm) ;
\draw[line width=0.3mm,blue] (-2.65cm,-1.35cm) arc (180:270:0.5) ;
\draw[line width=0.3mm,blue] (-2.65cm,2cm) arc (180:90:0.5) ;
\draw[line width=0.3mm,blue] (-2.15cm,2.5cm) -- (-1.25cm,2.5cm) ;
\draw[line width=0.3mm,blue] (-2.15cm,-1.85cm) -- (-1.25cm,-1.85cm) ;
\draw[line width=0.3mm,blue] (0.85cm,2.0cm) arc (180:90:0.5) ;
\draw[line width=0.3mm,blue] (0.85cm,-1.35cm) arc (-180:-90:0.5) ;
\draw[line width=0.3mm,blue] (0.85cm,-1.05cm) -- (0.85cm,1.7cm) ;
\draw[line width=0.3mm,blue] (5.3cm,-1.35cm) arc (0:-90:0.5) ;
\draw[line width=0.3mm,blue] (5.3cm,2cm) arc (0:90:0.5) ;
\draw[line width=0.3mm,blue] (4.8cm,2.5cm) -- (1.35cm,2.5cm) ;
\draw[line width=0.3mm,blue] (4.8cm,-1.85cm) -- (1.35cm,-1.85cm) ;
\draw[line width=0.3mm,blue] (5.3cm,-1.35cm) -- (5.3cm,2cm) ;
    \draw[line width=0.3mm,<->] (-0.8cm,1.85cm) -- (0.9cm,1.85cm); 
    \draw[line width=0.3mm,<->] (-0.8cm,-1.2cm) -- (0.9cm,-1.2cm);
       \node at (0.05cm,-1.25cm) [below] {analogy};
       \node at (0.05cm,1.9cm) [above] {analogy};
    \draw[line width=0.3mm,<->] (3.75cm,1.85cm) -- (2.1cm,1.85cm); 
    \draw[line width=0.3mm,<->] (3.95cm,-1.2cm) -- (2.1cm,-1.2cm); %
       \node at (3cm,-1.25cm) [below] {$\sigma_w\leftrightarrow z_w$};
       \node at (3cm,1.9cm) [above] {$\sigma_w\leftrightarrow z_w$};
    \draw[line width=0.3mm,<-] (-1.6cm,-0.65cm) -- (-1.6cm,1.4cm); 
    \draw[line width=0.3mm,<-] (1.6cm,-0.65cm) -- (1.6cm,1.4cm); 
\draw[line width=0.3mm,<-] (4.6cm,-0.65cm) -- (4.6cm,1.4cm); 
\node[red] at (4.6cm,0.375cm){$\xcancel{\phantom{xxx}}$};
\node[rotate=90, anchor=south] at (-1.9cm, 0.375cm) {determines};
\node[rotate=90, anchor=north] at (1.9cm, 0.375cm) {determines};
\end{tikzpicture}
\caption{The prescriptions for single-valued and equivariant iterated Eisenstein integrals $ \mathbb{I}^{\rm sv}$ and $ \mathbb{I}^{\rm eqv}$ are in a similar relation as the proposed coaction $\Delta$ in (\ref{eq:conj}) and the equivariantized map $\Delta^{\rm eqv}$ in (\ref{eq:DeltaEQvar}). The absence of a third vertical arrow between $\Delta^{\rm eqv} \mathbb I^\mm(\epsilon_k;\tau)$ and $\mathbb I^{\rm eqv}(\epsilon_k;\tau)$
is due to the different placements of the substitution $\sigma_w \rightarrow z_w$ in the concatenation order of the series composing $\Delta \mathbb I^\mm$ and ${\rm sv}\, \mathbb I$, see section \ref{noaeqv} below for further details.}
\label{fig.eqv}
\end{figure}

By minor modifications of the
computations in (\ref{eq.314a}) to (\ref{eq.318}), we then find that the equivariantized map (\ref{eq:DeltaEQvar}) and (\ref{deqv.02}) obeys the simplified composition rule
\begin{equation}
    \Delta^{{\rm eqv}} \big(\gamma\big[\Ip^\mm(\epsilon_k;\tau)\big] \big) = \left(\gamma\otimes  \gamma_{\text{\sout{path}}}\right) \Delta^{\rm eqv}\Ip^\mm(\epsilon_k;\tau)
\label{eqvdelta}
\end{equation}
which resembles the interplay (\ref{MPLdisk}) of the motivic coaction and discontinuities of MPLs. The notation $\Delta^{{\rm eqv}}$ and the terminology is chosen to emphasize that the equivariantized map (\ref{eq:DeltaEQvar}) is obtained from the proposed coaction (\ref{eq:conj}) through the same type of substitution $\sigma_w \rightarrow z_w$ that formally recovers $\Ieqv (\epsilon_k;\tau)$ from $\mathbb I^{\rm sv} (\epsilon_k;\tau)$, see figure \ref{fig.eqv}. However, we will see below that in contrast to the proposed coaction $\Delta$ (see section \ref{sec:tauzero}), the equivariantized map $\Delta^{\rm eqv}$ in (\ref{eq:DeltaEQvar}) and (\ref{deqv.02}) does not commute with the regularized limits $\tau \rightarrow 0$ and $\tau \rightarrow i\infty$.

By analogy with the discussion of  $\Ieqv (\epsilon_k;\tau)$ in \cite{Brown:2017qwo2}, the definition (\ref{eq:DeltaEQvar}) of $ \Delta^{{\rm eqv}} \Ip^{{\mm}}(\epsilon_k;\tau)$ depends on the choice of the arithmetic part $z_w$ of the zeta generator $\sigma_w$. One can follow the choice of $\Ieqv (\epsilon_k;\tau)$ in \cite{Dorigoni:2024oft} and employ the canonical $z_w$ from Theorem 5.4.1 (vi) of \cite{Dorigoni:2024iyt}.

\subsubsection{Incompatibility with the regularized $\tau \rightarrow 0$ limit}
\label{sec:choices}

On the left-hand side of (\ref{eqvdelta}) for $\gamma = {\cal S}$, the $S$-cocycle $\mathbb S^\mm(\epsilon_k)$ and the $\mathrm{SL}(2)$  transformation $U_S^{\pm 1} $ in $\Ip^\mm(\epsilon_k;-\frac{1}{\tau}) = \mathbb S^\mm(\epsilon_k) U_S^{-1} \Ip^\mm(\epsilon_k; \tau) U_S$ produce $\mathbb Q[(2\pi i)^{-1}]$-linear combinations of motivic MZVs where $\Delta^{{\rm eqv}}$ acts through the motivic coaction by (\ref{deqv.02}). Accordingly,
\begin{align}
\Delta^{{\rm eqv}} \mathbb S^\mm(\epsilon_k)
= (\Delta^{{\rm eqv}} \mathbb{M}^{\mm}_\sigma)^{-1}
\mathbb{X}^{\mm}_S U_S^{-1} (\Delta^{{\rm eqv}} \mathbb{M}^{\mm}_\sigma ) U_S
= (\mathbb{M}^{\dr}_\sigma)^{-1}
\mathbb S^\mm(\epsilon_k) U_S^{-1}  \mathbb{M}^{\dr}_\sigma U_S
\label{deqvS}
\end{align}
by construction matches the motivic coaction (\ref{mmvcoactioncorollary}) of the $S$-cocycle, using
$\Delta^{{\rm eqv}} \mathbb{M}^{\mm}_\sigma = \mathbb{M}^{\mm}_\sigma \mathbb{M}^{\dr}_\sigma$ and $\Delta^{{\rm eqv}} \mathbb{X}^{\mm}_S = \mathbb{X}^{\mm}_S$ in intermediate steps. We have used (\ref{deqvS}) in verifying the composition rule (\ref{eqvdelta}) for $\gamma = {\cal S}$, and the analogous computation for $\gamma = {\cal T}$ only necessitates that $\Delta^{{\rm eqv}}$ commutes with powers of $2\pi i$ by (\ref{deqv.02}).

However, (\ref{deqvS}) is incompatible with the regularized $\tau \rightarrow 0$ limit on the right-hand side of (\ref{eq:DeltaEQvar}) which results in
\begin{align}
\lim_{\tau \rightarrow 0} \big( \Delta^{\rm eqv}\Ip^\mm (\epsilon_k;\tau) \big) =(\mathbb{M}^{\dr}_\sigma)^{-1}
\mathbb S^\mm(\epsilon_k) \mathbb{M}^{\dr}_z (\mathbb{M}^{\dr}_\sigma)^{-1} U_S^{-1}  \mathbb{M}^{\dr}_\sigma  U_S
\label{nc:deqvS}
\end{align}
after inserting (\ref{eq:Sdr}) for $\mathbb S^\dr(\epsilon_k) $. Since the left-hand side of (\ref{eq:DeltaEQvar}) would reduce to $\Delta^{\rm eqv}\mathbb S^\mm(\epsilon_k) $ if we had taken $\tau \rightarrow 0$ prior to evaluating $\Delta^{\rm eqv}$, we conclude from the mismatch between (\ref{deqvS}) and (\ref{nc:deqvS}) that the equivariantized map does not commute with the regularized $\tau \rightarrow 0$ limit.

\subsubsection{Incompatibility with the regularized $\tau \rightarrow i\infty$ limit}

In the same way as the regularized $\tau \rightarrow i \infty$ limit of $\Ieqv (\epsilon_k;\tau)$ yields the series $(\sv \, \mathbb M_z)^{-1} \sv \, \mathbb M_\sigma$, the equivariantized map specializes to 
\begin{align}
\lim_{\tau \rightarrow i \infty} \big( \Delta^{\rm eqv}\Ip^\mm (\epsilon_k;\tau) \big) =
(\mathbb{M}^{\dr}_\sigma)^{-1}\mathbb{M}^{\dr}_z
\label{cn:deqvS}
\end{align}
using $\textbf{1}^\mm=\mathbb I^\mm (\epsilon_k;\tau \rightarrow i \infty)$ as well as $\textbf{1}^\dr=\mathbb I^\dr (\epsilon_k;\tau \rightarrow i \infty)$ on the right-hand side of (\ref{eq:DeltaEQvar}). This is clearly incompatible with the result $\Delta^{\rm eqv}\textbf{1}^\mm$ of performing the regularized $\tau \rightarrow i \infty$ limit on the left side of (\ref{eq:DeltaEQvar}) before evaluating the equivariantized map.

\subsubsection{No simple derivation of $\Ieqv (\epsilon_k;\tau)$ from an equivariantized antipode}
\label{noaeqv}

One may finally wonder if the derivation of the single-valued map from the coaction and the antipode in section \ref{sec:3.3.3} can be adapted to obtain $\Ieqv (\epsilon_k;\tau)$ by the replacement $\Delta \rightarrow \Delta^{\rm eqv}$ in the main formulae. We will see that this is not the case
for a naive substitution $\Delta \rightarrow \Delta^{\rm eqv}$ where the analogues of the steps in section \ref{sec:3.3.3} do not produce the correct expression (\ref{mainiequv}) for $\Ieqv (\epsilon_k;\tau)$.
The difference will turn out to lie in the placement of $\mathbb M_\sigma$ and $ \mathbb M_z$ which spoils the equivariance properties.

An intuitive first step would be to adapt (\ref{svde.02}) to
\begin{align}
\textbf{1} = \mu \circ
(\ant^{\rm eqv} \circ \Pi^\dr \otimes \textbf{1}) \circ \Delta^{\rm eqv}
\label{nogo.01}
\end{align}
which results in an equivariantized version $\ant^{\rm eqv}$ of the antipode (\ref{altsv.06}) with $( \mathbb M_z^\dr)^{-1}$ instead of $( \mathbb M_\sigma^\dr)^{-1}$ as its rightmost factor,
\begin{align}
\ant^{\rm eqv} \mathbb I^\dr(\epsilon_k;\tau) = \mathbb M_\sigma^\dr \big( \mathbb I^\dr(\epsilon_k;\tau) \big)^{-1} ( \mathbb M_z^\dr)^{-1}  \, .
\label{nogo.02}
\end{align}
A natural second step would be the adaptation of the single-valued map (\ref{svde.08}) to the operation
\begin{align}
\mu \circ
(\tilde \ant^{\rm eqv} \circ \Pi^\dr \otimes \textbf{1}) \circ \Delta^{\rm eqv} \mathbb I^{\mm}(\epsilon_k;\tau)
&= (\mathbb M_\sigma^\dr)^{-1}
\big( 
\tilde \ant^{\rm eqv}\mathbb I^{\dr}(\epsilon_k;\tau)
\big)\mathbb M_z^\dr \mathbb I^{\dr}(\epsilon_k;\tau) \, ,
\label{nogo.03}
\end{align}
where $\tilde \ant^{\rm eqv}$ is obtained from $\ant^{\rm eqv}$ by the same sign and complex conjugation as $\tilde \ant$ is obtained from $\ant$ in (\ref{altde.09}). Inserting $\tilde \ant^{\rm eqv}\mathbb I^{\dr}(\epsilon_k;\tau) =  \big( (  \mathbb M_\sigma^\dr  )^T \big)^{-1} 
\overline{ \mathbb I^\dr(\epsilon_k;\tau)^T  }
(  \mathbb M_z^\dr  )^T $ and dropping the $\dr$ superscripts on the right-side of (\ref{nogo.03}) leads to the result
\begin{align}
(\sv \, \mathbb M_\sigma)^{-1} \,\overline{ \mathbb I(\epsilon_k;\tau)^T}\,
  (\sv \, \mathbb M_z) \, \Ip(\epsilon_k;\tau)
\neq \Ieqv (\epsilon_k;\tau)
\label{nogo.04}
\end{align}
of the operation $\mu \circ
(\tilde \ant^{\rm eqv} \circ \Pi^\dr \otimes \textbf{1}) \circ \Delta^{\rm eqv}$
which has the restriction to the arithmetic part of the zeta generators in the second factor of ${\rm sv}\, \mathbb M_\bullet$ instead of the first one as $\Ieqv (\epsilon_k;\tau)$ in~(\ref{mainiequv}).

We have seen that the equivariantized map $\Delta^{\rm eqv}$ does not commute with regularized limits for $\tau$ and does not allow for a construction of $\Ieqv$ in parallel to $\mathbb I^{\rm sv}$, see figure~\ref{fig.eqv}. 
This is similar to the known constructions of $\Ieqv$ and $\mathbb{I}^{\rm sv}$ where modular properties were not achieved at the same time as other desired properties.

\section{Summary and outlook}

In this paper, we have proposed a coaction for iterated Eisenstein integrals as they appear in genus-one string scattering calculations. Our proposal~(\ref{eq:conj}) passes several consistency checks related to shuffle multiplication, taking derivatives and limits of the modular parameter $\tau$. It is hampered by lack of an intrinsic definition of a de Rham version of the iterated Eisenstein integrals. Nevertheless, our proposal has consequences for multiple modular values, expressed in~\eqref{ourSm.01}, that we could verify on large data sets and that are consistent with results of Brown~\cite{brown2017multiple}.

In section~\ref{sec:5}, we have studied compositions of our proposed coaction $\Delta$ from~\eqref{eq:conj} with modular transformations and find that
it does not transform equivariantly.
This is not surprising since the proposal~\eqref{eq:conj} was guided by the generating series $\mathbb{I}^{\rm sv}$ of single-valued iterated Eisenstein integrals that are also known not to transform equivariantly under ${\rm SL}(2,\mathbb Z)$. 
This is in contrast to the generating series $\mathbb{I}^{\rm eqv}$ of equivariant iterated Eisenstein integrals. The modified map $\Delta^{\rm eqv}$ in~\eqref{gaisv} was constructed to attain the equivariant composition law (\ref{eqvdelta}) with modular transformations but lacks other properties of $\Delta$ related to taking limits (see section~\ref{sec:3}).
We therefore prefer the map $\Delta$ in~\eqref{eq:conj} over its equivariant alternative $\Delta^{\rm eqv}$ in the quest for motivic coactions at genus one.

To show that $\Delta$ is the correct motivic coaction requires settling some of the open questions raised in section~\ref{sec_motivicderahm}. In particular, we leave it as an open problem to provide a proper definition of the objects $\eedr{}{}{}$ in the formulae for $\Delta$ in the light of de Rham periods. In this work, we merely present the $\eedr{}{}{}$ as iterated Eisenstein integrals appearing in the second entry of the coaction while preserving their essential properties.

The main physics motivation for the map $\Delta$ stems from scattering amplitudes in particle physics and string theory that involve combinations of elliptic multiple polylogarithms and elliptic multiple zeta values. Upon conversion to iterated Eisenstein integrals, elliptic multiple zeta values additionally introduce polynomials in the ratios of periods $\tau$. Thus, to connect to the physics literature, one would also need to investigate extensions of the proposal $\Delta$ to act on $\tau$ (noting that the motivic coaction itself only acts on periods rather than ratios of periods).

A natural task for follow-up work is to investigate the choices in the de Rham projection at genus one such that our map $\Delta$ agrees with the motivic coaction and at the same time extends to determine $\Delta \tau$. In this way, one would reconcile the mathematical framework of motivic coaction and periods with the physics intuition and literature on elliptic polylogarithms and multiple zeta values.

\subsection{Outlook}

The main motivation of this work is to set the stage for investigating more general coaction formulae beyond genus zero in future work. The availability of explicit coaction formulae for the special functions in scattering amplitudes is essential to find or delimit universal symmetries of physical theories under the cosmic Galois group \cite{Cartier:1988, Brown:2015fyf}. Amplitudes and related observables in string theory and a variety of quantum field theories were found to signal cosmic Galois symmetry through their stability under the motivic coaction. However,
most of the currently known examples stem from the polylogarithmic contributions to the respective amplitudes. Hence, a refined analysis of these unifying symmetries hinges on coaction formulae for the geometries beyond the sphere that govern Feynman integrals and string amplitudes.

A more immediate generalization of this work concerns coactions of elliptic MPLs \cite{Lev, Levrac, BrownLev, Broedel:2014vla, Broedel:2017kkb,Broedel:2018iwv,Broedel:2018izr,Broedel:2018qkq, Broedel:2019tlz, Enriquez:2023emp} instead of the iterated Eisenstein integrals obtained from their special values. In adapting our approach to the coaction of elliptc MPLs, combinations of zeta generators and Tsunogai derivations are expected to offer new insights on the MZVs and more general period integrals in the de Rham entry. Following the strategy of this work, the single-valued counterparts of elliptic MPLs known as elliptic modular graph forms in the string-theory literature \cite{DHoker:2018mys, DHoker:2020tcq, Basu:2020pey, Basu:2020iok, Dhoker:2020gdz, Hidding:2022vjf} should single out the relevant generating series and non-commuting variables to derive coaction formulae for elliptic MPLs. It will be a crucial stepping stone to formulate elliptic modular graph forms in terms of equivariant iterated integrals as initiated for rational points in \cite{Drewitt:2023con, Duhr:2025lvr}.

A more ambitious line of rewarding follow-up research concerns coactions and single-valued or equivariant versions of iterated integrals on higher-genus Riemann surfaces. The recent mathematics and physics literature offers several formulations of higher-genus polylogarithms \cite{BEFZ:2110, Ichikawa:2022qfx, BEFZ:2212, DHoker:2023vax, Baune:2024biq, DHoker:2024ozn, Baune:2024ber, DHoker:2025szl, Baune:2025sfy} and generalizations of modular graph functions \cite{DHoker:2013fcx, DHoker:2014oxd, Pioline:2015qha, DHoker:2017pvk, DHoker:2018mys, DHoker:2020uid}. Given that the proposed coaction and single-valued map at genus one draw key information from derivatives in the modular parameter $\tau$ and degenerations $\tau \rightarrow i\infty$, it is pressing to control the analogous operations at higher genus. In the string-theory motivated formulation evolving around single-valued Green functions, (non-)separating degenerations were investigated in \cite{Pioline:2015qha, DHoker:2017pvk, DHoker:2018mys, Dhoker:2020gdz, DHoker:2023vax}, and the method of complex-structure variation \cite{Verlinde:1986kw, DHoker:1988pdl, DHoker:2014oxd, DHoker:2025dhv} is a major stepping stone towards the differential structure in the moduli.

The most daunting challenge on the long run is to extend the explicit constructions of coaction formulae to integrals on higher-dimensional varieties. Recent precision computations in particle physics and gravity unravelled a wealth of K3 and Calabi-Yau surfaces, with rapid progress in the study of their periods and iterated integrals. We hope that the approach to coaction formulae put forward in this work can be uplifted not only to Riemann surfaces beyond genus one but also to higher-dimensional varieties.

\section*{Acknowledgements}

We are grateful to Ruth Britto, Francis Brown, Emiel Claasen, Daniele Dorigoni, Mehregan Doroudiani, Joshua Drewitt, Cl\'ement Dupont, Hadleigh Frost, Martijn Hidding, Deepak Kamlesh, Martin Raum, Carlos Rodriguez, Leila Schneps, Aleksander Shmakov, Yoann Sohnle, Yi-Xiao Tao and Bram Verbeek for combinations of helpful discussions and collaboration on related topics. Moreover, we are indebted to Matija Tapu$\check{\rm s}$kovi\'c for numerous valuable discussions, important pointers to the literature and essential guidance concerning the mathematical challenges related to this work. FP thanks Nordita for the possibility to conduct large parts of the work presented here during her stay there and for financial support through a Nordita visiting PhD fellowship. AK and FP thank the University of Uppsala for hospitality and support during part of this work. This work is funded by the European Union (FP by ERC Consolidator Grant LoCoMotive 101043686 and AK/OS by ERC Synergy Grant MaScAmp 101167287). Views and opinions expressed are however those of the author(s) only and do not necessarily reflect those of the European Union or the European Research Council. Neither the European Union nor the granting authority can be held responsible for them. The research of OS is also supported by the strength area ``Universe and mathematical physics'' which is funded by the Faculty of Science and Technology at Uppsala University. FP and OS thank the Banff International Research Station in Oaxaca for their hospitality and the organizers \& participants of the workshop ``Beyond Elliptic Polylogarithms'' for valuable discussions. The research of all authors was supported by the Munich Institute for Astro-, Particle and BioPhysics (MIAPbP) which is funded by the Deutsche Forschungsgemeinschaft (DFG, German Research Foundation) under Germany's Excellence Strategy -- EXC-2094 -- 390783311.

\begin{appendix}

\section{Examples for multiple modular values}
\label{appendixmultiple}

Following the presentation
in \cite{Brown2019}, this appendix gathers representative examples of MMVs in the conventions of (\ref{eq:mmvdef}).

\subsection{Modular depth two}

All MMVs $\MMV{j_1&j_2}{k_1&k_2}$ of modular depth two and degree $k_1{+}k_2\leq 12$ are $\mathbb Q[2\pi i]$-linear combinations of MZVs. Starting from degree $k_1{+}k_2=14$, one additionally encounters L-values of weight-$w$ holomorphic cusp forms $\Delta_w$ 
\begin{align}
\Lambda(\Delta_{w}, s) = (-i)^s \int^{i \infty}_0 \dd \tau \Delta_w(\tau) \tau^{s-1}
\label{defLval}
\end{align}
with $s\geq w$ outside the critical strip if $j_1{+}j_2$ is odd and related new periods $c(\Delta_{w}, s)$ if $j_1{+}j_2$ is even \cite{brown2017multiple, Brown2019, Dorigoni:2021ngn}. The admixtures of both $\Lambda(\Delta_{w}, s)$ and $c(\Delta_{w}, s)$ are projected out from the generating series of this work including (\ref{eq:limitS}) by the Pollack relations among $\ep_k^{(j)}$.

\subsubsection*{Degree $8$}
\begin{align}
    \MMV{0&0}{4&4} &= -\frac{2 \pi^2 \zeta_3^2 }{9} \, ,   \notag \\ \MMV{1&0}{4&4} &= -\frac{8 i\pi^5 \zeta_3}{405}  +\frac{5 i\pi^3 \zeta_5}{27}\, ,    \notag \\ \MMV{2&0}{4&4} &= -\frac{209 \pi^8}{364500}  +\frac{4 \pi^2 \zeta_3^2}{9} \, .
    \label{w8mmvs}
\end{align}
\subsubsection*{Degree $10$}

\begin{align}
    \MMV{0&0}{4&6} &= -\frac{503 \pi^{10}}{25515000}  +\frac{4 \pi^2 \zeta_{3,5}}{75}\, ,    \notag \\
    \MMV{1&0}{4&6} &=-\frac{4 i\pi^7 \zeta_3}{14175}  -\frac{i\pi^5 \zeta_5}{135} +\frac{7 i\pi^3 \zeta_7  }{90}\, ,    \notag \\
    \MMV{2&0}{4&6} &=-\frac{437  \pi^{10}}{53581500} -\frac{4 \pi^4 \zeta_3^2 }{315}  +\frac{4 \pi^2 \zeta_3\zeta_5}{15} \, .
    \label{w10mmvs}
\end{align}

\subsubsection*{Degree $12$}

\begin{align}
\MMV{0& 0}{4& 8} &= -\frac{83 \pi^{12} }{45841950} +  \frac{2 \pi^2 \zeta_5^2}{49}  + \frac{2 \pi^2 \zeta_{3, 7} }{147}\, ,   \notag \\
\MMV{0& 1} {4& 8} &= \frac{ i \pi^9 \zeta_3 }{297675}  - 
  \frac{10  i \pi^3 \zeta_9}{567}\, ,     \notag \\
\MMV{0& 2} {4& 8} &= \frac{83  \pi^{12}}{
   334884375 } - \frac{ 2 \pi^4 \zeta_3 \zeta_5 }{
   1575}  - \frac{2 \pi^4 \zeta_{3, 5} }{7875}\, .  
\label{w12mmvs}
   \end{align}
The appearance of
$\zeta_{3, 5}$ in the third line is an
example of ``transference of periods'' \cite{brown2017multiple, Brown2019}.

\subsubsection*{Degree $14$}

\begin{align}
\MMV{0& 0}{4& 10} &=
- \frac{ 1024 \pi^{14} c(\Delta_{12}, 12)}{652995} 
- \frac{4  \pi^2 \rho^{-1} (f_3 f_9) }{27} -\frac{32683519272816894175283003  \pi^{14} }{920158689212087733618892738056000} \, ,  \notag\\
\MMV{0& 1} {4& 10} &= \frac{ i \pi^{11} \zeta_3}{5051970} - \frac{11 i \pi^3 \zeta_{11} }{1080}   - \frac{ 128 i \pi^{13} \Lambda(\Delta_{12}, 12) }{1913625} \, .
\label{sec:lvals}
\end{align}
The $L$-value of the Ramanujan cusp form $\Delta_{12}(\tau) = q \prod_{n=1}^{\infty}(1{-}q^n)^{24}$ is defined by (\ref{defLval}). The new period $c(\Delta_{12}, 12)$ different from MZVs and $L$-values is only well-defined up to adding a rational number \cite{Brown2019}. Finally,
\begin{align}
\rho^{-1} (f_3 f_9)
&= - \frac{ 48 \zeta_{1, 1, 4, 6}}{691} 
- \frac{223  \zeta_{3, 9} }{691}
+  \frac{ 24 \pi^2 \zeta_{3, 7}}{691} 
+ \frac{4 \pi^4 \zeta_{3, 5}}{3455}  
- \frac{4 \pi^6 \zeta_3^2}{24185} 
+ \frac{4 \zeta_3^4}{691} 
- \frac{  4 \pi^4 \zeta_3 \zeta_5}{10365}+ \frac{54 \zeta_5 \zeta_7 }{691} \notag \\
&\quad  + \frac{28  \pi^2 \zeta_5^2}{691} + 
  \frac{80 \pi^2 \zeta_3 \zeta_7}{691}   - \frac{
  1598 \zeta_3 \zeta_9}{2073} 
  -\frac{ 1801374852589837913812703 \pi^{12}}{
  6491419324247532512302594272000}
\end{align}
belongs to those MZVs in the conjectural $\mathbb Q$-basis at weight 12 that do not admit any representative below depth four \cite{Blumlein:2009cf}. We have made use of the conjecture that the space of motivic MZVs is in bijection with real MZVs.

\subsection{Modular depth three}

At modular depth three, MMVs beyond $\mathbb Q[2\pi i]$-linear combinations of MZVs occur starting from degree 16 and again do not enter the generating series of this work.

\subsubsection*{Degree 12}

\begin{align}
    \MMV{0&0&0}{4&4&4} &=\frac{4 i\pi^3 \zeta_3^3}{81}\, ,    \notag \\
    \MMV{1&0&0}{4&4&4} &=-\frac{368 \pi^{12}}{47840625}-\frac{11 \pi^6\zeta_3^2}{1215} \, ,  
    +\frac{313 \pi^4\zeta_3\zeta_5 }{2025}
    +\frac{313 \pi^4 \zeta_{3,5} }{10125}\, ,  \notag \\
    \MMV{2&0&0}{4&4&4} &=\frac{481 i\pi^9 \zeta_3}{1093500}-\frac{4 i\pi^3 \zeta_3^3}{27}  +\frac{2 i\pi^7\zeta_5}{729}  -\frac{661 i\pi^5\zeta_7}{12150}\, .
    \label{w12mmvsd3}
\end{align}

\subsubsection*{Degree 14}

\begin{align}
    \MMV{0&0&0}{4&4&6} &=
    -\frac{i \pi^{11} \zeta_3}{
  1275750} + \frac{4i \pi^9  \zeta_5}{42525} - \frac{4i \pi^7 \zeta_7}{3375} - 
 \frac{4i\pi^5\zeta_9}{15} + \frac{14573 i \pi^3 \zeta_{11}}{5400} -\frac{8i \pi^3 \zeta_{3,3,5}}{225}\, ,   \notag \\
\MMV{0&0&1}{4&4&6} &= -\frac{37379 \pi^{14}}{185659897500} + \frac{11 \pi^4 \zeta_5^2}{1764} + \frac{\pi^4 \zeta_3\zeta_7}{315}+\frac{11\pi^4 \zeta_{3,7}}{5292}\, ,   \notag \\
\MMV{0&0&3}{4&4&6} &=
\frac{17231 \pi^{14}}{48223350000} + \frac{\pi^8 \zeta_3^2}{14175} - \frac{2\pi^6 \zeta_3\zeta_5}{567} - \frac{64 \pi^6 \zeta_{3,5}}{70875}\, .
\label{w14mmvsd3}
\end{align}
The irreducible $\zeta_{3,3,5}$ has already been identified as a part of an MMV in \cite{saad2020multiple}.

\section{Examples for coactions of iterated Eisenstein integrals} 
    \label{app_ExamplesEisenstein}

This appendix gathers some more examples of the coaction $\Delta$ of iterated Eisenstein integrals that complement the closed formulae at depth one and two in section \ref{sec:3.1}. The subsequent results are obtained by expanding the generating series in our proposal (\ref{eq:conj}) solely in terms of Tsunogai derivations and extracting the coefficients of a given word $\epsilon_{k_1}^{(j_1)} \ldots \epsilon_{k_\ell}^{(j_\ell)}$.
Starting from degree $\sum_{i=1}^\ell k_i=14$,
our proposal (\ref{eq:conj}) only generates coaction formulae for those linear combinations of iterated Eisenstein integrals that are accompanied by words in $\epsilon_{k}^{(j)} $ that are independent under Pollack relations.

The examples below are unaffected by Pollack relations, i.e.\ the subsequent coaction formulae are valid for individual iterated Eisenstein integrals: modular depth two and degree $\leq 12$ in section \ref{app:B.1} as well as modular depth three and degree $\leq 14$ in section \ref{app:B.2}.

\subsection{Modular depth two}
\label{app:B.1}

The following coactions at modular depth two are special cases of the closed formula (\ref{clform}):

\subsubsection*{Degree 8} 

    \begin{align}
    \Delta    \eem{2&0}{4&4}{\tau}&= \eem{2&0}{4&4}{\tau}+\eem{2}{4}{\tau}\eedr{0}{4}{\tau}+\eedr{2&0}{4&4}{\tau} + \frac{1}{3} \zeta_3^\dr \eem{0}{4}{\tau}\, ,    \notag \\
        \Delta    \eem{2&1}{4&4}{\tau}&= \eem{2&1}{4&4}{\tau}+\eem{2}{4}{\tau}\eedr{1}{4}{\tau}+\eedr{2&1}{4&4}{\tau} + \frac{1}{3} \zeta_3^\dr \eem{1}{4}{\tau}\, ,   \notag \\
    \Delta    \eem{2&2}{4&4}{\tau}&= \eem{2&2}{4&4}{\tau}+\eem{2}{4}{\tau}\eedr{2}{4}{\tau}+\eedr{2&2}{4&4}{\tau}\, . 
    \end{align}

\subsubsection*{Degree 10}

    \begin{align}
    \Delta    \eem{2&0}{4&6}{\tau}&=  \eem{2&0}{4&6}{\tau}+\eem{2}{4}{\tau}\eedr{0}{6}{\tau}+\eedr{2&0}{4&6}{\tau}+ \frac{1}{210}\,  \eem{0}{4}{\tau}\zeta^\dr_3 +\frac{1}{3}\eem{0}{6}{\tau} \zeta_3^\dr\, ,    \notag \\
    \Delta    \eem{1&3}{4&6}{\tau}&= \eem{1&3}{4&6}{\tau}+\eem{1}{4}{\tau}\eedr{3}{6}{\tau}+\eedr{1&3}{4&6}{\tau}- \frac{1}{840} \eem{2}{4}{\tau}\zeta_3^\dr\, ,  \\
    \Delta    \eem{0&4}{4&6}{\tau}&= \eem{0&4}{4&6}{\tau}+\eem{0}{4}{\tau}\eedr{4}{6}{\tau}+\eedr{0&4}{4&6}{\tau}+ \frac{1}{210}\,   \eem{2}{4}{\tau}\zeta_3^\dr-\frac{1}{5} \eem{0}{4}{\tau}\zeta_5^\dr\, ,   \notag \\
    \Delta    \eem{2&4}{4&6}{\tau}&= \eem{2&4}{4&6}{\tau}+\eem{2}{4}{\tau}\eedr{4}{6}{\tau}+\eedr{2&4}{4&6}{\tau} +\frac{1}{3} \eem{4}{6}{\tau}\zeta_3^\dr-\frac{1}{5} \eem{2}{4}{\tau}\zeta_5^\dr
    \, . \notag
  \end{align}

\subsubsection*{Degree 12} 

Some of the examples at $(k_1,k_2)=(4,8)$ illustrate the appearance of $\eem{j}{k}{\tau}$
under the coaction with $k \neq k_1,k_2$
(which is $k=k_2{-}k_1{+}2$ or $k=k_1{-}k_2{+}2$ in the general case (\ref{clform}), (\ref{aritd2}) and $k=6$ in the present case)
\begin{align}
\Delta    \eem{2&0}{4&8}{\tau}&=  \eem{2&0}{4&8}{\tau} +  \eem{2}{4}{\tau} \eedr{0}{8}{\tau}   +  \eedr{2&0}{4&8}{\tau} + \frac{1}{168} \eem{0}{6}{\tau} \zeta^\dr_3 + \frac{1}{3} \eem{0}{8}{\tau} \zeta^\dr_3\, ,  
\notag \\
\Delta    \eem{1&3}{4&8}{\tau}&=  \eem{1&3}{4&8}{\tau} +  \eem{1}{4}{\tau}  \eedr{3}{8}{\tau} +  \eedr{1&3}{4&8}{\tau} -\frac{1}{560} \eem{2}{6}{\tau} \zeta^\dr_3\, ,  
\\
\Delta    \eem{0&6}{4&8}{\tau}&=  \eem{0&6}{4&8}{\tau} +   \eem{0}{4}{\tau} \eedr{6}{8}{\tau} +  \eedr{0&6}{4&8}{\tau} +  \frac{1}{168} \eem{4}{6}{\tau} \zeta^\dr_3 - \frac{1}{7} \eem{0}{4}{\tau} \zeta^\dr_7\, .
\notag
\end{align}
Among the coactions $\Delta \eem{j_1&j_2}{6&6}{\tau}$ at $k_1=k_2=6$, the only instances of $\zeta^\dr$ occur for
    \begin{align}
    \Delta    \eem{4&j}{6&6}{\tau}=\, \, & \eem{4&j}{6&6}{\tau}+\eem{4}{6}{\tau}\eedr{j}{6}{\tau}+\eedr{4&j}{6&6}{\tau}+\frac{1}{5}\eem{j}{6}{\tau}\zeta_5^\dr
    \end{align}
at $j=0,1,2,3$ as well as the $ \Delta    \eem{j&4}{6&6}{\tau}$ related by shuffle.

\subsection{Modular depth three}
\label{app:B.2}

To avoid cluttering, we peel off the ubiquitous (deconcatenation) products of ${\cal E}^\mot$ and ${\cal E}^\dr$ in (\ref{nozeta}) without any accompanying de Rham MZVs from the coactions at modular depth three
\begin{align}
\Delta \eem{j_1&j_2 &j_3}{k_1&k_2 &k_3}{\tau}
&=  \eem{j_1&j_2 &j_3}{k_1&k_2 &k_3}{\tau}
+  \eem{j_1&j_2 }{k_1&k_2 }{\tau} \eedr{j_3}{k_3}{\tau}
+  \eem{j_1}{k_1}{\tau} \eedr{j_2 &j_3}{k_2 &k_3}{\tau} \notag \\
&\quad
+  \eedr{j_1&j_2 &j_3}{k_1&k_2 &k_3}{\tau} + \delta \eem{j_1&j_2 &j_3}{k_1&k_2 &k_3}{\tau}
\label{shortmd3}
\end{align}
and spell out the shorter expressions for $\delta \eem{j_1&j_2 &j_3}{k_1&k_2 &k_3}{\tau}$ in the remainder of this section. 

\subsubsection*{Degree 12}

\begin{align}
     \delta \eem{j&j&j}{4&4&4}{\tau}&=0 \, , \ \ \ \ \ \ j=0,1,2 \, ,  \\
   \delta \eem{1&0&0}{4&4&4}{\tau}&=-\frac{1}{4320}
     \zeta^\dr_3 \eem{0}{4}{\tau}-\frac{7}{720} \zeta_3^{\dr}  \eem{0}{6}{\tau}\, ,   \notag \\ 
     \delta \eem{1&1&0}{4&4&4}{\tau}&= \frac{1}{4320}\zeta_3^\dr \eem{1}{4}{\tau}+\frac{7}{720} \zeta_3^{\dr}  \eem{1}{6}{\tau}\, ,   \notag \\
     \delta \eem{2&0&0}{4&4&4}{\tau}&=\frac{1}{3}
     \zeta^\dr_3\eem{0}{4}{\tau} \eedr{0}{4}{\tau} +\frac{1}{3}\zeta_3^\dr \eem{0&0}{4&4}{\tau}
     -\frac{7}{360} \zeta_3^{\dr}  \eem{1}{6}{\tau}\, ,  
     \notag \\
     \delta \eem{2&2&0}{4&4&4}{\tau}&=\frac{1}{18} \left(\zeta_3^\dr\right)^2  \eem{0}{4}{\tau}+ \frac{7}{360} \zeta_{3}^\dr  \eem{3}{6}{\tau} +\frac{1}{3} \zeta_3^\dr  \eem{2&0}{4&4}{\tau}\, . \notag
\end{align}

\subsubsection*{Degree 14}

\begin{align}
    \delta \eem{0&2&4}{4&4&6}{\tau}&=
    \frac7{360} \zeta^\dr_3 \eem{5}{8}{\tau} - \frac{1}{3}\zeta^\dr_3 \eem{0}{4}{\tau}\eedr{4}{6}{\tau}  -\frac1{5} \zeta^\dr_5 \eem{0&2}{4&4}{\tau}\nn\\
    &\quad +\frac1{75} 
    \left(\zeta_{3,5}^\dr +5 \zeta_3^\dr \zeta_5^\dr\right)
    \eem{0}{4}{\tau}\,,\\
\delta \eem{0&4&2}{4&6&4}{\tau}&=
    \frac1{180} \zeta^\dr_3 \eem{5}{8}{\tau} +\frac1{210}\zeta^\dr_3 \eem{2}{4}{\tau} \eedr{2}{4}{\tau}
    - \frac{1}{3}\zeta^\dr_3 \eem{0&4}{4&6}{\tau}\nn\\
    &\quad  +\frac1{210}\zeta^\dr_3 \eem{2&2}{4&4}{\tau} -\frac1{5} \zeta^\dr_5 \eem{0}{4}{\tau}\eedr{2}{4}{\tau}\nn\\
    &\quad 
    -\frac1{1260} \left(\zeta_3^\dr\right)^2 \eem{2}{4}{\tau} - \frac1{75}     
    \zeta_{3,5}^\dr
    \eem{0}{4}{\tau}\,,\\
    \delta \eem{4&0&2}{6&4&4}{\tau}&=
    -\frac1{40} \zeta^\dr_3 \eem{5}{8}{\tau} -\frac1{210}\zeta^\dr_3 \eem{2}{4}{\tau} \eedr{2}{4}{\tau} - \frac1{210}\zeta^\dr_3 \eem{2&2}{4&4}{\tau}\nn\\
    &\quad - \frac{1}{3}\zeta^\dr_3 \eem{4&0}{6&4}{\tau}  +\frac1{5}\zeta^\dr_5 \eem{0&2}{4&4}{\tau} + 
    \frac1{5} \zeta^\dr_5 \eem{0}{4}{\tau}\eedr{2}{4}{\tau}\\
    &\quad 
    +\frac1{1260} 
    \left(\zeta_3^\dr\right)^2
    \eem{2}{4}{\tau} - \frac1{15}     
     \zeta_3^\dr \zeta_5^\dr
     \eem{0}{4}{\tau} \,.\nn
\end{align}

These results are consistent with the shuffle relation~\eqref{shufflmul} using
\begin{align}
    \delta \eem{0&2}{4&4}{\tau}= -\frac13\zeta^\dr_3 \eem{0}{4}{\tau}\,.
\end{align}
Moreover, we used the $f$-alphabet isomorphisms
\begin{align}
\rho^{-1}(f_3f_5)\big\vert_\dr= \frac15 \zeta_{3,5}^\dr + \zeta_3^\dr\zeta_5^\dr\,,\quad
\rho^{-1}(f_5f_3)\big\vert_\dr =-\frac15\zeta_{3,5}^\dr \,,
\end{align}
where $|_\dr$ indicates the subsequent projection to de Rham MZVs.

\section{Examples for antipodes of iterated Eisenstein integrals}
\label{appendixapd}

This appendix lists simple examples of the antipode of iterated Eisenstein integrals that follow from the generating series in (\ref{altsv.06}). At modular depth one and two, we have
\begin{align}
\ant \eedr{j_1}{k_1}{\tau} = -  \eedr{j_1}{k_1}{\tau}
\end{align}
as well as
\begin{align}
\ant \eedr{j_1 &j_2}{k_1 &k_2}{\tau} &=  \eedr{j_2 &j_1}{k_1 &k_2}{\tau} + \delta_{j_1,k_1-2} \frac{\zeta^\dr_{k_1-1}}{k_1{-}1} \eedr{j_2}{k_2}{\tau} - \delta_{j_2,k_2-2} \frac{\zeta^\dr_{k_2-1}}{k_2{-}1} \eedr{j_1}{k_1}{\tau}  \notag \\
  &\quad - \theta_{k_1<k_2} \zeta^\dr_{k_1-1} \stfdr{j_1&j_2}{k_1&k_2}{\tau}+  \theta_{k_2<k_1} \zeta^\dr_{k_2-1} \stfdr{j_2&j_1}{k_2&k_1}{\tau} \, ,
\end{align}
where
\begin{align}
\stfdr{j_1&j_2}{k_1&k_2}{\tau} &= 
\frac{(-1)^{j_1}  (k_2 {-} k_1 {+} 1)!  j_2! (k_2 {-} j_2 {-} 
      2)!\, {\rm BF}_{k_2}}{(k_1 {-} 
        1)   (k_2 {-} 1)! (k_2 {-} j_1 {-} j_2 {-} 2)! (j_1 {+} j_2 {-} k_1 {+} 
         2)! \, {\rm BF}_{k_2 - k_1 + 2}} \eedr{j_1 {+} j_2 {-} k_1 {+} 2}{k_2 {-} 
      k_1 {+} 2}{\tau} \, .
\label{nwaritd2}
 \end{align}
The corrections involving $\zeta^\dr_{2\ell+1}$ can be formally obtained from those in the coaction formula (\ref{clform}) at modular depth two through the formal replacements $({\cal E}^\mm,{\cal E}^\dr) \rightarrow (-{\cal E}^\dr,0)$ and $\zeta^\dr_{2\ell+1} \rightarrow -\zeta^\dr_{2\ell+1}$. These substitution rules translate the relevant contributions to the respective generating series $(\mathbb{M}^{\dr}_\sigma)^{-1}\Ip^\mm (\epsilon_k;\tau)\mathbb{M}^{\dr}_\sigma$ and $\mathbb M_\sigma^\dr ( \mathbb I^\dr(\epsilon_k;\tau) )^{-1} ( \mathbb M_\sigma^\dr)^{-1}$ into one another.

\end{appendix}


\providecommand{\href}[2]{#2}\begingroup\raggedright\endgroup

\end{document}